\documentclass[traditabstract]{aa}
\usepackage{amsmath}
\usepackage{txfonts}
\usepackage{ifpdf}

\ifpdf
\usepackage[pdftex]{graphicx}
\usepackage{epstopdf}
\usepackage[pdftex,colorlinks=false,pdfstartview=Fit,urlcolor=pdfwww,
  bookmarks=true,bookmarksnumbered=true]{hyperref}

\else
\usepackage[dvips]{graphicx}
\usepackage[dvips]{hyperref}

\fi
\usepackage{url}
\usepackage{framed}
\usepackage{natbib,ifthen}

\providecommand{\sorthelp}[1]{}

\bibpunct{(}{)}{;}{a}{}{,} 

\newcommand{\fnurl}[1]{\footnote{\url{#1}}}

\newcommand{\artdeco}{\texttt{artDeco}}
\newcommand{\Artdeco}{\texttt{ArtDeco}}
\newcommand{\Madam}{\texttt{Madam}}
\newcommand{\lmax}{\ensuremath{\ell_{\text{max}}}}
\newcommand{\kmax}{\ensuremath{k_{\text{max}}}}

\newcommand{\nside}{\ensuremath{N_{\text{side}}}}
\newcommand{\npsi}{\ensuremath{N_{\text{psi}}}}
\newcommand{\ndof}{\ensuremath{N_{\text{dof}}}}
\newcommand{\Planck}{\textit{Planck}}

\newcommand{\Cn}{\tens{C}_{\rm n}}
\newcommand{\Ct}{\tens{C}_{\rm t}}

\newcommand{\Cb}{\tens{C}_{\rm b}}
\newcommand{\Cwn}{\tens{C}_{\rm wn}}

\newcommand{\Amat}{\tens{A}}
\newcommand{\Pmat}{\tens{P}}
\newcommand{\Zmat}{\tens{Z}}
\newcommand{\Fmat}{\tens{F}}
\newcommand{\Smat}{\tens{S}}
\newcommand{\Xmat}{\tens{X}}
\newcommand{\Gmat}{\tens{G}}
\newcommand{\unimat}{\tens{I}}
\newcommand{\deltal}{$\Delta\ell$}

\newcommand{\axlm}{$a_{X\ell m}$}
\newcommand{\aslm}{$a_{s\ell m}$}

\newcommand{\nsided}{$N_{\text{side}}^{\text{3D}}$}
\newcommand{\npsid}{$N_\text{psi}^\text{3D}$}
\newcommand{\chid}{$\chi^2$}

\newcommand{\muK}{$\mu$K}

\begin{document}

\title{Impact of beam deconvolution on noise properties in CMB measurements:
Application to Planck LFI}

\author
{E.~Keih\"anen\inst{1},
K.~Kiiveri\inst{1,3},
V.~Lindholm\inst{1,3},
 M.~Reinecke\inst{2}
 \and
 A.-S.~Suur-Uski\inst{1,3}}

\institute{
Department of Physics,
Gustaf H\"allstr\"ominkatu 2,
00014 University of Helsinki, Finland \\
\email{elina.keihanen@helsinki.fi}
\and
Max-Planck-Institut f\"ur Astrophysik, Karl-Schwarzschild-Str.~1, 85741 Garching, Germany\\
\email{martin@mpa-garching.mpg.de}
\and
Helsinki Institute of Physics,
Gustaf H\"allstr\"ominkatu 2, 00014 University of Helsinki, Finland}

\abstract{
We present an analysis of the effects of beam deconvolution on noise properties in CMB measurements.
The analysis is built around the \artdeco\ beam deconvolver code.
We derive a low-resolution noise covariance matrix that describes the residual noise
in deconvolution products, both in harmonic and pixel space.
The matrix models the residual correlated noise that remains in time-ordered data after destriping,
and the effect of deconvolution on this noise.
To validate the results, we generate noise simulations that mimic the data from the \Planck\ LFI instrument.
A $\chi^2$ test for the full 70 GHz covariance in multipole range $\ell=0-50$ yields a mean reduced $\chi^2$ of 1.0037.
We compare two destriping options, full and independent destriping, when deconvolving subsets of available data.
Full destriping leaves substantially less residual noise, but leaves data sets intercorrelated.
We also derive a white noise covariance matrix that provides an approximation of the full noise at high multipoles,
and study the properties on high-resolution noise in pixel space through simulations.
}


\authorrunning{E. Keih\"anen et al.}
\titlerunning{Impact of beam deconvolution on noise properties}

\maketitle

\section{Introduction}

Several methods have been proposed for deconvolution mapmaking for CMB 
(cosmic microwave background)  measurements
\citep
{armitage-wandelt-2004,armitage-caplan2009,harrison-etal-2011,keihanen2012}.
The aim of deconvolution mapmaking is to produce a map that is free from effects of beam asymmetry.
For instance,  the image of a point source in a deconvolved map appears symmetric,
instead of being elongated according to the beam shape.
More importantly from the point of view of cosmology,
deconvolution mapmaking produces a map with a unique effective beam window
that takes the same form in all map pixels. 
This happens at the cost of more complicated noise structure.
A simple binning operation keeps white noise uncorrelated: 
white noise in the input time-ordered information (TOI) translates into uncorrelated
noise in the binned map.
This is not the case for deconvolved maps. The deconvolution operation creates correlations between pixels.
Correlated noise present already in the input TOI complicates the noise structure further.

In this work we study the impact of beam deconvolution on noise properties in CMB 
maps. We build our treatment around one particular deconvolution code, \artdeco.
The \artdeco\ code \citep{keihanen2012} was developed
for beam deconvolution of absolute CMB measurements.
The code takes as input the time-ordered data stream along with pointing information 
and a beam description, and produces as primary output the harmonic coefficients \axlm\ ($X=T,E,B$) of the sky.
From these coefficients one can further construct a sky map that is free from beam asymmetry effects.
The computation is carried out in harmonic space.  In doing so the method breaks the assumption
shared by pixel-based mapmaking methods
that the sky signal is constant within a pixel.

We investigate two kinds of input noise: pure white noise and destriped $1/f$ noise.
The starting point in the latter is a situation where TOI is first destriped with
a general destriper with noise prior. One source of such data is the \Madam\ map-making code
 \citep{keihanen2005,keihanen2010},
which we use in this work.
The code subtracts an estimate of the correlated noise component from the TOI
and yields a cleaned TOI stream where residual noise is dominated by white noise.
This cleaned TOI is then provided as input to the \artdeco\ beam deconvolver.

The noise in the output map is correlated for two reasons:
\begin{itemize}
\item There is residual correlated noise in the TOI stream that cannot be entirely removed by destriping techniques.
\item The deconvolution process itself changes the noise properties.
\end{itemize}
We aim at deriving a noise covariance matrix (NCVM) that captures both phenomena.

We validate our results using simulated \Planck\ LFI data.
However, we aim at keeping the method more general.
The required conditions are as follows:
\begin{itemize}
\item The experiment records the absolute signal, not differential.
\item The experiment offers full or nearly full sky coverage.
\item Beam shapes are known.
\item Sky sampling is dense compared to beam size.
\item Noise is piecewise stationary and its spectrum is known.
\item The data is cleaned of correlated noise through destriping or
another method with equivalent results.
\end{itemize}
We do not require a particular scanning pattern, or a particular noise spectrum.

Our calculations are carried out in harmonic space,
which is the natural domain for beam deconvolution.
We construct a harmonic noise covariance matrix that describes the
noise properties of the harmonic \axlm\ coefficients, which
are the primary output of \artdeco.
Because the problem is huge, it is not possible to construct the covariance matrix for all multipoles.
In practice, the noise covariance matrix approach is limited to multipoles $\ell\lessapprox100$.

We further produce a pixel noise covariance matrix,
which describes the noise properties of deconvolved sky maps, constructed from the $a_{Xlm}$ coefficients
through spherical harmonic transform.
A noise covariance matrix is an essential ingredient in many CMB power spectrum estimation methods.
A method for producing a noise covariance matrix for un-deconvolved maps has been presented in 
\cite{keskitalo2010}.  Its application to \Planck\ LFI data is presented in \cite{planck2014-a07}.

The paper is structured as follows. 
In Sect. \ref{sec:highell_simulations} we perform a series of full-resolution simulations,
 to gain a general view of  the impact of deconvolution on residual noise.
In Sect. \ref{sec:whitenoise} we focus on white noise. We derive a white noise covariance matrix,
and compare its predictions with Monte Carlo simulations.
The main result of this paper is the derivation of a full low-resolution noise covariance matrix,
presented in Sect. \ref{sec:full_ncvm}. In Sect. \ref{sec:validation} we perform a thorough validation of the 
harmonic covariance matrix through Monte Carlo simulations.  
We show results from \chid\ tests and from noise bias comparison.
In Sect. \ref{sec:pixelspace} we perform analysis in pixel space.
We summarise our results in Sect. \ref{sec:conclusions}.


\section{Simulations}
\label{sec:highell_simulations}

We start by performing a series of full-resolution noise simulations to assess the impact of
 deconvolution.
Our simulations mimic the data from the LFI instrument of \Planck\ experiment.
We created a time-ordered data stream, which we filled with simulated noise,
and fed it as input to the \artdeco\ deconvolver.
The simulations were run on Sisu, the Cray XC40 supercomputer of CSC, Finland.

\subsection{Generating noise}

We regenerated the LFI detector pointing with LevelS software \citep{reinecke2006}.
The simulated mission covers 4 years of data.
We used the internal noise generator of the \Madam\ map-making code
to generate a noise time stream. 
The noise properties mimic those of the real Planck data.
We considered the three LFI channels: 30 GHz, 44 GHz, and 70 GHz.
The noise parameters used in simulations 
were taken from \cite{planck2014-a03}.
For convenience, the parameters are listed in Table \ref{tab:noise_parameters}.


\begin{table}
\caption{Noise parameters used in simulations:
Knee frequency ($f_{\rm knee}$), slope ($\beta$), and white noise rms ($\sigma$).
From \cite{planck2014-a03}.
}
\label{tab:noise_parameters}
\begin{tabular}{l rr rr rr}
 & \multicolumn{2} {c}{$f_{\rm knee}$} & \multicolumn{2}{c} {$\beta$} & \multicolumn{2}{c} {$\sigma$}  \\
\noalign{\vskip-3pt}
& \multispan2\hrulefill \hskip5pt & \multispan2\hskip5pt \hrulefill \hskip5pt & \multispan2 \hskip5pt\hrulefill \\
Horn & M & S & M & S &  M & S \\
\hline\hline
  \noalign{\vskip 4pt}
  \omit{\bf 70\,\,GHz}\hfil\cr
  \noalign{\vskip 4pt}
LFI-18 & 14.82 & 17.78  & -1.060 & -1.180  & 4.553 & 4.146  \cr
 LFI-19 & 11.72 & 13.71  & -1.207 & -1.111 & 5.144 & 4.926  \cr
LFI-20 &  7.96 & 5.67 & -1.198 & -1.298 & 5.212 & 5.507     \cr
LFI-21 & 37.89 & 13.27 & -1.247 &-1.205 & 4.003 & 4.971  \cr
 LFI-22 & 9.68  & 14.80 & -1.424 & -1.237 & 4.356 & 4.715   \cr
 LFI-23 & 29.65  & 59.03 & -1.073 & -1.211  & 4.476 & 4.790   \cr
  \noalign{\vskip 5pt}
  \omit{\bf 44\,\,GHz}\hfil\cr
  \noalign{\vskip 4pt}
 LFI-24 & 26.78  & 88.30  & -0.942 & -0.908 & 3.159  & 2.734   \cr
LFI-25 & 20.07  & 46.37  & -0.845 & -0.904 & 2.834  & 2.698   \cr
LFI-26 & 64.42 & 68.19  & -0.918 & -0.758 & 3.295 & 2.887   \cr
   \noalign{\vskip 5pt}
  \omit{\bf 30\,GHz}\hfil\cr
  \noalign{\vskip 4pt}
 LFI-27   & 174.53 &  108.79 & -0.927 & -0.907 & 1.605  & 1.729  \cr
LFI-28   & 130.08 & 43.08 & -0.931 & -0.900  & 1.812 & 1.633  \cr
\hline\hline
\end{tabular}
\end{table}

We consider two types of noise.
In the first case we generated pure white noise, with variances taken from Table \ref{tab:noise_parameters}.
The variance is assumed constant in time, but varies between radiometers.
In the second series of simulations we generated realistic $1/f$ noise,
again with parameters from Table \ref{tab:noise_parameters}.
We destriped the noise stream with \Madam,
and stored the cleaned data on disk.
The data was then fed as input to \artdeco.

The destriping procedure follows the one applied to LFI mapmaking \citep{planck2014-a07}.
We applied the same destriping mask, and used the same HEALPix map resolution $N_{\rm side}$=1024 (3.44'), 
and same baseline length (1s for 70 GHz and 0.25 s for 30 GHz)
as in real \Planck\ mapmaking.
We refer to these simulations as realistic noise simulations,
in contrast to the white noise simulations.
In both simulations we applied realistic \Planck\ flags to the TOI stream.
This excludes roughly 8\% of the data.
\Madam\ has the capacity of storing the destriped TOI
on disk in ``4D map'' format.
The four dimensions in this context refer to the three angles
$\theta,\phi,\psi$ which define the detector pointing,
and pointing period index as fourth dimension.
Pointing period refers to a period of typical duration of 40 min,
where the \Planck\ scanning pattern follows a fixed ring on the sky.

\Madam\ generates a three-dimensional grid by discretising the 
detector pointing angles.
The pixelization obeys {\tt HEALPix} pixelization in $\theta$ and $\phi$,
and the $\psi$ angle is divided uniformly into \npsi\ bins.
Another parameter \nside\ controls the resolution in $\theta,\phi$.
The TOI is binned on the grid according to its pointing,
and the grid is stored by pointing period.
The data volume is compressed typically by a factor of 20 compared to the full TOI.
Radiometer-specific parameters required in subsequent analysis steps
are stored as keywords in the 4D object header.
For the present study we used resolution \nside=1024, \npsi=4096.
At this resolution the 4D maps take roughly 8 GB of disk space per detector.

In white noise simulations, we use \Madam\ only to generate the noise, 
to apply the flags, and to store the data stream in 4D map format. No destriping is applied.

The 4D map is an intermediate data object that is useful for many purposes.
In particular, it can further be compressed into the ``3D map'' format
that serves as input to \artdeco.  
This step is trivial, it consists of coadding all pointing periods on a single grid.
Further, we can construct a binned map from the same 4D map input.
The binning operation represents ``normal'' mapmaking without beam deconvolution.
The combined operation of destriping, 4D map production, and map binning,
is nearly equivalent to the operation of producing a destriped map directly by \Madam.
The only difference is the discretization of the $\psi$ angle,
which at the chosen resolution \npsi=4096 has a negligible effect on the final map.

To reduce the scatter in results, we produced several noise realizations for each case studied.
At 30 GHz, we generated 40 realizations for each case.  At 44 GHz and 70 GHz,
where the simulations are more demanding, we produced 10 realizations.
The spectra shown are averaged over available realizations.
A complete list of simulations is given in Table
\ref{tab:MC_parameters}.

\subsection{Destriping options}
\label{sec:destriping_options}

The full simulated TOI consists of four years of data for all LFI radiometers.
In cases where we deconvolve only a subset of data from a frequency channel,
we have two options for the destriping step.
Either we can use the full data set for destriping,
and then pass a subset of the cleaned TOI to deconvolution,
or we can use the same data subset in both processing steps.
We refer to the two options as {\em full destriping} and {\em independent destriping}, respectively.
Full destriping leads to a smaller residual noise level in the final data products,
as there is more information available for the destriping solution and removal of correlated noise.
On the other hand, independent destriping has the benefit over full destriping
that the residual noise in the final products is uncorrelated from one data set to another.

We compared the two destriping options for horn pair 18/23.
This represents one third of the full 70 GHz data set.
The horn pair data set was selected for closer examination
on the basis of low-resolution simulations (Sect. \ref{sec:validation}), 
where it showed as the ``worst case'' subset.

\subsection{Binned maps}
\label{sec:binnedmaps}

We compare the deconvolved spectrum to the {\em binned map spectrum},
 which we constructed as follows.  We bin the same destriped TOI into a sky map 
as
\begin{equation}
   \vec m= (\Pmat^{\rm T}\Cn^{-1}\Pmat)^{-1} \Pmat\Cn^{-1}\hat{\vec y}
\end{equation}
Here $\Pmat$ is the pointing matrix,
$\hat{\vec y}$ is the input time-ordered data (TOI),
 and $\Cn$ is a diagonal 
white noise covariance.
If several detectors are involved,
their data are appended to a single TOI stream.

Binning represents the usual mapmaking process, which does not attempt
to correct for the beam shapes.
We then compute the pseudo spectrum of the map with the {\tt Anafast} tool of the {\tt HEALPix} package.
We refer to the spectrum obtained this way as map spectrum.
A comparison of the map spectrum and deconvolved spectrum gives insight to the impact of the
deconvolution process on noise.
We perform the map-binning operation starting from the same 4D map objects
that serve as input to the deconvolution procedure.
That way we can be sure that we have exactly the same inputs in both procedures.

We deviate from the mapmaking procedure applied in actual \Planck\ analysis
in one aspect.
LFI mapmaking applies the horn-uniform radiometer weighting scheme,
when combining several radiometers into one map \citep{planck2014-a07}.
Horn-uniform weighting reduces leakage of temperature signal to polarization
through beam shape mismatch,
at the cost of slightly increased noise compared to the other option,
noise weighting.
In noise weighting, the radiometer weights are constructed as 
$w_j = 1/\sigma_j^2$
where $\sigma_j$ is the white noise rms for radiometer $j$, from Table \ref{tab:noise_parameters}.
In horn-uniform weighting the weights are made equal for the M and S radiometers of same horn.
Because the beam shapes are explicitly accounted for in deconvolution,
horn-uniform weighting provides no benefit over noise weighting, and \artdeco\ internally applies noise weighting.
Since our aim is to study the effects of beam deconvolution,
and not those of different radiometer weighting schemes,
we build also the binned maps with noise weighting.
Horn-uniform weighting is, however, applied in the destriping phase.
This way we have two mapmaking methods that take as input the same exact destriped TOI,
and differ only in the way they deal (or do not) with the beam.
The detector weighting scheme has a small effect on the residual white noise level. 
We return to this in Sect. \ref{sec:resolution}.

\subsection{Deconvolution}

\Artdeco\  yields as output the harmonic coefficients $a_{T\ell m}$, $a_{E\ell m}$, $a_{B\ell m}$.
with $\ell=0\ldots$\lmax. Here \lmax\  is an input parameter that defines the multipole range considered.
Another input parameter \kmax\ sets the level of beam asymmetry taken into account.
The harmonic beam expansion is constructed for $b_{s\ell k}$, where $0<\ell\le\ell_{\rm max}$ and $|k|\le k_{\rm max}$.
The maximum \lmax\ that can be reached is dependent on the beam width.
For the full-resolution simulations we chose \lmax=700 for 30 GHz, \lmax=1000 for 44 GHz, and \lmax=1500 for 70 GHz.
In all cases we used \kmax=6.

The \artdeco\ deconvolution operation can formally be written as
\begin{equation}
  \vec a = (\Amat^{\dagger}\Cn^{-1}\Amat)^{-1}\Amat^{\dagger}\Cn^{-1}\hat {\vec y}  \label{deconv}
\end{equation}
The solution is built on the assumption that the noise in the input TOI
is white, or at least dominated by white noise.
Matrix $\Amat$ is given by \citep{keihanen2012},
\begin{equation}
   A^j_{s\ell m} = \sum_k b^\ast_{s\ell k}D^{\ell\ast}_{mk}(\omega_j),
\end{equation}
where $b$ is the harmonic representation of the beam,
$D$ is a Wigner function, which takes as argument the pointing angles
$\theta$, $\phi$, and $\psi$,
and $\omega_j$ is a short for the triplet of pointing angles for sample $j$.

The spin index $s$ takes values 0 (temperature) and $\pm2$ (polarization).
Index $k$ takes values from 0 to $\pm\kmax$. 
Indices $\ell,m$ are the the usual indices of spherical
harmonics, $\ell$ being related to angular scale and $m$ to orientation.
In this notation, the output harmonic coefficients are written as \aslm\ and the input TOI stream as $y_j$.
The $E,B$ coefficients are further constructed as
\begin{eqnarray}
a_{E\ell m} &=& -\frac12 (a_{2\ell m}+a_{-2\ell m})  \label{alm_conversion} \\
a_{B\ell m} &=& -\frac1{2i} (a_{2\ell m}-a_{-2\ell m}) \nonumber
\end{eqnarray}

In this work we are interested in the impact of deconvolution on noise.
Since both destriping and deconvolution are linear processes,
we can examine the signal and noise components in isolation.
From now on we assume that the input TOI consists of pure noise (white and correlated).
We compute the {\em deconvolved spectrum} from the output \axlm\ coefficients as
\begin{equation}
  \hat C_{XY\ell} = \frac{1}{2\ell+1} \sum_{m=-\ell}^\ell a_{X\ell m}^{\ast} a_{Y\ell m}   \label{MC_bias}
\end{equation}
where $X,Y$ stand for $T,E,B$.
We further average the spectrum over the available noise realizations.

\subsection{Beams}

Throughout this paper we use \Planck\ LFI beams \citep{planck2014-a05}.
We included in the deconvolution process the main and intermediate beam components.
The contribution of far sidelobes is assumed to be removed before the destriping step
 \citep{planck2014-a03}.

Table \ref{tab:beam_parameters} lists some characteristics of the beams per frequency.
The FWHM and ellipticity values are based on effective FEBeCoP beams \citep{mitra2010},
and include the main beam only.  
The values are taken from \cite{planck2014-a05}, and are shown here to give an idea of typical 
beam shape at each frequency.
Ellipticity is defined as  the ratio of beam widths along the two main axes of the beam. 
A symmetric beam will have ellipticity equal to 1.

For simulations we used the combined main and intermediate radiometer scanning beams.
The characteristics of these can be found in \cite{planck2014-a05}.
The last column of Table \ref{tab:beam_parameters} gives the beam efficiency at each frequency.
These are computed from of the harmonic beam expansion
of the noise-weighted frequency beam.  Beam efficiency is proportional to  the monopole component of the expansion,
and normalised so that the full $4\pi$ beam has efficiency 1. 
The squared value gives the effect in the spectral domain.
The missing fraction reflects the power absorbed by far sidelobes.


\begin{table}
\caption{Table of beam characteristics.
FWHM and ellipticity are based on the FEBeCoP effective beams 
 \citep{planck2014-a05}.
Beam efficiency includes the main and intermediate beam components.
}
\label{tab:beam_parameters}
\begin{tabular}{l rrr}
Frequency & FWHM & ellipticity & efficiency \\
\hline\hline
  \noalign{\vskip 4pt}
30 GHz & 32.193'' & 1.318 &  0.99031  \cr
44 GHz & 27.000'' & 1.035 &  0.99771 \cr
70 GHz & 13.213'' & 1.223 &  0.99124  \cr
\hline\hline
\end{tabular}
\end{table}



\begin{figure}
\includegraphics[width=1.0\columnwidth]{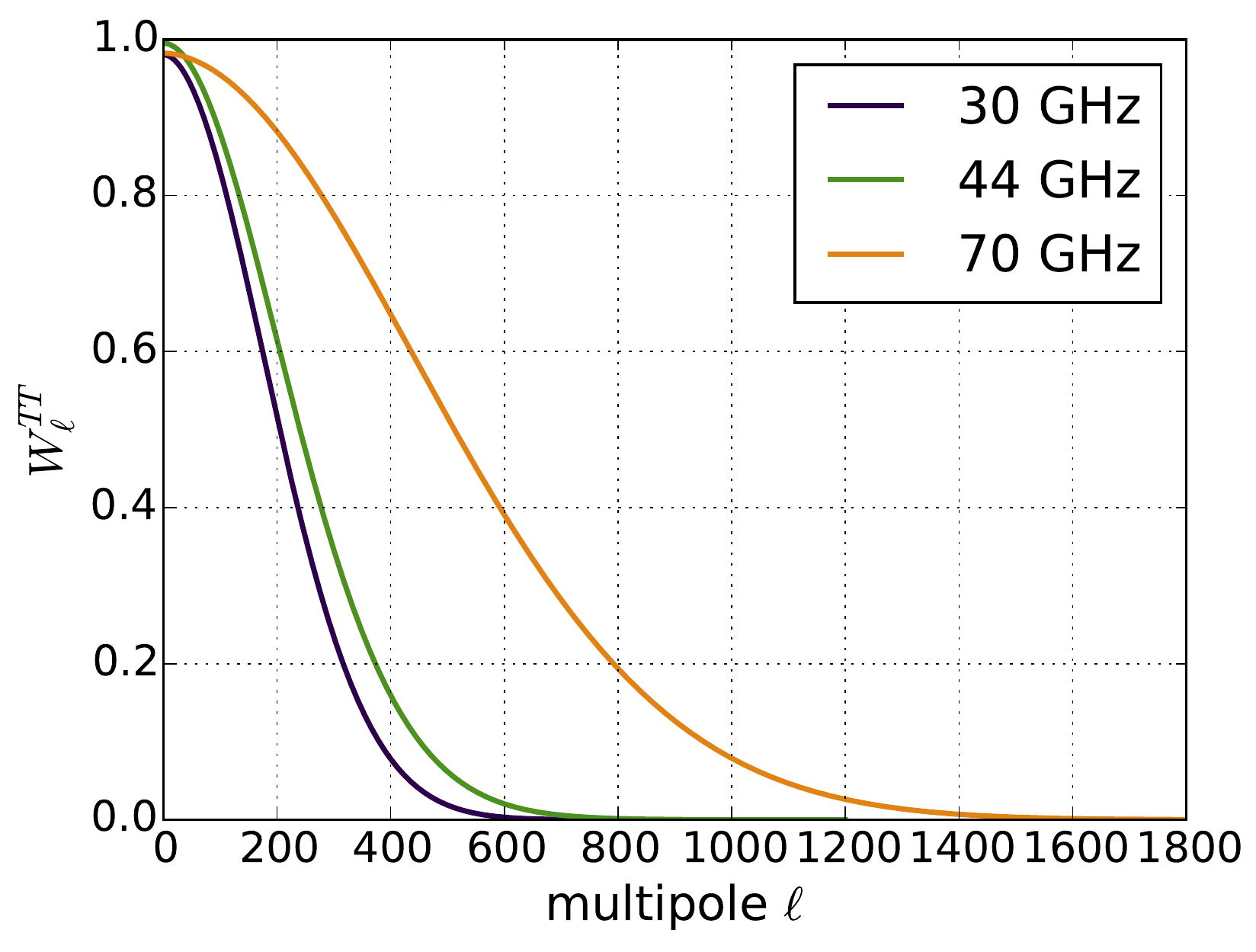}
\caption{Squared symmetrized TT beam window for 30, 44, 70 GHz.
{\it Solid}: definition of Eq. (\ref{beamwindow1}). {\it Dashed}: definition of Eq. (\ref{beamwindow2})}
\label{fig:beam_window}
\end{figure}

For comparison purposes we constructed a {\em symmetrized beam window} as follows.
First we computed the weighted sum individual radiometer beams, 
\begin{equation}
\bar b_{\ell m} = \left[\sum_{j} \frac{1}{\sigma_j^2}\right]^{-1} \sum_{j} \frac{1}{\sigma_j^2}  b^j_{\ell m} 
\end{equation}
where $j$  labels the radiometers, and $\sigma_j$ are their respective white noise rms.
The $m=0$ component of the beam expansion,
\begin{equation}
W_{\ell[m=0]} = \sqrt{\frac{4\pi}{2\ell+1}}\bar b_{\ell 0}  \label{beamwindow1}
\end{equation}
represents a beam that is symmetrized by averaging over the azimuthal angle
around the beam centre.  
Factor $4\pi/(2\ell+1)$ serves to normalize the window function to unity at $\ell=0$ for a beam with efficiency equal to 1.
The power spectrum of the beam expansion provides an alternative way of constructing a symmetrized beam,
as proposed by  \cite{wu2001},
\begin{equation}
W^2_{\ell [{\rm ms}]} =   \frac{4\pi}{2\ell+1} \sum _m |\bar b_{\ell m} |^2.  \label{beamwindow2}
\end{equation}
where ``ms" stands for mean of squares.
The beam window of Eq. (\ref{beamwindow2}) is always larger than that from Eq. (\ref{beamwindow1}), thus representing a narrower beam.
For LFI beams the difference is small.
The squared window functions $W_{l}^2$ for both definitions are plotted in Fig. \ref{fig:beam_window}.
The squared window can directly be compared with power spectra.

\subsection{Results from frequency simulations}


\begin{figure}
\includegraphics[width=0.95\columnwidth]{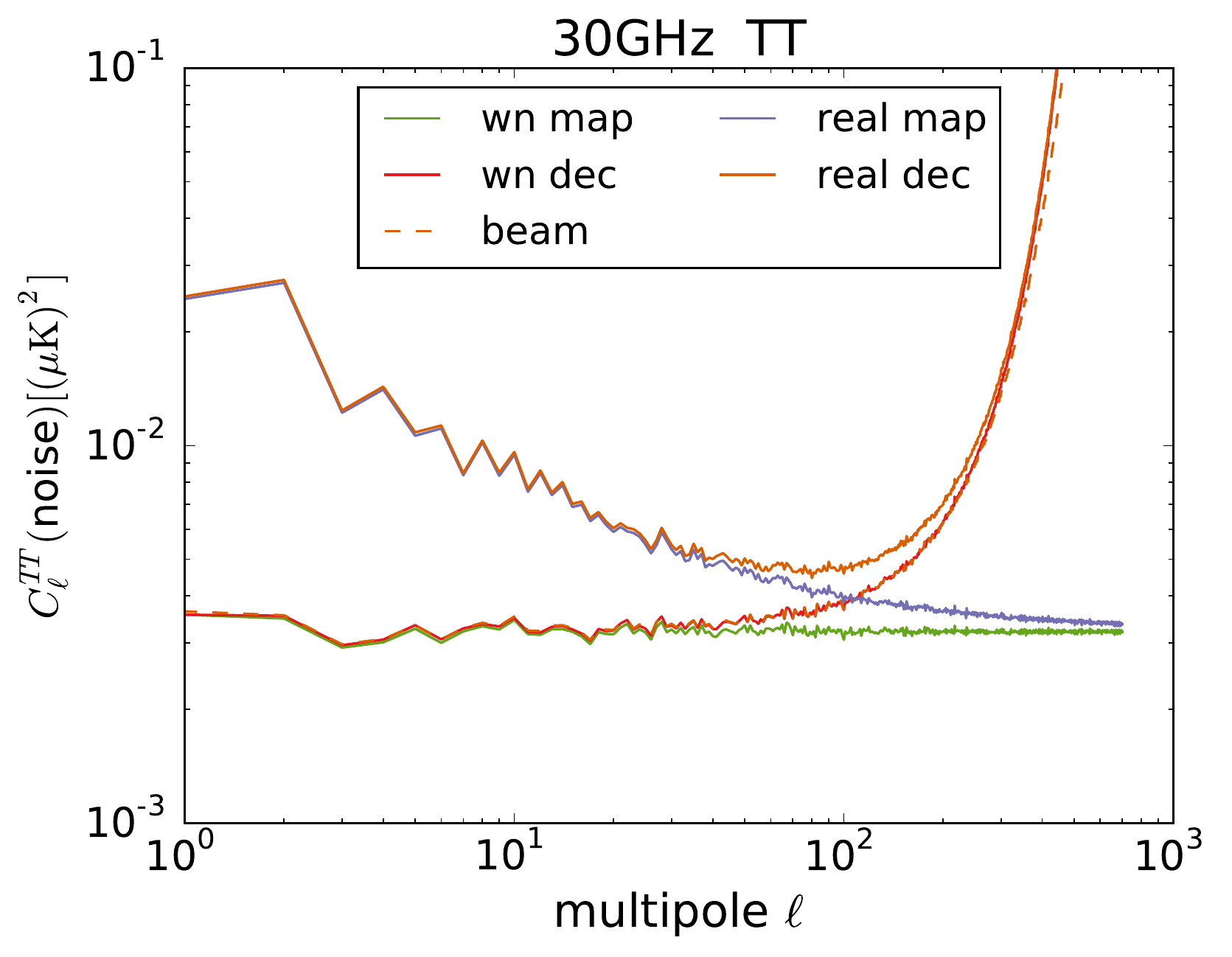}
\caption{TT spectrum at 30 GHz.  We show the deconvolved spectrum and map spectrum 
for white noise and realistic noise simulations.
The spectra are averaged over 40 noise realizations to reduce scatter.
Unsmoothed deconvolved spectra rise towards high multipoles, roughly as proportional
to the square of the inverse window function.
The dashed line shows the uniform white noise spectrum divided by the squared beam window.
}
\label{fig:highell_spectra}
\end{figure}

Figure \ref{fig:highell_spectra} shows the 30 GHz TT spectrum for white noise and realistic noise inputs.
All spectra have been averaged over 40 noise realizations.
We show in the same plot the deconvolved spectrum and the binned map spectrum.
The main effect from deconvolution is evident.
While the white noise map spectrum is uniform, 
the deconvolved spectrum rises steeply towards higher multipoles.
As a first approximation, the effect of deconvolution is to scale the
binned map spectrum by the inverse of the symmetric beam window squared (shown in dashed line type).
This is, however, only true as a first approximation. The real deconvolution spectrum rises even more 
steeply due to beam asymmetry. We investigate this further in Sect. \ref{sec:whitenoise}.

The primary deconvolution output is the harmonic coefficients $a_{Xlm}$.
For many practical purposes one may want to transform the result into an ordinary sky map.
The procedure of constructing a sky map from the deconvolved $a_{Xlm}$
coefficients involves a smoothing operation that brings the noise spectrum back down
to the level of un-deconvolved noise. We discuss this procedure
in Sect.~\ref{sec:pixelspace}.
In this and the following three sections, where we do analysis in harmonic space,
we always show the unsmoothed raw noise.
In harmonic space the smoothing is a simple $\ell$-dependent scaling operation which does not change
the relative differences between spectra.
The relative differences between spectra are thus more important than their absolute level.


\begin{figure}
\includegraphics[width=0.95\columnwidth]{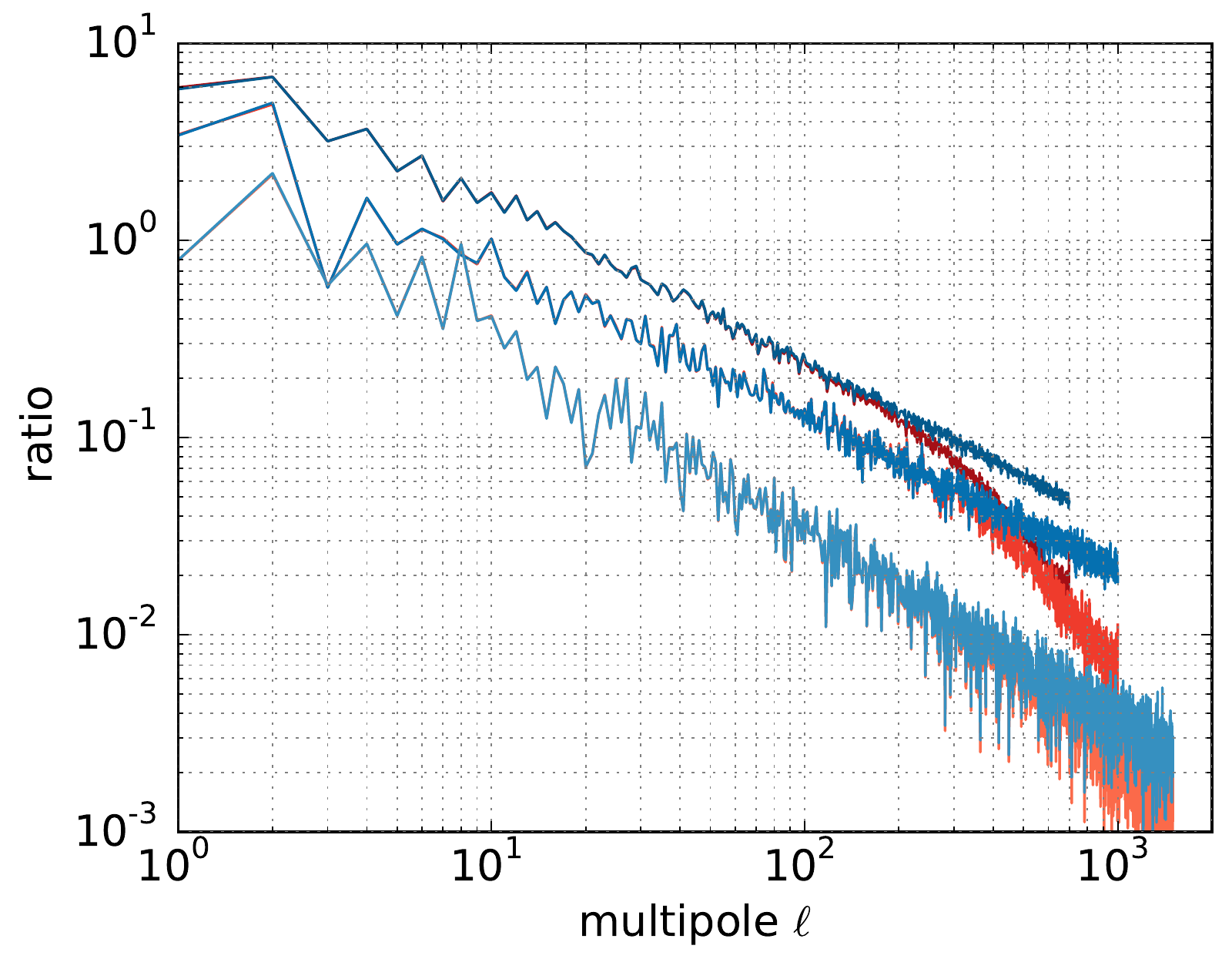}
\caption{Ratio of the residual correlated noise spectrum and the white noise spectrum,
for deconvolution (red) and ordinary map-making (blue).
From top down: 30 GHz, 44 GHz, 70 GHz.
}
\label{fig:residual_oof}
\end{figure}

Residual correlated noise dominates the lower multipoles,
but decreases towards higher multipoles.
Fig. \ref{fig:residual_oof} shows the ratio of correlated residual noise spectrum
and white noise spectrum.
The quantity shown 
is computed as $(C_{\ell\rm real}-C_{\ell\rm wn})/C_{\ell\rm wn}$,
where $C_{\ell\rm real}$ and $C_{\ell\rm wn}$ are the spectra from realistic
and white noise simulations, respectively.

An interesting and perhaps surprising observation is that deconvolution leaves relatively less correlated residual 
 noise on top of the white noise component at high multipoles, than does normal mapmaking.
This has important consequences.
The level of correlated residual noise is a measure of the error we make 
if we neglect the correlated component when estimating the residual noise.
In Sect. \ref{sec:whitenoise}
we derive a model for the deconvolved white noise spectrum,
which we use as an approximation for the full noise at high multipoles.
The accuracy of the approximation can be read from Fig. \ref{fig:residual_oof}.
Further,  MC simulations involving only white noise spectrum are much cheaper
than simulations with realistic noise,
as the destriping step is avoided. 

\begin{figure}
\includegraphics[width=0.95\columnwidth]{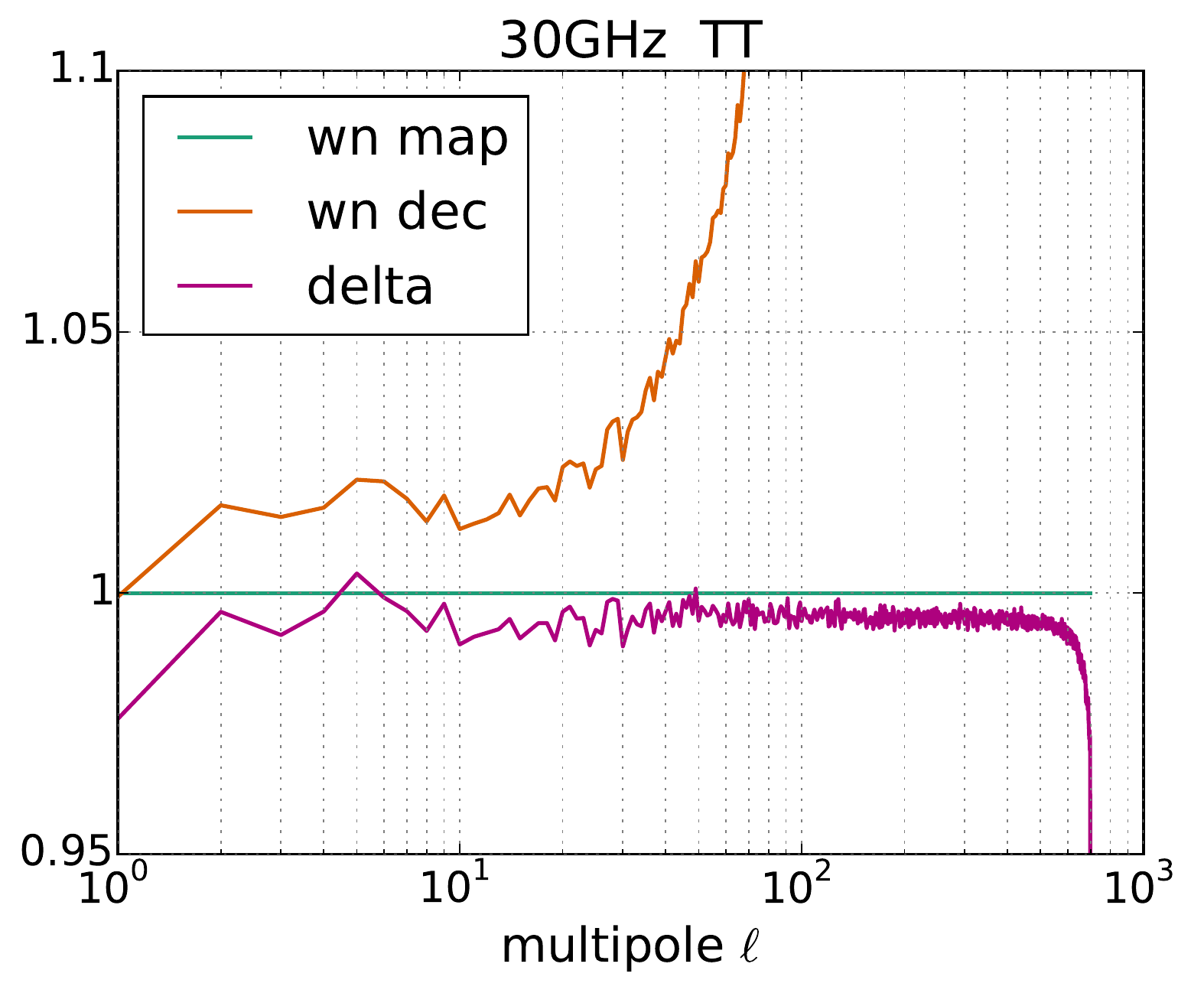}
\caption{Ratio of the deconvolved 30 GHz white noise $TT$ spectrum, 
and the corresponding map spectrum.
Orange: deconvolution with realistic beam. 
Purple: deconvolution with a delta beam.  
The deconvolved spectrum rises above the reference level due to missing beam power,
which deconvolution compensates for.}
\label{fig:highell_whitenoise_30GHz}
\end{figure}

We now look closer at the white noise spectrum.
The spectra vary strongly due to the limited number of realizations.
However, when comparing spectra derived from the same simulation with different methods, 
we can make use of the fact that we have the same noise realizations in both cases.
We take the white noise map spectrum as reference level,
and divide the other spectra by it.
This reduces the scatter in the spectra and shows the relative differences clearly.
Spectral ratios constructed this way are plotted in Figure \ref{fig:highell_whitenoise_30GHz}.
The white noise map spectrum is uniform by construction.
The shape of the deconvolved white noise spectrum reflects the beam window function.
The deconvolved spectrum rises above the binned map spectrum  at the lowest multipoles by approximately 2\%.
This is the effect of the missing power lost into the far sidelobe component of the beam.
(The $y=$99\% efficiency of Table \ref{tab:beam_parameters} translates
to an $1-y^2\approx$2\% effect in spectral domain.)
Deconvolution scales the signal up to correct for the missing power,
and the same increase is observed in the noise spectrum.

The purple line on Fig. \ref{fig:highell_whitenoise_30GHz}
is the result of a simulation where we deconvolved for 
a perfect delta beam (``deconvolution without deconvolution'').
Even if the beam is a delta peak, 
the deconvolution procedure is not identical to the combined map-binning and {\tt anafast} procedure.
Deconvolution performs a maximum-likelihood fit of the harmonic coefficients to the TOI,
while {\tt anafast} computes a harmonic expansion over the sky map, without taking into account the varying hit counts.
The two procedures are identical only in the case where the distribution of observations is uniform
over the sphere.  
The spectrum from deconvolution with a delta beam is slightly below the reference
level of the white noise map spectrum, but the difference is small.
Above $\ell=600$ the spectrum drops, which is a typical boundary effect in deconvolution.
The drop occurs at a much higher multipole than the drop in the spectral ratio of Fig. \ref{fig:residual_oof}.
The exact mechanism behind the latter drop remains unclear.


\subsection{Horn pair LFI18/23}

\begin{figure}
\includegraphics[width=0.95\columnwidth]{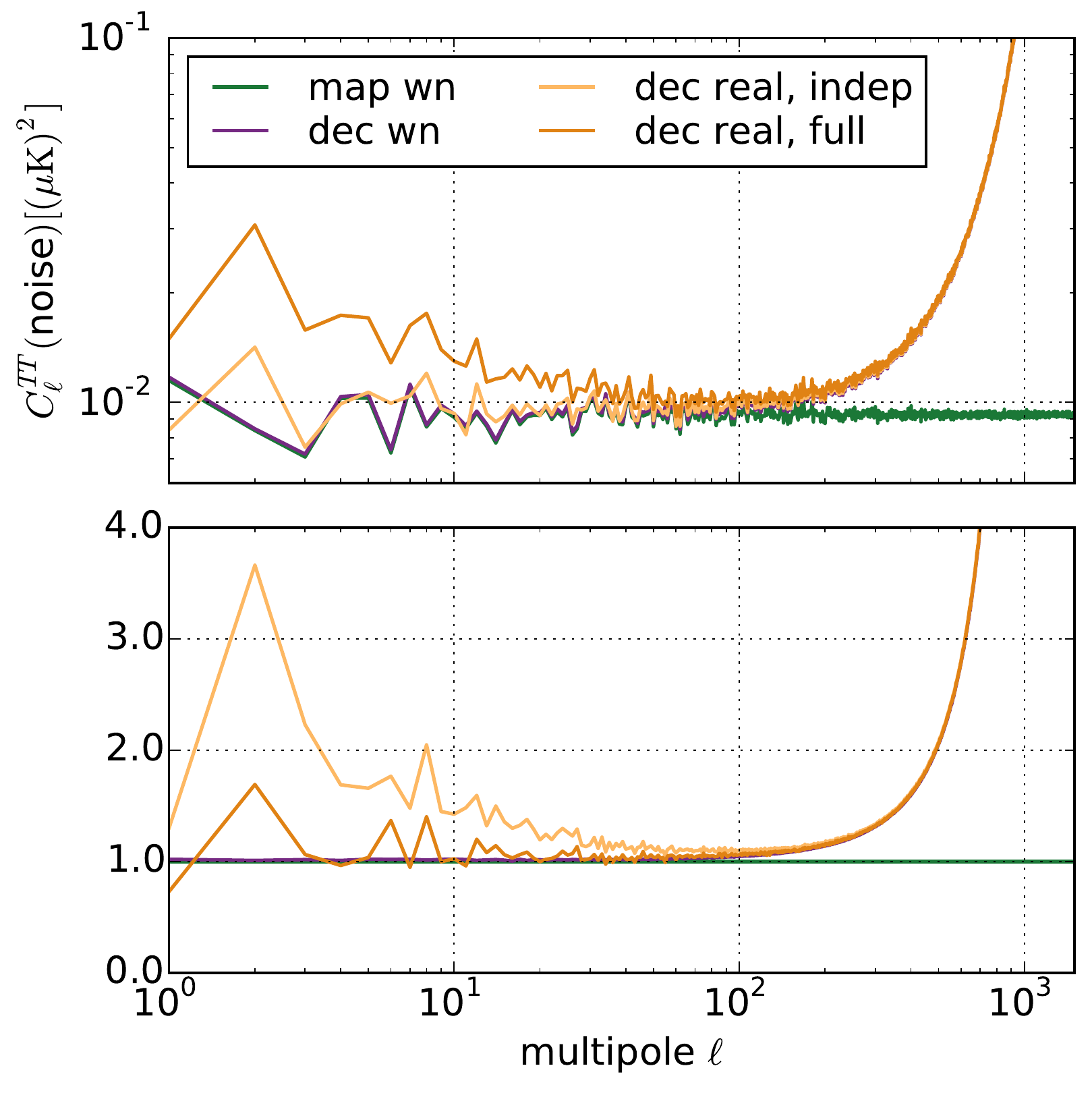}
\caption{$TT$ spectrum for horn pair LFI18/23.
We show the spectra for devonvolved white and realistic noise.
In the case of realistic noise 
we compare two destriping options: independent and full destriping.
The plot is based on 10 Monte Carlo realizations.
The lower panel shows the same spectra replotted in dimensionless units
 to bring out their relative differences;
The undeconvolved white noise map spectrum (green) is taken as reference level,
and all other spectra are divided by it.
}
\label{fig:highell_LFI1823}
\end{figure}

Finally we compare the two destriping options discussed in Sect. \ref{sec:destriping_options}.
We performed two simulations, where
we deconvolved the data from horn pair LFI18/23.
In the first set of simulations we used for destriping the same 18/23 horn-pair data set,
in the second one the full 70 GHz data set.
The results are shown in in Fig.  \ref{fig:highell_LFI1823}.
We show the deconvolved spectra for white noise and for realistic noise, with both destriping options.
The white noise map spectrum is shown in the same plot for reference.
The scatter is large due to the small number of realizations (10),
but the main conclusion is still clear. Full destriping leaves considerably less residual noise at low multipoles.

In the lower panel we  replot the spectra the same way we did for full 30 GHz data
in Fig. \ref{fig:highell_whitenoise_30GHz}.
We divided the spectra by the white noise map spectrum
to reduce scatter, and to bring out the differences more clearly.


\section{White noise covariance}
\label{sec:whitenoise}

\subsection{White noise covariance}

We proceed to study the properties of deconvolved noise in more detail.
The deconvolution operation of Eq. (\ref{deconv}),
when operating on a noise time stream $\vec n$,
yields a vector of harmonic coefficients whose
properties are described by the {\em harmonic noise covariance matrix} given by
\begin{equation}
 \tens{C} = (\Amat^{\dagger}\Cn^{-1}\Amat)^{-1}\Amat^{\dagger}\Cn^{-1} 
 \langle \vec n \vec n^{T}\rangle 
 \Cn^{-1}\Amat(\Amat^{\dagger}\Cn^{-1}\Amat)^{-1}\text{.}  \label{artcov}
\end{equation}
It describes the noise correlation between elements $a_{slm}$.

In this section we study the simpler case where the input TOI consists of pure white noise, 
with diagonal variance 
\begin{equation}
 \Cn = \langle \vec n \vec n^{T}\rangle\text{.}
\end{equation}
The harmonic noise covariance matrix simplifies into
\begin{equation}
\tens{C} = (\Amat^{\dagger}\Cn^{-1}\Amat)^{-1}\text{.} \label{whitenoise_ncvm}
\end{equation}
The noise covariance is thus given by the inverse of the {\em deconvolution matrix}
$(\Amat^{\dagger}\Cn^{-1}\Amat)$. 
This is a product we can extract from \artdeco.
The size of the matrix rapidly increases with increasing \lmax.
The rank of the matrix with polarization is $3(l_{\rm max}+1)^2$.
The memory requirement increases as square of the rank,
and the CPU time needed for inversion in the third power, or $l_{\rm max}^6$.
The full matrix inversion rapidly becomes prohibitive.
In practice we can only construct it for a limited multipole range.

\subsection{Noise bias}


\begin{figure*}
\includegraphics[width=2.0\columnwidth]{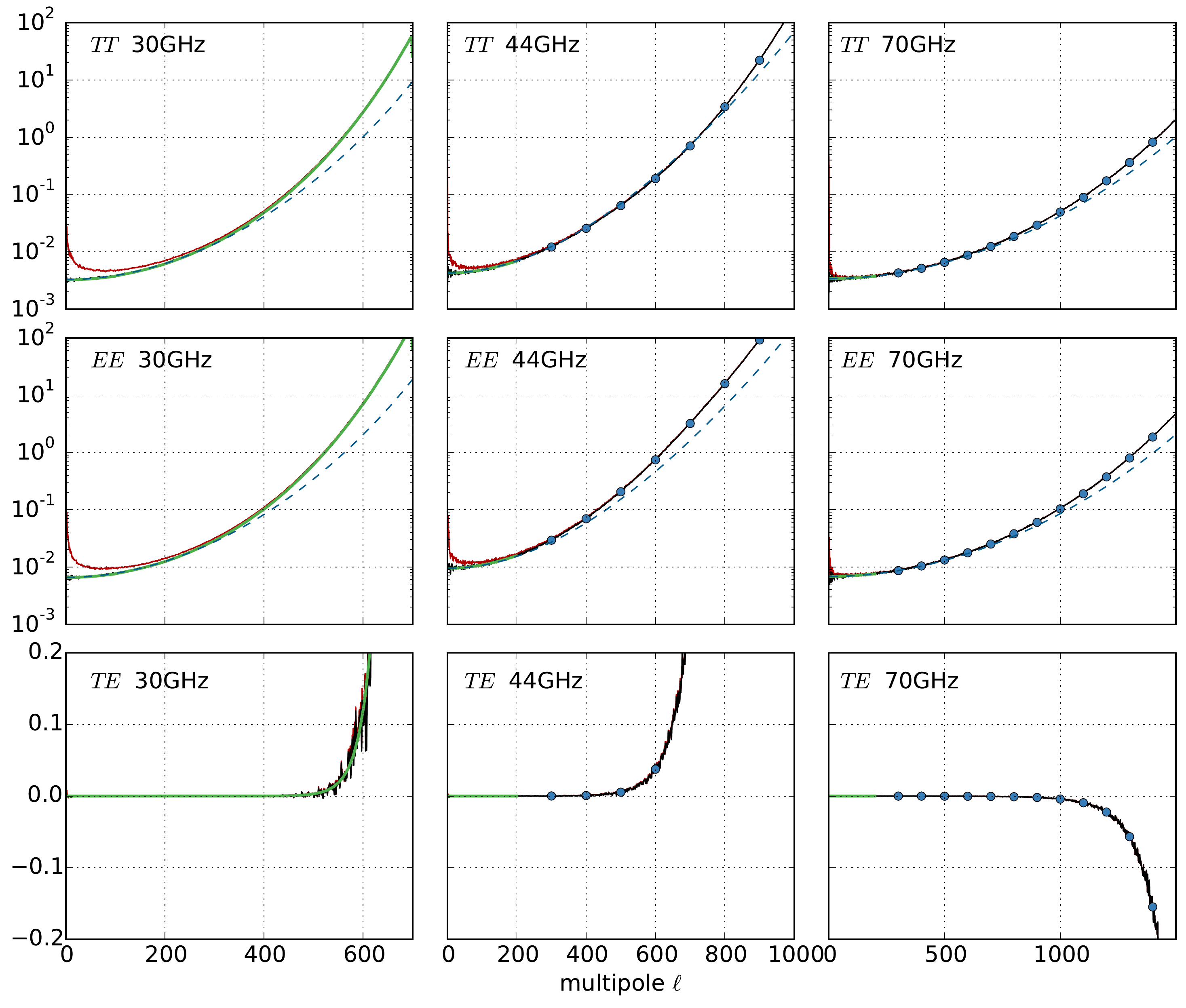}
\caption{$TT$, $EE$, and $TE$ deconvolved noise spectrum for 30, 44, and 70 GHz.
{\em Black}: white noise MC simulation.
{\em Red}: realistic MC simulation.
{\em Dashed}:  estimate based on inverse symmetrized beam window.
{\em Green}:  noise bias from white noise covariance matrix. 
At 30 GHz we construct the bias for the complete multipole range,
by a procedure that assembles the spectrum in pieces.
At 44 and 70 GHz we construct the bias for range $\ell$=0--200,
and individual values for multipoles $\ell=300,400\ldots1400$}
\label{fig:highell_whitenoise}
\end{figure*}


\begin{figure}
\includegraphics[width=0.95\columnwidth]{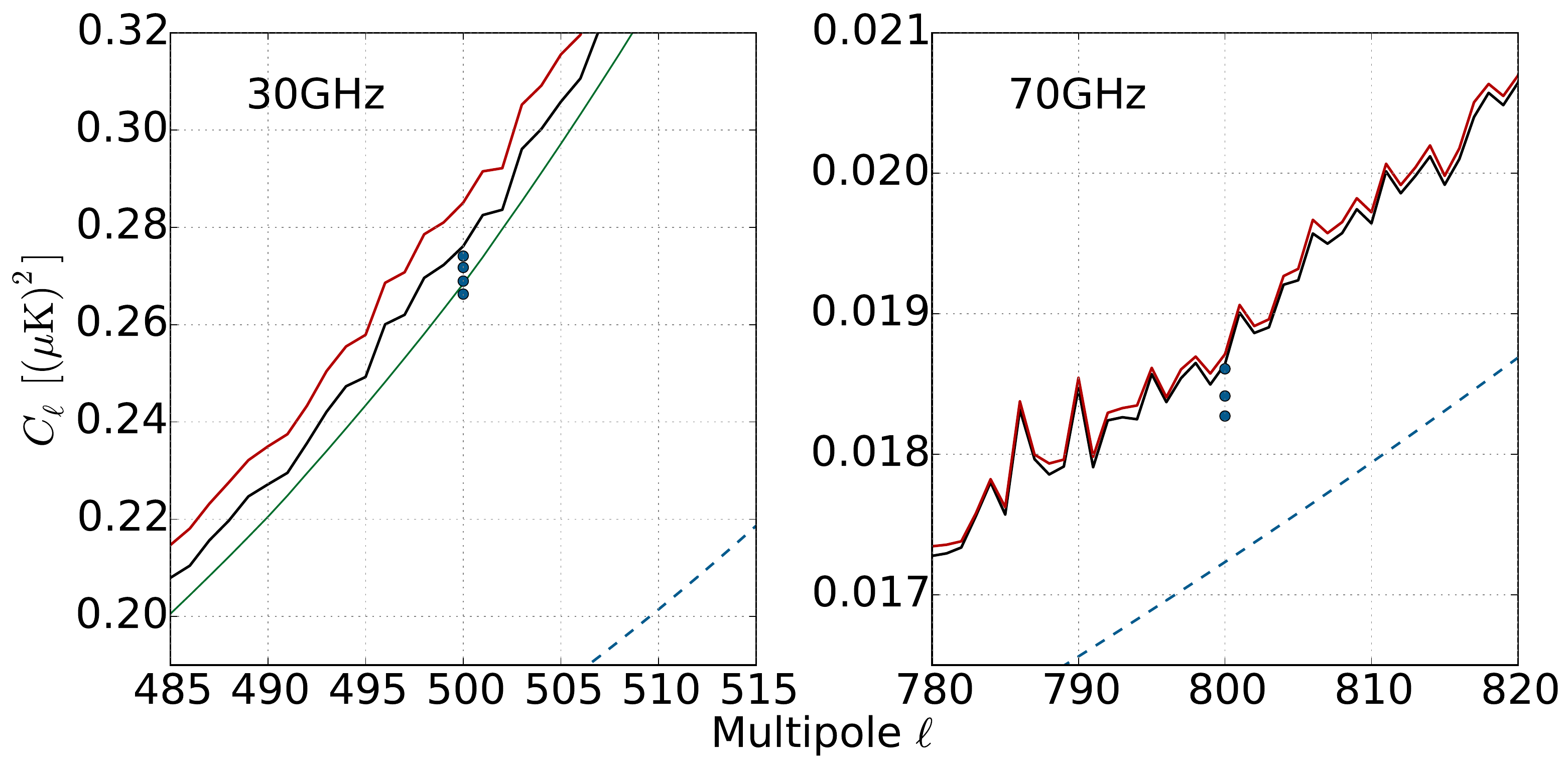}
\caption{Zoom to $TT$ bias at 30 GHz and 70 GHz.
{\em Red}: MC with realistic noise. 
{\em Black}: MC with white noise.
{\em Green} (30 GHz): piecewise constructed noise bias.
{\em Dashed}: estimate based on the beam window.
The circles indicate the bias for $\ell=$500 or 800,
constructed from a partial NCV matrix of width (from down up)  \deltal=5, 10, 20, 40 (30 GHz),
 or \deltal=5, 10, 20 (70 GHz) 
}
\label{fig:zoom_whitenoise}
\end{figure}


\begin{figure}
\includegraphics[width=0.95\columnwidth]{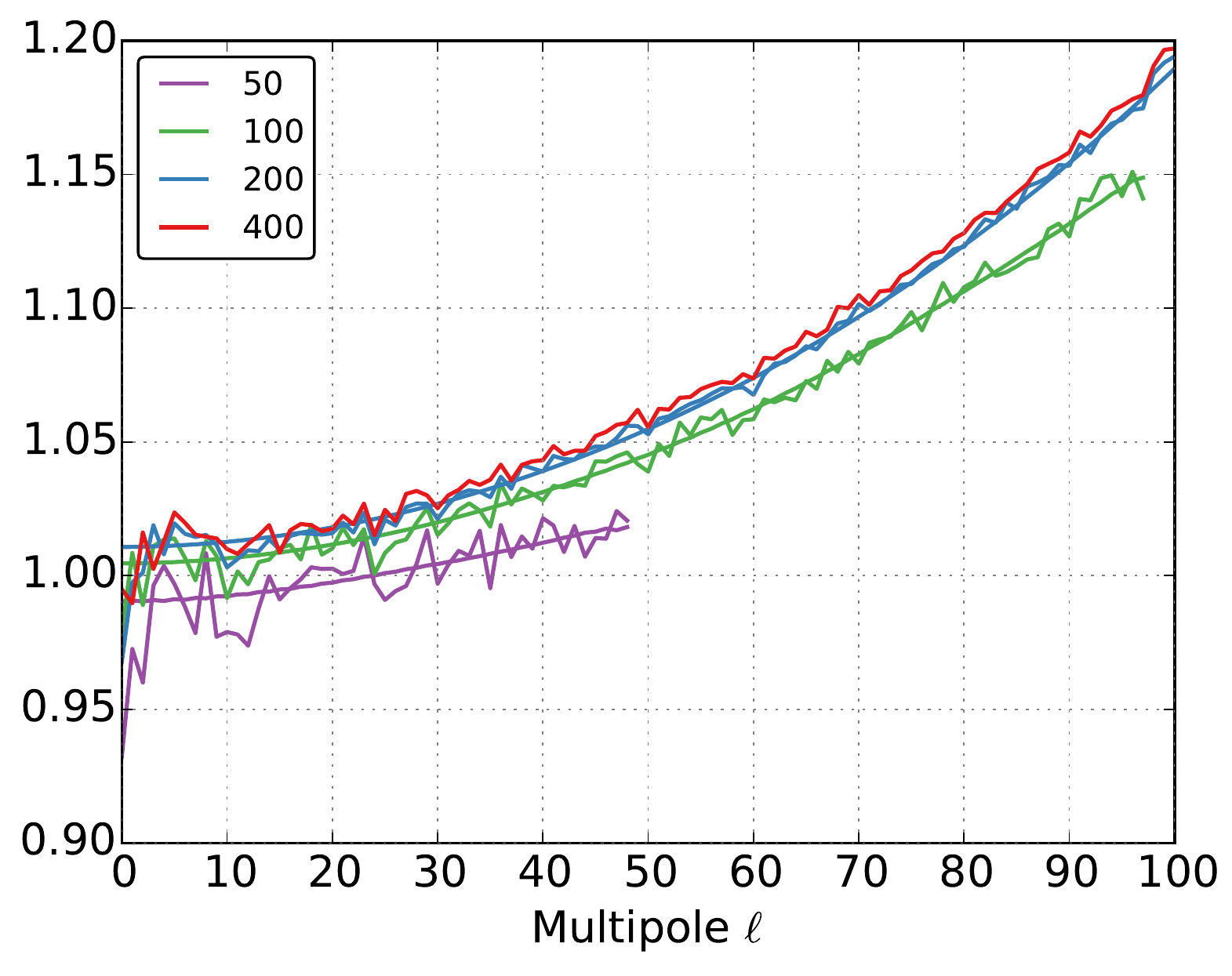}
\caption{Effect of \lmax\ parameter on 30 GHz $TT$ white noise spectrum.
We deconvolve the same data set with \lmax=50, 100, 200, 400.
To reduce scatter, we divide each spectrum by the white noise map spectrum,
and show them in dimensionless units.
The plot is based on 40 Monte Carlo realizations.
The smooth lines in same colour show the predicted noise bias, 
constructed from harmonic white noise covariance matrix,
for \lmax=50, 100, 200.
To bring them into units comparable with the Mont Carlo spectra, we divide them by the white noise level 
$3.2180\cdot10^{-15}$ K$^2$ (from Table \ref{tab:whitenoise}).
The last 2-5 unreliable multipoles are excluded for clarity.
}
\label{fig:bias_resolution}
\end{figure}

We want to compare the noise spectra from simulations
to the covariance matrix.
We derive from the covariance matrix a noise bias,
a prediction for the expectation value of the noise spectrum of Eq. (\ref{MC_bias}).

We define $g_{Xs}$ as factors that convert 
the spin harmonics $a_{s\ell m}$ into $a_{X\ell m}$, $X=T,E,B$.
We write Eq. (\ref{alm_conversion}) as
\begin{equation}
  a_{X\ell m} = \sum_s g_{Xs} a_{s\ell m}
\end{equation}
Now 
\begin{eqnarray}
g_{T,0} &=& 1 \nonumber  \\
g_{E,2} = g_{E,-2} &=&-1/2  \\
g_{B2} = -g_{B,-2} &=& i/2 \nonumber  \\ 
g_{T,2} = g_{T,-2} = g_{E,0} = g_{B,0} &=& 0 \nonumber .
\end{eqnarray}
The noise bias estimate is obtained from the noise covariance $\tens C$  as
\begin{equation}
  C_{XY\ell} = \sum_{ss'mm'\ell'\ell''}  \frac{1}{2\ell+1} g_{Xs} g_{Ys'}^\ast \delta_{\ell \ell'}\delta_{\ell \ell''}\delta_{mm'}
                     \tens{C}_{\ell'ms,\ell''m's'}
\end{equation}
Formally we can write
\begin{equation}
  C_{XY\ell} = {\rm Tr} [\tens{C} \tens G^{XY\ell}]   \label{noisebias_ncvm}
 \end{equation}
where
\begin{equation}
  \tens{G}^{XY\ell}_{\ell'ms,\ell''m's'} = \frac{1}{2\ell+1} g_{Xs} {g^{\ast}}_{Ys'} \delta_{\ell \ell'}\delta_{\ell\ell''}\delta_{mm'}
\end{equation}
The noise bias is extracted from a block-diagonal part of the NCVM,
where each $3\times3$ block corresponds to a pair of $(\ell,m)$ indeces,
and $s$ index takes all allowed values ($0,\pm2$).

Figure \ref{fig:highell_whitenoise} shows the $TT$, $EE$, and $TE$, noise spectra for three LFI frequencies. 
The $BB$ spectra are quite similar to $EE$, and are not shown.
We show the deconvolved white noise spectrum from MC simulation,
along with a noise bias estimate 
obtained from the white noise covariance.  We show also the realistic noise spectra,
which rise above the white noise spectra at low multipoles.
The white noise component dominates the noise at high multipoles.
Consequently, the white noise covariance provides a good model for the full noise at these high multipoles.
We show also the raw estimate based on the symmetric beam window.
It provides a reasonable approximation at intermediate multipoles,
but underestimates the noise at high multipoles.

Because of the large size of the deconvolution matrix, we cannot compute the exact noise bias for the full multipole range.
We constructed the 30 GHz bias in pieces.  First we constructed and inverted the matrix for range $\ell$ = 0 -- 110 to obtain the noise bias 
for multipoles $\ell$ = 0 -- 100. We dropped the last 10 multipoles to eliminate boundary effects.
Similarly, noise bias for multipoles $\ell$ = 100 -- 200 was constructed from a matrix computed for $\ell$ = 90 -- 210.
Multipoles $\ell$ = 200 -- 300 were constructed from two pieces in range $\ell$ = 190 -- 260 and $\ell$ = 240 -- 310.
Because the number of $m$ multipoles increases with increasing $\ell$,
we had to make the pieces narrower as we proceeded towards higher multipoles.
Range $\ell$ = 300 -- 400 was composed from pieces of 20 multipoles each, range $\ell$ = 400 -- 600 from pieces of 10, and
finally $\ell$ = 600 -- 700 from pieces of 5 multipoles.  In all cases above $\ell$ = 300 we included 5 extra multiples
at both ends, which we then dropped when combining the pieces.
The whole computation took 20\,000 CPU hours.
The estimate constructed this way is far more accurate than the one
based on symmetric beam window, as can be seen from Fig. \ref{fig:highell_whitenoise}. 
However, the bias still underestimates the MC noise spectrum by 2 -- 3\%
because it neglects correlations between distant multipoles. 
At 30 GHz this is not a major problem,
because we make an error of similar magnitude anyway by neglecting the correlated 
noise component (see Fig \ref{fig:residual_oof}).

The computation becomes increasingly heavy towards higher multipoles. 
We therefore adopted a different procedure for 44 GHz and 70 GHz.
We constructed the bias in one piece for multipoles $\ell$ = 0-200.
At higher multipoles we concentrated the computational resources on a number of individual multipoles,
for which we estimated the bias as accurately as possible.
We optimized the matrix inversion routine in such a way that it only computes the diagonal elements
actually needed for the bias.
We constructed the white noise covariance for the multipoles from $\ell_0$ - \deltal\ to $\ell_0+$\deltal,
and picked the central value $l_0$ for plotting.  Values of $\ell_0$ were taken to be multiples of 100.

At 44 GHz we used \deltal=30 for $\ell_0$ = 300--500,  and \deltal=20 for $l_0$ = 400 -- 800.
A fortunate coincidence comes to help here. The correlations between distant multipoles become weaker with increasing $\ell$,
so that even though we had to make the pieces narrower, the error in the bias rather stayed constant.
The noise bias values obtained for $\ell_0$ are shown by circles in Fig. \ref{fig:highell_whitenoise}. 
The differences between predicted noise bias and MC spectra are of the order of 1\%, 
and in many cases the predicted value lies 
within the statistical scatter of the MC spectrum.

At 70 GHz the computation is more demanding. We applied a similar procedure as at 44 GHz.
We constructed the full covariance for $\ell=0-200$,
then evaluated the bias for multiples of 100 above that. We used \deltal=30 for $l_0$ = 300--500, 
\deltal=20 for $l_0$ = 600 -- 800,
and \deltal=10 for $l_0$ = 900 -- 1400. We reach an accuracy of 1 -- 3\%.

The importance of parameter \deltal\ is illustrated in Fig. \ref{fig:zoom_whitenoise}.
We computed the noise bias for $\ell=500$ at 30 GHz, and for $\ell=800$ at 70 GHz, 
for increasing values of \deltal. At 30 GHz we show the result for \deltal = 5, 10, 20, 40.
The values approach the Monte Carlo result as \deltal\ increases, as expected.
The computational burden rapidly increases with increasing \deltal.
While computing the \deltal=20 point took 94 minutes on 960 cores, 
the \deltal=40 point already took three hours on 1920 cores.

At 70 GHz we computed the bias for \deltal = 5, 10, 20.  
The last point took 4 hours on 1920 cores.
Again the results are approaching the MC result,
but the three data points are not sufficient to conclude if there is a real convergence.
Also the statistical scatter in the MC spectrum, based on 10 realizations,
is too large for us to give an accurate estimate of the remaining error,
other than saying that it is of the order of 1\%.

The 1-3\% error in bias estimation may still be too large for many applications.
To compute the bias more accurately, we would need a more efficient matrix inversion routine,
or one that constructs the diagonal of the inverse.
Although more efficient methods surely exist, they are not likely to bring us very far,
since the largest matrices we invert already have rank over 200\,000.
Should higher accuracy be required, the best approach is probably 
to produce a larger number of Monte Carlo simulations
and to extract the bias from them.
This will require significant optimization of the pipeline.

The realistic noise simulations at 70 GHz took 15 min of wall-clock time per MC cycle on 960 cores,
corresponding to 250 CPU hours.  Roughly half of that went into noise generation and destriping in Madam,
another half into the deconvolution process.  The cost of white noise simulations was 52 CPUh per realization.
At 30 GHz, the cost was 52 CPUh for realistic and 41 CPUh for white noise simulations.

The limiting factor in our current pipeline is the disk space, rather than CPU time,
since we are writing the intermediate 4D map products on disk.
The pipeline could be made a lot more efficient by if the step of writing disk I/O could be avoided. 
In case of white noise simulations, this could be done by generating a noise 3D map directly in \artdeco\ memory,
the same way we generate a noise TOI  in \Madam.
This remains a topic for future development.

At 44 GHz, white noise simulations give sub-percent accuracy above multipoles $\ell=800$, as can be seen
from Fig. \ref{fig:residual_oof}.
At 70 GHz, this accuracy is reached already around $\ell=400$.
At 30 GHz, the contribution from the correlated component is above 2\% over the whole multipole range.

\subsection{Resolution}
\label{sec:resolution}

The auto spectra ($TT$, $EE$, $BB$) of a binned white noise map become uniform,
when averaged over a large number of realizations.
The level of the spectrum is dependent on the hit count distribution.
For a given total number of hits and fixed noise rms per sample,
the lowest spectral level is obtained when the hits are uniformly distributed over the sky.
Downgrading the resolution of a white noise map in general reduces the spectral level, 
as the distribution becomes more uniform.

The spectral levels can be predicted from a known hit distribution,
as described in \cite{planck2014-a07}.
We tabulate the predicted white noise levels for our simulation parameters 
for different resolutions in Table \ref{tab:whitenoise}. 
We show the results for two radiometer weighting schemes: 
noise weighting and horn-uniform weighting.
Horn-uniform weighting was used in actual LFI mapmaking.


\begin{table}
\begin{center}
\caption{White noise level for ordinary mapmaking,
for noise weighting (nw) and horn-uniform weighting (huw),
and for various map resolutions (\nside).
The average size of a \nside=1024 (64) pixel is 3.44' (55') }
\label{tab:whitenoise}
\begin{tabular}{r rr rr}
 & \multicolumn{2} {c}{$C_\text{TT}$ ($10^{-15} {\rm K}^2$) } & \multicolumn{2}{c} {$C_\text{EE}, C_\text{BB}$ ($10^{-15} {\rm K}^2$)}  \\
\noalign{\vskip-3pt}
& \multispan2\hrulefill \hskip5pt & \multispan2\hskip5pt \hrulefill \hskip5pt  \\
\nside & nw & huw & nw & huw  \\
\hline\hline
  \noalign{\vskip 4pt}
  \omit{\bf 30\,\,GHz}\hfil\cr
  \noalign{\vskip 4pt}
64     & 3.1740  & 3.1740      &    6.3356 &  6.3633   \cr
128   & 3.1910   &   3.1914   &  6.3666  &    6.3944 \cr
256     &  3.2003 &  3.2015    &  6.3837  &    6.4117  \cr
512     &  3.2067  &  3.2083   &  6.3971 &    6.4251 \cr
1024     & 3.2180  &  3.2199   & 6.4294  &    6.4576 \cr
\noalign{\vskip 5pt}
  \omit{\bf 44\,\,GHz}\hfil\cr
  \noalign{\vskip 4pt}
 64 & 4.2407 & 4.2917 & 9.2814 & 9.3898  \cr
 128 & 4.2544 & 4.3058 & 9.3170 & 9.4259  \cr
  256 & 4.2620 & 4.3135 & 9.3364 & 9.4455  \cr
   512 & 4.2666 & 4.3182 & 9.3491 & 9.4583  \cr
   1024 & 4.2738  & 4.3254 & 9.3758 & 9.4852  \cr
   \noalign{\vskip 5pt}
  \omit{\bf 70\,GHz}\hfil\cr
  \noalign{\vskip 4pt}
 64     &   3.3202   &  3.3327     &    6.6476     & 6.7021 \cr
128    &   3.3291  &    3.3432   &       6.6626   &  6.7172  \cr
256    &  3.3334   &      3.3482  &    6.6697    &    6.7243   \cr
512    &  3.3360  &   3.3511    &     6.6745    &    6.7292   \cr
1024  &  3.3381  &    3.3535   &       6.6808   &   6.7355  \cr
\hline\hline
\end{tabular}
\end{center}
\end{table}

In deconvolution the situation is different.
The resolution of the input 4D map does not have
a significant effect on residual noise.
Instead, resolution dependence enters through parameter \lmax.
This is demonstrated in Fig. \ref{fig:bias_resolution}.
We deconvolved the same 30 GHz data set with \lmax = 50, 100, 200, 400.
We show also the corresponding noise bias estimates for \lmax = 50,100, 200,
constructed from the noise covariance matrix.
For \lmax=400 we were unable to compute the exact noise bias, due to 
the huge size of the covariance matrix.
For plotting purposes the spectra have been divided by the spectrum of an \nside=1024 (3.44') white noise binned map.
The noise bias are scaled to the same dimensionless units by dividing them by the theoretical white noise level 
of $3.2180\cdot10^{-15}$ K$^2$, taken from Table \ref{tab:whitenoise}.
The noise level at low multipoles increases with increasing \lmax,
as there are more multipoles to be solved from same amount of data.

The result has relevance for the following section,
where we construct the full low-resolution noise covariance matrix for multipole range $\ell=0-50$.
In order to have an exact match between the covariance matrix and the actual \axlm coefficients,
it is required that both have been computed with same \lmax.


\section{Full noise covariance}
\label{sec:full_ncvm}

We now consider the more complicated case where the input consists of $1/f$ type noise,
which has been destriped to remove the correlated component.
We aim at deriving a full noise covariance matrix,
which takes into account the correlated noise component.
We assume a destriping process that involves a noise prior.
One such implementation is the \Madam\ code presented in \cite{keihanen2010}.
The map binned from the destriped TOI approaches that of generalised least squared (GLS) mapmaking methods 
\citep{poutanen2006,ashdown2007a,ashdown2007b,ashdown2009}
when the baseline length approaches one sample.

\subsection{Destriping}

In the destriping approach the initial noise stream is modelled as
\begin{equation}
  \vec n'=\Fmat \vec b+\vec n_\text{wn}\text{,}   \label{noisemodel}
\end{equation}
where $\vec n_\text{wn}$ represents white noise, and the correlated component is modelled
by a sequence of baselines $\vec b$.
Matrix $\Fmat$ formally spreads the noise baselines into TOI.

The destriping solution gives an estimate for the baseline vector $\vec b$ as
\begin{equation}
  \hat {\vec b} = (\Fmat^{T}\Cn^{-1}\Zmat\Fmat+\Cb^{-1})^{-1}\Fmat^{T}\Cn^{-1}\Zmat \vec y\text{,}
\end{equation}
where
\begin{equation}
  \Zmat = \unimat-\Cn^{-1} \Pmat(\Pmat^{T}\Cn^{-1}\Pmat)^{-1}\Pmat^{T}\Cn^{-1}\text{.}
\end{equation}
Here $\Pmat$ is the pointing matrix, and $\Cb$ is the  {\it a priori} covariance of the baselines,
constructed from the known noise spectrum.
The cleaned TOI is constructed as
\begin{equation}
  \hat{\vec y} = \vec y-\Fmat\hat{\vec b} = 
  \vec y-\Fmat(\Fmat^{T}\Cn^{-1}\Zmat\Fmat+\Cb^{-1})^{-1}\Fmat^{T}\Cn^{-1}\Zmat \vec y\text{.}  \label{toiclean}
\end{equation}

Consider then the residual noise $\vec n$ in a TOI stream that has 
been processed according to Eq.~(\ref{toiclean}).
We aim at deriving a formula for $\langle \vec n \vec n^{T}\rangle $,
and finally for the covariance of Eq.~(\ref{artcov}),
where $\vec n$ is now the destriped noise stream.
We assume that the original noise stream $\vec n'$ obeys the model (\ref{noisemodel}), so that
\begin{equation}
  \langle \vec n'\vec n'^{\rm T}\rangle = \Cn +\Fmat \Cb\Fmat^{\rm T}\text{.}
\end{equation}

The baseline length, hidden inside matrix $\Fmat$, is a key parameter in the calculation. 
The baseline length plays a double role in the noise covariance.
On the one hand, it limits the
frequencies that are taken into account by the covariance matrix.
The frequencies above the inverse of the baseline length are neglected.
If we choose too long a baseline length, we will underestimate the residual noise.
For optimal results, we should thus choose a baseline length as short as possible.

On the other hand, the baseline length in $\Fmat$ represents the baseline length that was used in destriping.
If we select a shorter baseline for covariance computations, we will not model the destriping process correctly.
However, simulations with \Planck\ LFI data \citep{planck2014-a07} show that
the destriping result converges already around a baseline length of one second.
There is correlated noise at higher frequencies, but destriping is unable to remove it, and
making the baseline shorter does not change the results.
We can turn this around and argue that we do not make a significant error if we assume a shorter baseline
length than was actually used for destriping,
as long as the actual baseline length was short enough that the optimal solution was already reached.

All this put together, we argue that we get the optimal noise covariance when 
we make as short a baseline as possible.
We bring this to the extreme and set the baseline length to one TOI sample.
With this assumption, matrix $\Fmat$ reduces into a unity matrix, $\Fmat=\unimat$.
The destriped TOI (\ref{toiclean}) simplifies into
\begin{equation}
  \hat{\vec y} =  \vec y-(\Cn^{-1}\Zmat+\Cb^{-1})^{-1} \Cn^{-1}\Zmat \vec y 
                     = (\Cb\Cn^{-1}\Zmat+1)^{-1} \vec y\text{.}  \label{signaltoi}
\end{equation}

\subsection{Covariance matrix}

We can rewrite Eq. (\ref{signaltoi}) for the destriped noise component
\begin{equation}
  \vec n = (\Cb\Cn^{-1}\Zmat+1)^{-1} \vec n'\text{.}
\end{equation}
We reformulate this with the aim of constructing the time domain covariance $\langle\vec n \vec n^T\rangle$.
Inserting $\Zmat$ we obtain
\begin{equation}
   \vec n = \Cn[ \Cn+\Cb-\Cb\Cn^{-1} \Pmat(\Pmat^{\rm T}\Cn^{-1}\Pmat)^{-1} \Pmat^{\rm T} ]^{-1} \vec n'  \,.
\end{equation}
We apply the Woodbury formula to get
\begin{eqnarray}
   \vec n &=& \Cn \big[ \Ct^{-1}+\Ct^{-1}\Cb\Cn^{-1}\Pmat 
   [ \unimat -(\Pmat^{\rm T}\Cn^{-1}\Pmat)^{-1} \Pmat^{\rm T}\Ct^{-1}\Cb\Cn^{-1}\Pmat]^{-1}  \nonumber \\
  &&  \hskip 30mm \times\, (\Pmat^{\rm T}\Cn^{-1}\Pmat)^{-1} \Pmat^{\rm T}\Ct^{-1}  \big] \vec n'\label{SHnoise}
\end{eqnarray}
where 
\begin{equation}
   \Ct = \Cn+\Cb
\end{equation}
denotes the full (white+correlated)  time domain noise covariance of un-destriped TOI.
We bring $(\Pmat^{\rm T}\Cn^{-1}\Pmat)^{-1} $ inside the brackets
and note that $\Ct^{-1}\Cb\Cn^{-1}=\Cn^{-1}-\Ct^{-1}$. Equation (\ref{SHnoise}) simplifies to
\begin{equation}
  \vec n = \Cn\Ct^{-1}\vec n' +
  \Cb\Ct^{-1}\Pmat [\Pmat^{\rm T}\Ct^{-1}\Pmat]^{-1} \Pmat^{\rm T}\Ct^{-1} \vec n'   \,.
\end{equation}
Assuming that the true noise obeys the noise model, we have $\Ct= \langle\vec n' \vec n'^{\rm T}\rangle$,
and we obtain for the time domain covariance of the destriped noise
\begin{eqnarray}
 \langle\vec n \vec n^T\rangle
 &=& \Cn\Ct^{-1}\Cn  +\Pmat[\Pmat^{\rm T}\Ct^{-1}\Pmat]^{-1}\Pmat^{\rm T}   \\
  &&\hskip 5mm -\Cn\Ct^{-1}\Pmat[\Pmat^{\rm T}\Ct^{-1}\Pmat]^{-1}\Pmat^{\rm T}\Ct^{-1} \Cn   \,.  \nonumber
\end{eqnarray}
Inserting this into Eq. (\ref{artcov})
we finally obtain for the harmonic noise covariance matrix
\begin{eqnarray}
 \tens{C}
 &=&  (\Amat^{\dagger}\Cn^{-1}\Amat)^{-1} 
          \Amat^{\dagger}\Ct^{-1}\Amat  
          (\Amat^{\dagger}\Cn^{-1}\Amat)^{-1} \nonumber \\
  && \!\! + (\Amat^{\dagger}\Cn^{-1}\Amat)^{-1} 
           \Amat^{\dagger}\Cn^{-1} \Pmat
           (\Pmat^{\rm T}\Ct^{-1}\Pmat)^{-1}
           \Pmat^{\rm T}\Cn^{-1}\Amat 
            (\Amat^{\dagger}\Cn^{-1}\Amat)^{-1}\nonumber \\
  && \!\! - (\Amat^{\dagger}\Cn^{-1}\Amat)^{-1}
           \Amat^{\dagger}\Ct^{-1}\Pmat
          (\Pmat^{\rm T}\Ct^{-1}\Pmat)^{-1}
           \Pmat^{\rm T}\Ct^{-1} \Amat  
           (\Amat^{\dagger}\Cn^{-1}\Amat)^{-1}\text{.}  \label{artdeco_ncvm}
\end{eqnarray}
Equation (\ref{artdeco_ncvm}) is the main result of this section.

\subsection{Cross-correlation}

As discussed in Sect. \ref{sec:destriping_options},
there are two destriping options  available when deconvolving data subsets.
Either we destripe the full data set (full destriping)
or only the deconvolved data set (independent destriping).
In the former case, the common destriping step generates correlation
in residual noise between data sets.
This effect too we can assess through the noise covariance formalism.
We can construct a cross-covariance matrix
\begin{equation}
\tens{C}_{12} = \langle \vec a_1 \vec a_2^T \rangle
\end{equation}
where $\vec a_1$ and $\vec a_2$ are the harmonic coefficients obtained 
by deconvolution of the two data sets. 
The cross-covariance matrix is constructed in very much the same way as
the covariance matrix of Eq. (\ref{artdeco_ncvm}),
only now we have on both sides two destriping operations $\Amat_1$ and $\Amat_2$.
Formally both operate on the same full data stream,
but yield zeros when operating to a part of TOI that does not belong to the data set.
If the two data sets do not overlap, we have
\begin{equation}
 \Amat_1^{\dagger}\Ct^{-1}\Amat_2=0.
 \end{equation}  
With this assumption, the cross-covariance matrix 
is given by
\begin{eqnarray}
 \tens{C}_{12}
 &=& (\Amat_1^{\dagger}\Cn^{-1}\Amat_1)^{-1} \big[
           \Amat_1^{\dagger}\Cn^{-1} \Pmat
           (\Pmat^{\rm T}\Ct^{-1}\Pmat)^{-1}
           \Pmat^{\rm T}\Cn^{-1}\Amat_2 
           \nonumber \\
  && - 
           \Amat_1^{\dagger}\Ct^{-1}\Pmat
          (\Pmat^{\rm T}\Ct^{-1}\Pmat)^{-1}
           \Pmat^{\rm T}\Ct^{-1} \Amat_2  \big]
           (\Amat_2^{\dagger}\Cn^{-1}\Amat_2)^{-1}  \,. \label{ncvm_cross}
\end{eqnarray}
The auto-covariance for the full destriping case is given by the same formula (\ref{artdeco_ncvm})
as for independent destriping, only $\Pmat$ is taken to be the pointing matrix of the full data set,
while in case of independent destriping it is interpreted as the submatrix that contains the rows 
for the data set in question.

\subsection{Implementation}
\label{sec:implementation}

Equations (\ref{artdeco_ncvm}) and (\ref{ncvm_cross}) appear complicated,
but actually consist
of only five ingredient matrices that appear in different combinations:
$\Amat^{\dagger}\Cn^{-1}\Amat$, $\Amat^{\dagger}\Ct^{-1}\Amat$,
$\Amat^{\dagger}\Cn^{-1}\Pmat$, $\Amat^{\dagger}\Ct^{-1}\Pmat$,
and $\Pmat^{\dagger}\Ct^{-1}\Pmat$.
We can further reduce the computation load as we describe in the following.
As described in \cite{keihanen2012}, \artdeco\ performs deconvolution through
a process where the TOI is first compressed into a ``3D map''.
The data is binned on a 3-dimensional grid, spanned by the pointing angles  $\theta,\phi,\psi$.
In $\theta$ and $\phi$, \artdeco\ are pixelized as in  {\tt HEALPix},
and  $\psi$ is divided uniformly into $N_\text{psi}$ cells.
We can formally write the deconvolution operation $\Amat\Cn^{-1}$ as
\begin{equation}
\Amat^\dagger\Cn^{-1} = \Gmat\Smat^{\rm T}\Cn^{-1}   \,,
\end{equation}
where $\Smat^{\rm T}$, operating on a TOI, represents the operation of binning 
the TOI into a 3D map, and $\Gmat$ the remaining part of the deconvolution
operation.

Matrix $\Pmat^{\rm T}$ represents the operation of binning a TOI into
a two-dimensional sky map. This too can be split into a two-step operation
where one first constructs a 3D map, and from that a sky map.
We write
\begin{equation}
\Pmat^{\rm T} = \Xmat^{\rm T}\Smat^{\rm T}   \,.
\end{equation}
An element of a 3D map is identified through a combined index $qi$,
where $q$ refers to sky pixel, and $i=0\ldots N_{\rm psi}-1$ to the bin defined 
by beam rotation angle $\psi_i$.
The elements of a sky map are referred to as $Ip$, $Qp$, or $Up$.
In this notation, the $\Xmat$ matrix is given by
\begin{eqnarray}
\Xmat_{qi,Ip} &=& \delta_{q\in p}  \nonumber \\
\Xmat_{qi,Qp} &=& \cos(2(\psi_i+\psi_{\rm pol}))\, \delta_{q\in p} \\
\Xmat_{qi,Up} &=& \sin(2(\psi_i+\psi_{\rm pol}))\, \delta_{q\in p} \nonumber
\end{eqnarray}
where $\psi_{\rm pol}$ is the detector-specific angle of polarization sensitivity.
To add to the generality, we allow the 3D map pixel resolution to be higher than that of the sky map.
In this notation $\delta_{q\in p}=$ if pixel $q$ belongs into the larger pixel $p$.
However, in our computations we always use same resolution for both,
so that $\delta$ becomes the usual Kronecker delta.
Operation $\Xmat^{\rm T}$ is thus a trivial coaddition operation involving weighting
by the polarization angle.

The five ingredient matrices become
\begin{eqnarray}
\Amat^\dagger \Cn^{-1}\Amat &=& \Gmat^\dagger(\Smat^{\rm T}\Cn^{-1}\Smat)\Gmat \label{ingredients} \\
\Amat^\dagger \Ct^{-1}\Amat &=& \Gmat^\dagger (\Smat^{\rm T}\Ct^{-1}\Smat)\Gmat \nonumber \\
\Amat^\dagger \Cn^{-1}\Pmat &=& \Gmat^\dagger (\Smat^{\rm T}\Cn^{-1}\Smat)\Xmat \nonumber \\
\Amat^\dagger \Ct^{-1}\Pmat &=& \Gmat^\dagger (\Smat^{\rm T}\Ct^{-1}\Smat)\Xmat \nonumber \\
\Pmat^{\rm T} \Ct^{-1}\Pmat &=& \Xmat^{\rm T} (\Smat^{\rm T}\Ct^{-1}\Smat)\Xmat \nonumber  \,.
\end{eqnarray}
In the end, there remain three matrices to evaluate: $(\Smat^{\rm T}\Cn^{-1}\Smat)$, $(\Smat^{\rm T}\Ct^{-1}\Smat)$,
and $\Gmat$.  Operation $\Xmat$ is carried out algorithmically where needed. 
Matrix $\Gmat$ is part of the \artdeco\ deconvolver, and we can extract it from the code.
Matrix $(\Smat^{\rm T}\Cn^{-1}\Smat)$ represents the white noise covariance between elements of a 3D map.
It is diagonal, with elements given by $\sigma^2/N_{pi}$, where $\sigma^2$ is the white noise variance in TOI domain,
and $N_{pi}$ is the number of hits to bin $pi$. 

To proceed with matrix $(\Smat^{\rm T}\Ct^{-1}\Smat)$, we first
use the Woodbury formula to rewrite $\Ct^{-1}$ as
\begin{equation}
  \Ct^{-1} = (\Cn+\Cb)^{-1} = \Cn^{-1} -\Cn^{-1}(\Cb^{-1}+\Cn^{-1})^{-1}\Cn^{-1} \,.  \label{ctinv_exact}
\end{equation}
This formulation has the advantage that $\Cn$ appears as inverse, which makes dealing with flagged data easier.
The operation of Eq. (\ref{ctinv_exact}) can be interpreted as a filtering operation in  time domain.
To construct one column $(\Smat^{\rm T}\Ct^{-1}\Smat)$, we pick a TOI from a 3D map with one non-zero element,
apply the filter of Eq. (\ref{ctinv_exact}) to it, and compress the TOI back into a 3D map. 
The FFT technique is usually efficient in filtering, but since we are dealing with TOIs with very few non-zero elements,
we find more efficient a procedure where we first use FFT to construct the filter, but then to perform the actual filtering in time domain.
When dealing with  \Planck\ data we assume that noise is uncorrelated from one pointing period (typically one hour) to another,
and do the filtering by pointing period.

We further benefit from the fact that matrices of Eq. (\ref{ingredients}) are additive.
If we have two sections of TOI, with no noise correlation between,
we can compute the contribution from each section to the ingredient matrices independently,
and coadd the results.  In particular, several detectors are combined simply by coadding
their matrices.

We construct the ingredient matrices for each radiometer quadruplet and each half-year survey,
write them on disk,
and then coadd them in different combinations in a separate step.
That way we avoid doing repeated 
computations when dealing with different data combinations.

\subsection{Flagged data}

To add to the generality of the method, we allow for a case where part of the data is flagged as unusable.
{\tt Madam} handles flags by setting formally $\Cn^{-1}=0$ for the flagged samples.
This is easy to take into account in $(\Smat^{\rm T}\Cn^{-1}\Smat)$.
Matrix $(\Smat^{\rm T}\Ct^{-1}\Smat)$ requires one more manipulation step.
Eq. (\ref{ctinv_exact}) is expensive to construct,
since the middle matrix $\Cb^{-1}+\Cn^{-1}$ is no more stationary.
We therefore make the approximation
\begin{equation}
  (\Cb^{-1}+\Cn^{-1})^{-1} \approx (\Cb^{-1}+\Cwn^{-1})^{-1}  \,
\end{equation}
where $\Cwn$ is the original white noise variance, where flags have not been applied.
This can be evaluated efficiently using Fourier techniques.
We thus have for the last ingredient matrix
\begin{equation}
  \Smat^{\rm T}\Ct^{-1}\Smat \approx 
  \Smat^{\rm T}\Cn^{-1}\Smat 
  -\Smat^{\rm T}\Cn^{-1} (\Cb^{-1}+\Cwn^{-1})^{-1} \Cn^{-1} \Smat  \,.
\end{equation}
In absence of flags the formula is exact.
The approximation misestimates the correlation between distant 3D map elements,
but captures the main effect of flagging: the reduction in the amount of available data
and the consequent increase in residual noise.

\subsection{Destriping mask and detector weighting}
\label{sec:mask}

The mapmaking procedure of \Planck\ LFI involves a destriping mask \citep{planck2014-a07}.
The mask is used for the purpose of reducing errors arising from high signal gradients
and from bandpass mismatch between detectors. 

The proper way of treating the mask in the NCVM would be to introduce another white noise covariance $\tens{C}_\text{mask}$,
whose inverse $\tens{C}_\text{mask}^{-1}=0$ for samples that fall in the masked region.
The masked covariance is then applied in the destriping phase, while the normal white noise covariance is applied in the deconvolution phase.
Unfortunately, unlike in the case of flags, the different covariance terms appear in combinations that do not cancel out,
with the result that the noise covariance matrix of Eq. (\ref{artdeco_ncvm}) is no more valid, and not easily enhanced.
Note that replacing the pointing matrix $\Pmat$
by its masked version does not have the desired effect, since reducing $\Pmat$ has the effect of reducing the residual noise, 
as there is a smaller number of pixels to solve, while the effect of destriping mask is to increase the residual noise level.

The mask thus affects the residual noise in a way that is not captured by our noise covariance matrix.  
This has to be accepted as a source of uncertainty in noise modelling.

In principle, a similar limitation arises from detector weighting. 
The destriping procedure of \Planck\ LFI replaces the white noise covariance $\Cn$ by
a slightly different modified covariance, which is made equal between a pair of detectors. 
This is referred to as horn-uniform weighting.  
The effect from this is expected to be small.

\subsection{Inputs}

The required inputs for NCVM computation are detector pointing, harmonic beam expansion,
and noise spectrum. 
It is not required that noise parameters remain the same throughout the mission,
although in our simulations we have assumed so.  The TOI can be assembled of sections
with different noise spectra, and each section's contribution added to the ingredient matrices.

Four user parameters control the computation.
Parameters \lmax\ and \kmax\ control the deconvolution process and enter matrix $\Gmat$.
Parameters \nsided\ and \npsid\ set the resolution of the 3D map dimension of matrices $\Smat$ and  $\Gmat$. 
The meaning of \nsided\ is twofold. On the one hand it sets the internal resolution of the deconvolution calculation.
As a rule of thumb, 1.5\nsided\ should exceed \lmax.
Thus, for deconvolution with \lmax=50 one should have at least \nsided=32, and for \lmax=100 at least \nsided=64.
On the other hand,  \nsided\ also sets the destriping resolution.
The covariance matrix models a destriping procedure where the sky signal is assumed constant 
within a pixel of resolution \nside=\nsided.

The required computational resources increase rapidly with increasing \lmax\ and \nsided.
The complexity of the computation increases as \lmax$^5$,
and the size of the matrix as \lmax$^4$.
In practice we can produce a full covariance matrix for only the low multipoles $\ell\lessapprox100$.

\subsection{Constructing NCVM for Planck LFI frequencies}


\begin{table}
\caption{List of data combinations for which we construct the NCVM products.
We give the data range ({\em Survey}) and combination of horns ({\em Horns}) for the destriping phase 
and for the devonvolution phase.
All products have been computed with deconvolution parameters \lmax=50, \kmax=2, \nsided=64, \npsid=256. 
For each case listed we produce a harmonic noise covariance matrix,  
a \nside=16 (220') pixel-pixel covariance matrix, and a noise bias estimate.}
\label{tab:ncvm_products}
\begin{center}
\begin{tabular}{l l l l cc}
 \rule[-5pt]{0pt}{0pt}  & \multispan{2} Deconvolution && \multispan{2} Destriping  \\
\cline{2-3}
\cline{5-6} 
\rule[-5pt]{0pt}{14pt} Input type & Horns & Survey & & Horns & Survey  \\
\hline\hline
 \noalign{\vskip 5pt}
  \omit{\bf 30\,GHz}\hfil\cr
  \noalign{\vskip 4pt}
white  & 27--28 & full  &&  27--28 & full   \\
wn+1/f  & 27--28 & full  && 27--28 & full  \\
   \noalign{\vskip 5pt}
  \omit{\bf 44\,GHz}\hfil\cr
  \noalign{\vskip 4pt}
white  & 24--26 & full  &&  - & \\
wn+1/f  & 24--26 & full && 24--26 & full  \\
   \noalign{\vskip 5pt}
  \omit{\bf 70\,GHz}\hfil\cr
  \noalign{\vskip 4pt}
white  & 18--23 & full  && - & -  \\
wn+1/f  & 18--23 & full  &&  18--23 & full  \\
\multicolumn{2}{l}{\,\, Single years} \\
white  & 18--23 & yr1 && - & -\\
white  & 18--23 & yr2 &&  -&   \\
white  & 18--23 & yr3 && - & - \\
white  & 18--23 & yr4 && - & - \\
wn+1/f  & 18--23 & yr1  && 18--23 & full  \\
wn+1/f  & 18--23 & yr2  &&  18--23  & full  \\
wn+1/f  & 18--23 & yr3  &&  18--23  & full  \\
wn+1/f  & 18--23 & yr4  && 18--23  & full  \\
wn+1/f  & 18--23 & yr1  && 18--23  & yr1   \\
wn+1/f  & 18--23 & yr2  && 18--23  & yr2    \\
wn+1/f  & 18--23 & yr3  && 18--23  & yr3   \\
wn+1/f  & 18--23 & yr4  && 18--23   & yr4    \\
\multicolumn{2}{l}{\,\, Horn pairs} \\
white  &18/23 & full &&  - & -\\
white  &19/22 & full && - & - \\
white  & 20/21 & full && -  & - \\
wn+1/f  & 18/23 & full  && 18--23 & full \\
wn+1/f  & 19/22 & full  &&  18--23 & full  \\
wn+1/f  & 20/21 & full  &&  18--23  & full  \\
wn+1/f  & 18/23 & full  &&  18/23 & full  \\
wn+1/f  & 19/22 & full  &&  19/22 & full  \\
wn+1/f  & 20/21 & full  &&  20/21  & full \\
\end{tabular}
\end{center}
\end{table}

We have computed NCVM products for Planck LFI frequencies, 
for a number of radiometer and survey combinations.
A complete list is given in Table \ref{tab:ncvm_products}.
For each radiometer and survey combination we produced two sets of NCVM products:
one with the white noise matrix, 
given by Eq. (\ref{whitenoise_ncvm}),
and one with the full NCVM matrix given by Eq. (\ref{artdeco_ncvm}).
We used the same pointing, noise parameters, and beam, as in the simulations of Sect. \ref{sec:highell_simulations}.

We constructed the harmonic NCVM matrix for multipoles up to \lmax=50.
Beam elements with high $k$ only affect the high $\ell$ multipoles \citep{keihanen2012}.
We  chose \kmax=2 for this low-multipole computation.
Value \lmax=50\ requires a minimum pixel resolution of \nsided=32 (110').
To realistically model the mapmaking process of \Planck\ LFI,
we should have \nsided=1024 (3.44') , which is out of reach.
As a trade-off, we used \nsided=64 (55') and  \npsid=256.
The effect of this approximation on matrix quality is assessed in Sect. \ref{sec:validation}.

For all cases, the set of NCVM products includes:  a harmonic NCVM matrix with \lmax=50 (930 MB),
a noise bias estimate computed according to Eq. (\ref{noisebias_ncvm}), 
and a pixel-pixel covariance matrix at resolution \nside=16 (720 MB).
All products include polarization.  The construction of the pixel-pixel matrix from the harmonic matrix is 
described in Sect. \ref{sec:pixelspace}.

In addition to the full frequency matrices for 30, 44, and 70 GHz, 
we produced matrices for 70 GHz horn pairs 18/23, 19/22, 20/21,
and for individual years (years 1 -- 4) involving all 70 GHz radiometers.
For each of these, we considered both full and independent destriping.


\begin{figure*}
\includegraphics[width=18cm]{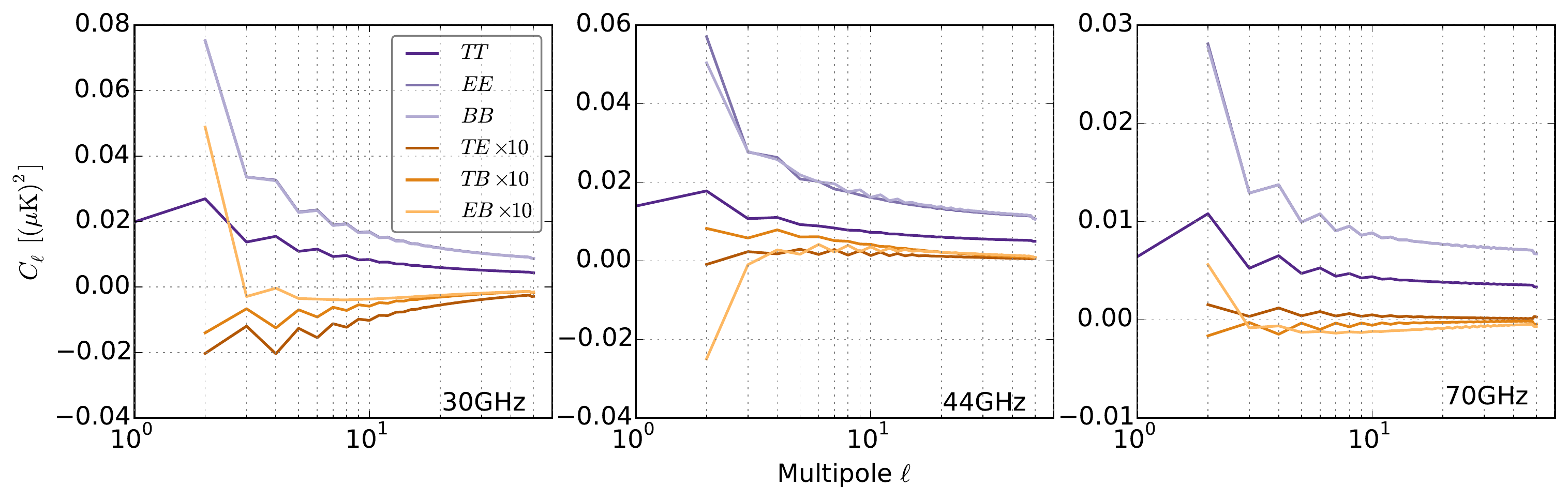}
\caption{Predicted noise bias for LFI frequencies: $TT$, $EE$, $BB$, $TE$, $TB$, $EB$.
The cross spectra have be multiplied by 10 to bring out the structure more clearly.}
\label{fig:bias_freq}
\end{figure*}


\begin{figure}
\includegraphics[width=0.95\columnwidth]{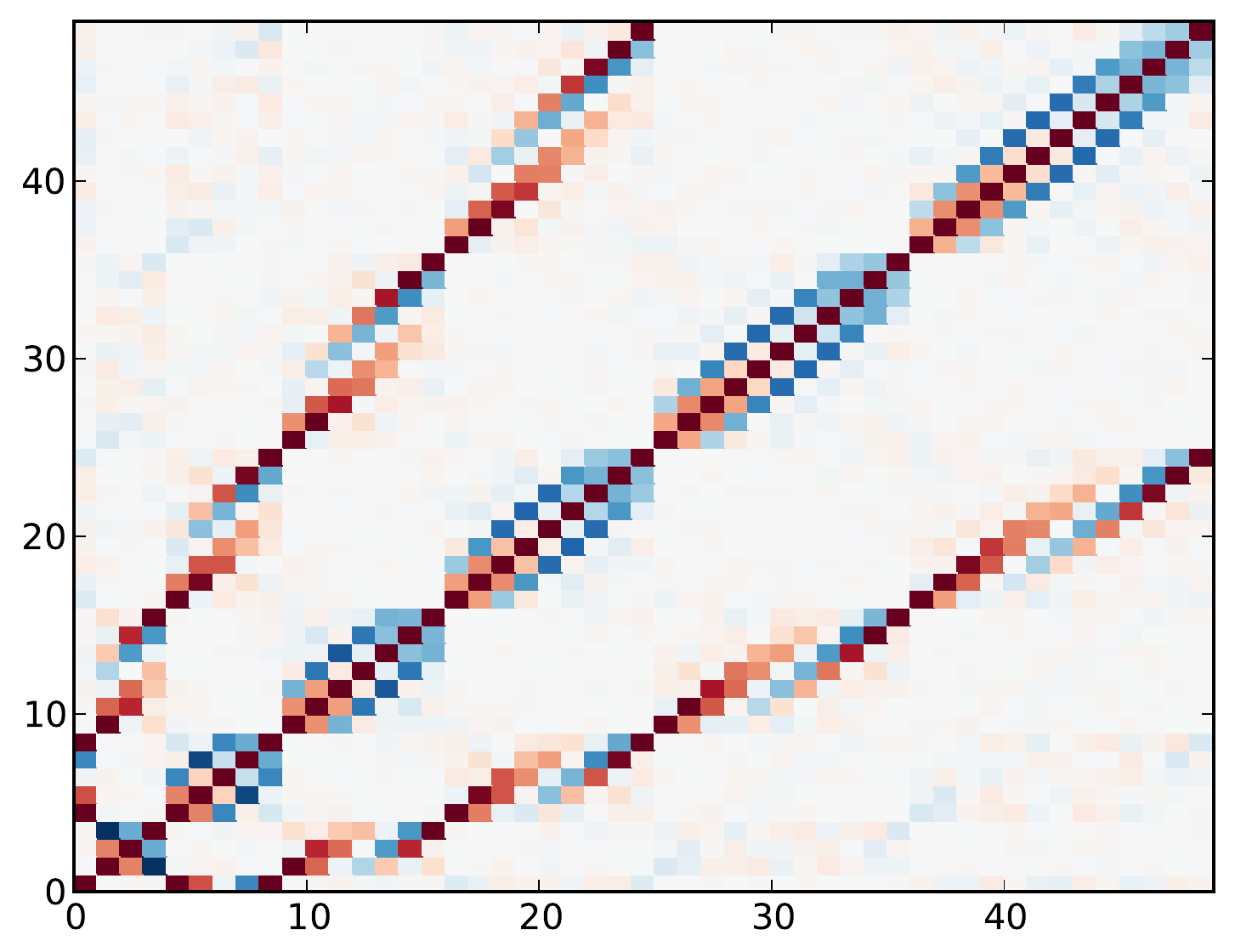}
\caption{Structure of the $TT$ block of the 30 GHz NCVM at low multipoles ($\ell$ = 0 -- 6).
Index on $x$ and $y$ axis is $\ell^2+\ell+m$. Positive correlation is shown in red colour, negative in blue.}
\label{fig:pcolor}
\end{figure}

The construction of the ingredient matrices as described in Sect. \ref{sec:implementation} 
took 2000 CPU hours for each year and horn pair, adding up to 24\,000 CPU hours for the whole 70 GHz data set.
From those we constructed the NCVM for different horn pair and year combinations.
The process of constructing the final NCVM from the ingredients was dominated by the inversion
of the middle matrix, and took about 1500 CPU hours per matrix.

The predicted noise bias for LFI frequencies is shown in Fig. \ref{fig:bias_freq}.
We show three auto spectra ($TT$, $EE$, $BB$) and three cross spectra ($TE$, $TB$, $EB$) .
The cross spectra are small, and are multiplied by 10 in the plot to show the structure more clearly.

Figure \ref{fig:pcolor} illustrates the matrix structure in $\ell,m$ space.
We show the real part $TT$ block of the 30 GHz NCVM for lowest multipoles ($\ell<7$).
The matrix elements are arranged according to the combined index $i=\ell^2+\ell+m$.
The strongest correlations appear between elements with same $\ell$,
or same $m$ and $\ell$ separated by an multiple of 2.


\section{Validation of the covariance matrix}
\label{sec:validation}

\subsection{Low-resolution simulations}


\begin{table*}
\caption{List of MC simulation cases.
The parameters from left to right are:
Simulation identifier; input data type;
combination of horns and survey included in deconvolution phase (``full'' refers to the full 4-year data set,
``yrN'' to single years N=1,2,3,4); deconvolution parameters \lmax\ and \kmax;
\nsided\ and \npsid\ resolution of the 3D map;
horn combination and survey in destriping phase; baseline length (s); destriping resolution;
destriping mask used (T/F); detector weighting (nw=noise weighting, huw=horn-uniform weighting);
number of Monte Carlo realizations.
}
\label{tab:MC_parameters}

\tiny
%
\begin{tabular}{ll rcrcrr p{-1pt} rcrrcl r}
\multispan{2} \rule[-5pt]{0pt}{0pt}  & \multispan{6} Deconvolution & \multispan{6} Destriping  & \\
\cline{3-8}
\cline{10-15} 
{\bf ID} \rule[-5pt]{0pt}{14pt}  & {\bf data} & {\bf Horns}  & {\bf Surv.} &  \lmax & \hbox{\!\!\kmax\!\!\!} & \nsided &\npsid  &&
 {\bf Horns} & {\bf Surv.} & {\bf Base}  & \nside &  {\bf Mask} & {\bf Weight} & {\bf MC}\\
\hline\hline
\noalign{\vskip4pt}
\multicolumn{2}{l} { \bf 30 GHz}   \\
white    & white      & 27--28 & full  & 50 & 2 & 64 & 256 &&     27--28 & full & -        & - & - &-  & 100 \\
ideal         & wn+1/f    & 27--28 & full  & 50 & 2 & 64 &  256 &&     27--28 & full & 0.063  & 64 & F & nw & 100 \\
real           & wn+1/f    & 27--28 & full  & 50 & 2 &  64 &  256 &&     27--28 & full & 0.25  & 1024 & T & huw & 100 \\
hres wn & white    & 27--28 & full  & 700 & 6 & 1024 & 4096 &&  27--28 & full & --        &  - & - & - & 40 \\
hres  real   & wn+1/f  & 27--28 & full  & 700 & 6 & 1024 & 4096 &&  27--28 & full & 0.25  & 1024 & T & huw & 40 \\
\hline
\noalign{\vskip4pt}
\multicolumn{2}{l} {\bf 44 GHz}   \\
white         & white      & 24--26 & full  & 50 & 2 & 64 & 256 &&     24--26 & full & -        & - & - &-  & 100 \\
ideal         & wn+1/f    & 24--26 & full  & 50 & 2 & 64 &  256 &&     24--26 & full & 0.25  & 64 & F & nw & 100 \\
real           & wn+1/f    & 24--26 & full  & 50 & 2 &  64 &  256 &&     24--26 & full & 1.0  & 1024 & T & huw & 100 \\
hres wn  & white    & 24--26 & full  & 1000 & 6 & 1024 & 4096 &&  24--26 & full & --        &  - & - & - & 10 \\
hres  real   & wn+1/f  & 24--26 & full  & 1000 & 6 & 1024 & 4096 &&  24--26 & full & 1.0  & 1024 & T & huw & 10 \\
\hline
\noalign{\vskip4pt}
\multicolumn{2}{l} {\bf 70 GHz}   \\
white         & white      & 18-23 & full  & 50 & 2 & 64 & 256 &&     18--23 & full & -        & - & - &-  & 100 \\
ideal         & wn+1/f    & 18-23 & full  & 50 & 2 & 64 &  256 &&     18--23 & full & 0.25  & 64 & F & nw & 100 \\
real           & wn+1/f    & 18-23 & full  & 50 & 2 &  64 &  256 &&     18--23 & full & 1.0  & 1024 & T & huw & 100 \\
hres wn  & white    & 18-23  & full  & 1500 & 6 & 1024 & 4096 &&  18--23  & full & --        &  - & - & - & 10 \\
hres  real   & wn+1/f  & 18-23  & full  & 1500 & 6 & 1024 & 4096 &&  18--23  & full & 1.0  & 1024 & T & huw & 10 \\
\multicolumn{2}{l}{\,\, Single years} \\
white   & white   & 18--23 & yrN  & 50 & 2 & 64 &  256 &&     18--23 & yrN & - & - & - & - & 100 \\
ideal indp   & wn+1/f    & 18--23 & yrN  & 50 & 2 & 64 &  256 &&     18--23 & yrN & 0.25  & 64 & F & nw & 100 \\
real indp     & wn+1/f    & 18--23 & yrN  & 50 & 2 &  64 &  256 &&     18--23 & yrN & 1.0  & 1024 & T & huw & 100 \\
ideal full   & wn+1/f    & 18--23 & yrN  & 50 & 2 & 64 &  256 &&     18--23 & full & 0.25  & 64 & F & nw & 100 \\
real full     & wn+1/f    & 18--23 & yrN  & 50 & 2 &  64 &  256 &&     18--23 & full & 1.0  & 1024 & T & huw & 100 \\
\multicolumn{2}{l}{\,\, Horn pairs} \\
white   & white   & 18/23 & full  & 50 & 2 & 64 &  256 &&     18--23 & full & - & - & - & - & 100 \\
white   & white   & 19/22 & full  & 50 & 2 & 64 &  256 &&     18--23 & full & -  & - & - &- & 100 \\
white   & white   & 20/21 & full  & 50 & 2 & 64 &  256 &&     18--23 & full & -  & - & - & -- & 100 \\
ideal indp   & wn+1/f    & 18/23 & full  & 50 & 2 & 64 &  256 &&     18/23 & full & 0.25  & 64 & F & nw & 100 \\
ideal indp   & wn+1/f    & 19/22 & full  & 50 & 2 & 64 &  256 &&     19/22 & full & 0.25  & 64 & F & nw & 100 \\
ideal indp   & wn+1/f    & 20/21 & full  & 50 & 2 & 64 &  256 &&     20/21 & full & 0.25  & 64 & F & nw & 100 \\
real indp     & wn+1/f    & 18/23 & full  & 50 & 2 &  64 &  256 &&     18/23 & full & 1.0  & 1024 & T & huw & 100 \\
real indp     & wn+1/f    & 19/22 & full  & 50 & 2 &  64 &  256 &&     19/22 & full & 1.0  & 1024 & T & huw & 100 \\
real indp     & wn+1/f    & 20/21 & full  & 50 & 2 &  64 &  256 &&     20/21 & full & 1.0  & 1024 & T & huw & 100 \\
ideal full   & wn+1/f    & 18/23 & full  & 50 & 2 & 64 &  256 &&     18--23 & full & 0.25  & 64 & F & nw & 100 \\
ideal full    & wn+1/f    & 19/22 & full  & 50 & 2 & 64 &  256 &&     18--23 & full & 0.25  & 64 & F & nw & 100 \\
ideal full    & wn+1/f    & 20/21 & full  & 50 & 2 & 64 &  256 &&     18--23 & full & 0.25  & 64 & F & nw & 100 \\
real full     & wn+1/f    & 18/23 & full  & 50 & 2 &  64 &  256 &&     18--23 & full & 1.0  & 1024 & T & huw & 100 \\
real full     & wn+1/f    & 19/22 & full  & 50 & 2 &  64 &  256 &&     18--23 & full & 1.0  & 1024 & T & huw & 100 \\
real full     & wn+1/f    & 20/21 & full  & 50 & 2 &  64 &  256 &&     18--23 & full & 1.0  & 1024 & T & huw & 100 \\
hres wn full & white        & 18/23  & full  & 1500 & 6 & 1024 & 4096 &&  18--23  & full & --        &  - & - & - & 10 \\
hres  real full  & wn+1/f  & 18/23  & full  & 1500 & 6 & 1024 & 4096 &&  18--23  & full & 1.0  & 1024 & T & huw & 10 \\
hres wn indp & white    & 18/23  & full  & 1500 & 6 & 1024 & 4096 &&  18/23  & full & --        &  - & - & - & 10 \\
hres  real indp  & wn+1/f  & 18/23  & full  & 1500 & 6 & 1024 & 4096 &&  18/23  & full & 1.0  & 1024 & T & huw & 10 \\
hres wn full & white         & 19/22  & full  & 1500 & 6 & 1024 & 4096 &&  18--23  & full & --        &  - & - & - & 10 \\
hres  real full   & wn+1/f  & 19/22  & full  & 1500 & 6 & 1024 & 4096 &&  18--23  & full & 1.0  & 1024 & T & huw & 10 \\
hres wn indp & white      & 19/22  & full  & 1500 & 6 & 1024 & 4096 &&  19/22  & full & --        &  - & - & - & 10 \\
hres  real indp   & wn+1/f  & 19/22  & full  & 1500 & 6 & 1024 & 4096 &&  19/22  & full & 1.0  & 1024 & T & huw & 10 \\
\hline
\noalign{\vskip4pt}
\multicolumn{2}{l}{\bf Special simulations} \\
delta beam   & wn       & 27--28 & full  & 700 & 2 & 1024 & 4096 &&  27--28 & full & -  & - & - & - & 40 \\
delta beam   & wn+1/f  & 27--28 & full  & 700 & 2 & 1024 & 4096 &&  27--28 & full & 0.25  & 1024 & T & huw & 40 \\
\lmax   & wn       & 27--28 & full  & \!\!\! 100--400 & 2 & 1024 & 4096 &&  27--28 & full & -  & - & - & - & 40 \\
\lmax   & wn+1/f  & 27--28 & full  & \!\!\! 100--400 & 2 & 1024 & 4096 &&  27--28 & full & 0.25  & 1024 & T & huw & 40 \\
delta beam  & white    & 18--23  & full  & 1500 & 2 & 1024 & 4096 &&  18--23  & full & --        &  - & - & - & 10 \\
delta beam  & wn+1/f  & 18--23  & full  & 1500 & 2 & 1024 & 4096 &&  18--23  & full & 1.0  & 1024 & T & huw & 10 \\
\lmax & white    & 18--23  & full  & \!\!\! 100--800 & 2 & 1024 & 4096 &&  18--23  & full & --        &  - & - & - & 10 \\
\lmax  & wn+1/f  & 18--23  & full  & \!\!\! 100--800 & 2 & 1024 & 4096 &&  18--23  & full & 1.0  & 1024 & T & huw & 10 \\
\nside1024   & wn+1/f    & 18, 23 & full  & 50 & 2 & 64 &  256 &&     18--23 & full & 0.25  & 1024 & F & nw & 50 \\
mask   & wn+1/f    & 18, 23 & full  & 50 & 2 & 64 &  256 &&     18--23 & full & 0.25  & 64 & T & nw & 50 \\
weight   & wn+1/f    & 18, 23 & full  & 50 & 2 & 64 &  256 &&     18--23 & full & 0.25  & 64 & F & huw & 50 \\
base   & wn+1/f    & 18, 23 & full  & 50 & 2 & 64 &  256 &&     18--23 & full & 1.0  & 64 & F & nw & 50 \\

\end{tabular}

\end{table*}

We validated our NCVM products through MC simulations.
A complete list of simulations is given in Table \ref{tab:MC_parameters}.
The simulations followed the same two-step procedure as the high-resolution simulations reported 
in Sect. \ref{sec:highell_simulations}.
We first generated and destriped a noise stream with the Madam destriping code,
and then ran the
 \artdeco\ deconvolution code, to yield an array $a_{Xlm}$ of harmonic coefficients.
The noise and beam parameters are the same as in Sect. \ref{sec:highell_simulations}.
This time we ran the deconvolution step with parameters \lmax=50, \kmax=2,
in line with the settings used in NCVM construction.
With this low an \lmax\ value the deconvolution step is fast, and the computation time is dominated 
by generation of noise rather than deconvolution.
We produced 100 noise realizations for each case studied.
As before, we considered two types of input noise: white and realistic $1/f$ noise.

We cannot expect the NCVM to perfectly model the residual noise, due to simplifications we have to do 
when constructing the matrix, either due to limitations of the formalism, or due to limited computing resources.
We can list four factors:
\begin{enumerate}
\item The covariance matrix assumes destriping resolution of \nside=64, while actual LFI mapmaking uses \nside=1024.
\item  LFI mapmaking involves a destriping mask, neglected by the NCVM. 
\item LFI mapmaking applies horn-uniform weighting in the destriping phase, NCVM computation intrinsically uses noise weighting.
\item NCVM assumes a baseline length of one sample, while the actual baseline length is 1.0 or 0.25 sec.
\end{enumerate}

We performed two series of simulations, which we refer to as ``ideal'' and ``realistic''.
The destriping parameters are given in Table \ref{tab:MC_parameters}.
The realistic simulations mimick the LFI mapmaking procedure as closely as possible:
we use a destriping mask, apply horn-uniform weighting,
and destripe at a high resolution \nside=1024.

In the ideal simulations, the destriping parameters follow the intrinsic assumptions of NCVM.
Destriping is performed at low resolution \nside=64, no destriping mask is applied,
and horn-uniform weighting is replaced by noise weighting. 
To exactly match the assumptions of NCVM computation, we ought to use a baseline length of one sample.
This becomes, however, computationally very demanding.  As a trade-off we selected a baseline length 
of a quarter of the one used in actual mapmaking.

The aim of the ideal simulation set is to verify that the NCVM is constructed correctly.
The realistic simulation set tests how accurately the matrix produced models the actual LFI data products,
given the approximations involved.

\subsection{Chi2 tests}


\begin{figure}
\centering
\includegraphics[width=8.8cm]{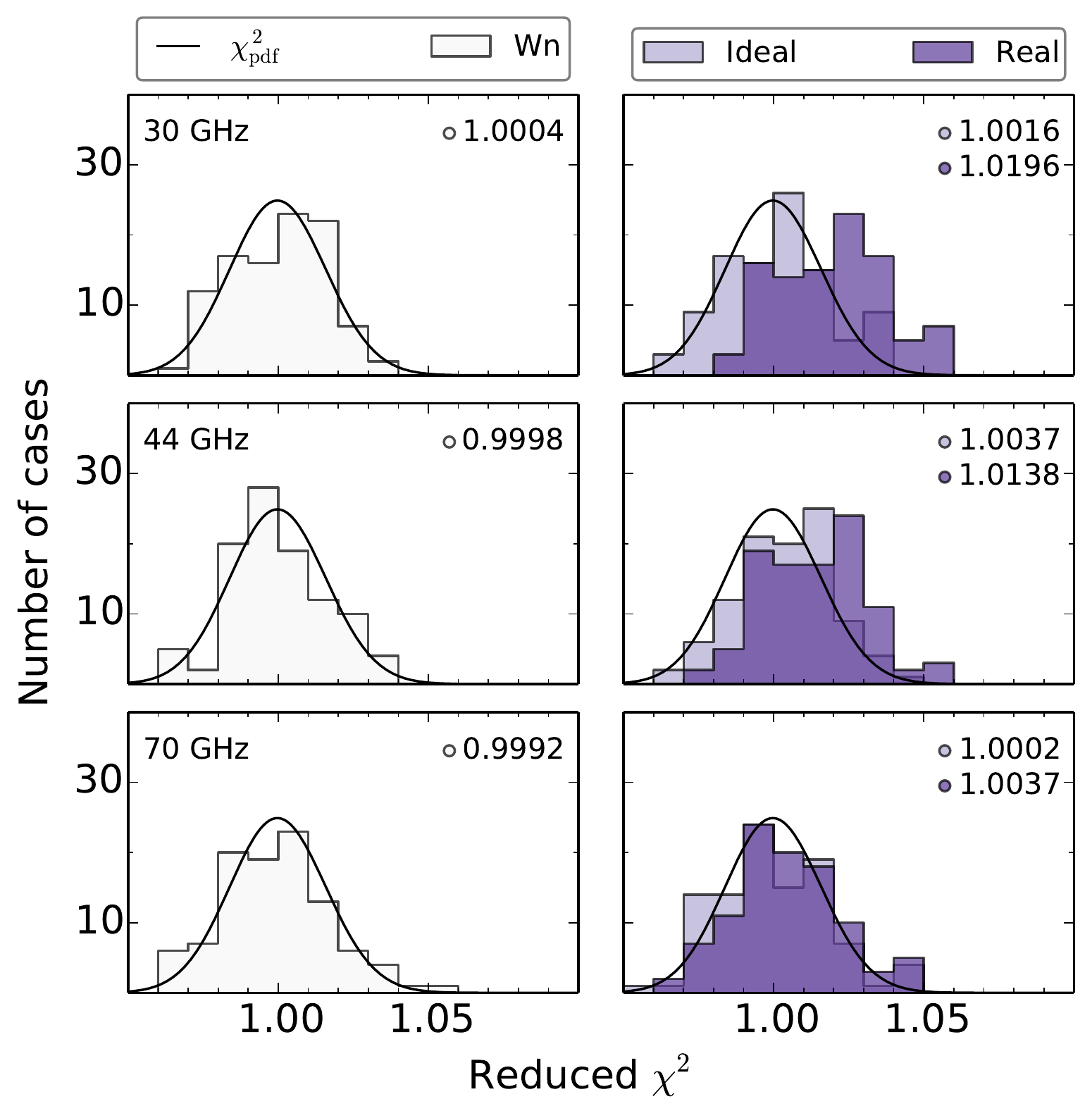}
\caption{Reduced \chid\ distribution for 30 GHz, 44 GHz, 70 GHz,
for white noise (left), ideal, and realistic simulation (right),
for 100 realizations. 
Also shown is the \chid\ distribution corresponding to an ideal result.
The numbers within the plot give the mean value of the distribution.}
\label{fig:chi2_frequency}
\end{figure}


\begin{figure}
\centering
\includegraphics[width=8.8cm]{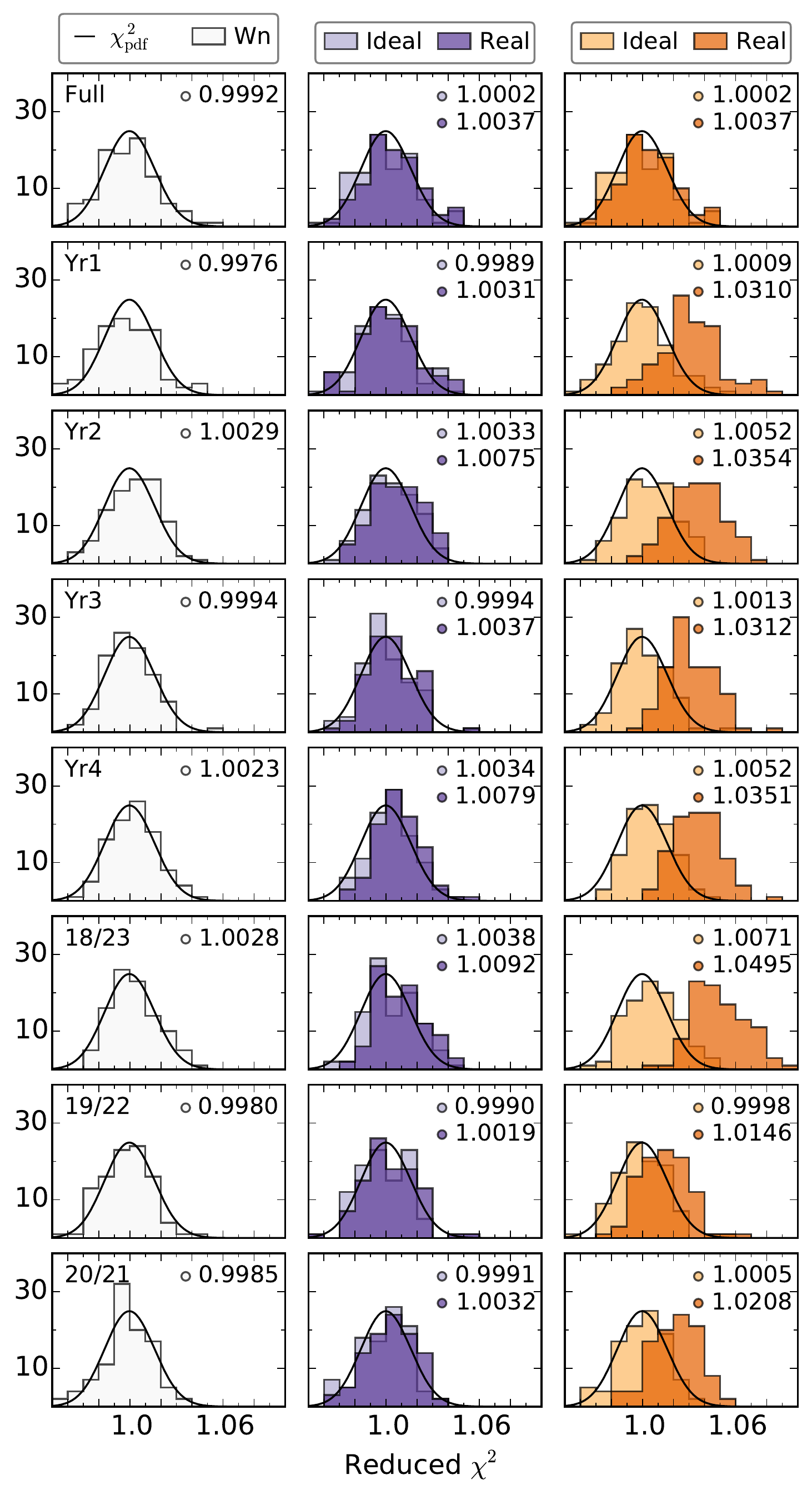}
\caption{Reduced \chid\ distribution for 70 GHz subsets.
From top to bottom: full (4 years) 70 GHz data, years 1 -- 4 (full frequency),
horn pairs 18/23, 19/22, 20/21 (4 years).
From left to right:
white noise only,
independent destriping ideal/realistic,
full destriping ideal/realistic.
 In the case of full mission, independent and full destriping options are identical.
}
\label{fig:chi2_70GHz_partial}
\end{figure}

%
\begin{figure}
\centering
\includegraphics[width=8.8cm]{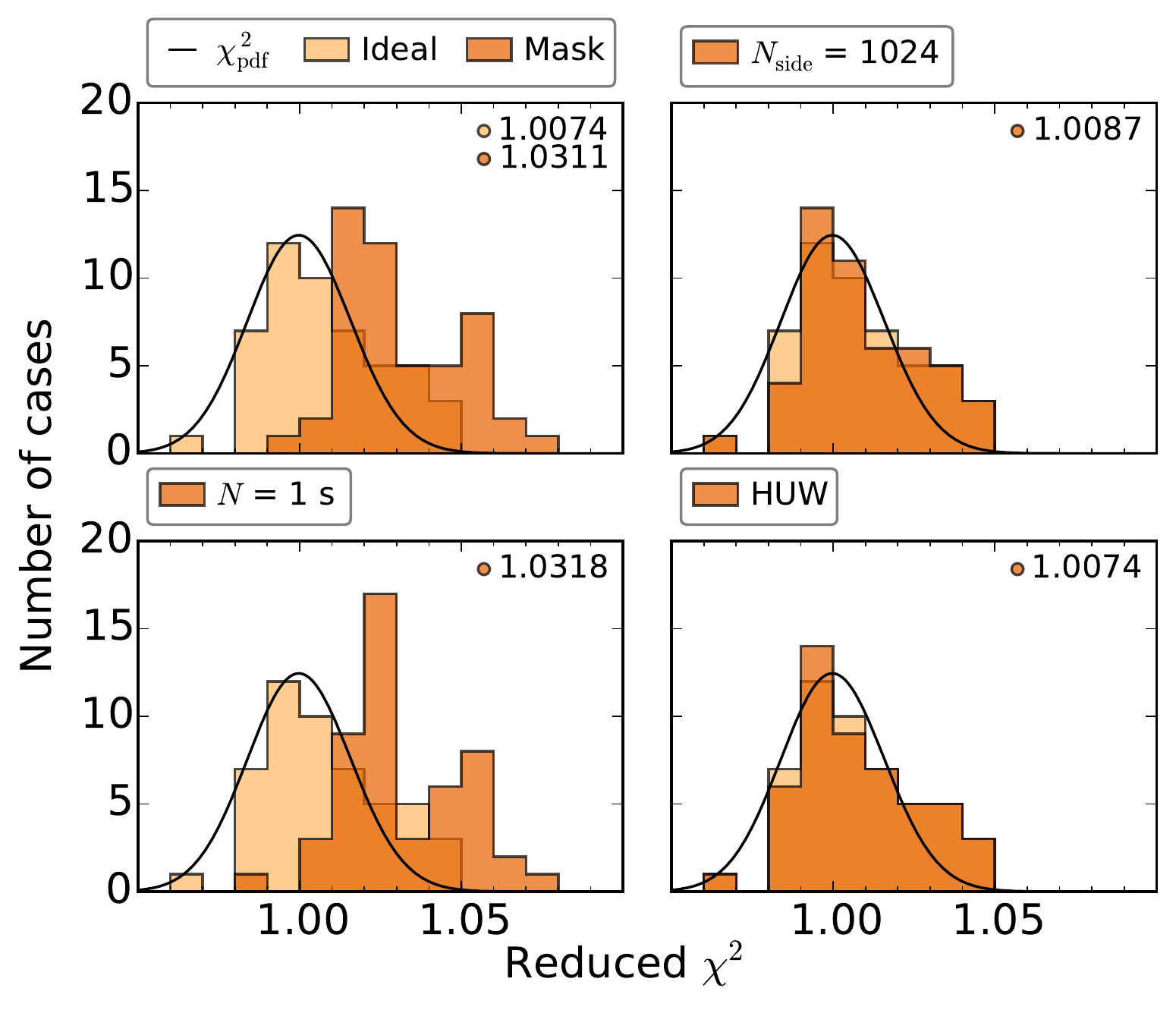}
\caption{Effect of various approximations on the \chid\ result for the worst LFI18/23 case.
We add to the ideal simulation one non-ideality at a time:
destriping mask, horn-uniform weighting, finite baseline length, and destriping resolution.
The result for the ideal simulation case is shown in the background.
}
\label{fig:chi2_worstcase}
\end{figure}

We validate the noise covariance matrices through \chid\ tests.
We generated 100 realizations of of harmonic vectors \axlm,
as described above.
For each vector and corresponding matrix we compute the value
\begin{equation}
\chi^2 =  \frac{\vec a^{\rm T} \tens{C}^{-1} \vec a}{\ndof}  \,.
\end{equation}
Here $\tens{C}$ is the covariance matrix, and \ndof\ is the number of degrees of freedom.
Multipole $\ell$ contributes $2\ell+1$ degrees of freedom, 
except for the monopole and dipole, which are identically zero for $E$ and $B$.
This yields in total \ndof=3(\lmax+1)$^2-8$ degrees of freedom, which for \lmax=50 gives \ndof=7795.
For a matrix that  perfectly describes the noise properties, the results obey the Chi distribution,
For large values of \ndof\ the distribution becomes Gaussian,
with expectation value 1 and standard deviation of $\sqrt{2/\ndof}$,
 which for \ndof=7795 is 0.016. 

We compared the white noise covariance matrix of Eq. (\ref{whitenoise_ncvm}) against white noise simulations, 
and the full noise matrix of Eq. (\ref{artdeco_ncvm})  against ideal and realistic simulations.
Results from tests at the three Planck LFI frequencies, 30, 44, and 70 GHz are shown in Fig. \ref{fig:chi2_frequency}.
We show a histogram of the \chid\ values for 100 realizations, and quote their mean.
The white noise matrix models the white noise residual nearly perfectly, \chid\ mean ranging from 0.9992 to 1.0004,
Comparison between the full matrix and ideal simulation gives slightly larger values,
from 1.0002 at 70 GHz to 1.0037 at 44 GHz, while the expected one-sigma deviation for 100 realizations is 0.0016.
Deviations from one are larger in the case of realistic simulation,
as expected, given the approximations involved.  At 30 GHz we find a \chid\ mean of 1.0196, at 70 GHz 1.0037.

Results for 70 GHz subsets of are shown in Fig. \ref{fig:chi2_70GHz_partial}.
In the first column we compare the white noise matrix against a pure white noise simulation.
Second and third column show results for the full matrix.  Results are shown for two destriping options: 
full 70 GHz destriping, and independent destriping.

In all cases, we find very good agreement with the white noise matrix and corresponding simulations.  
The same can be said about the full matrix and the ideal simulations.
The \chid\ mean is at maximum 1.0038. In the case of realistic simulation, deviations are again larger. 
The agreement is, however, still quite good for cases where deconvolution and destriping involve the same data 
(independent destriping), the maximal \chid\ value being 1.0079.

The case of full destriping, where the whole 70 GHz data set is destriped together,
but only a subset of the cleaned data is provided to deconvolution,
turns out to be more difficult to model with the covariance matrix.
While the ideal simulation still shows good agreement, the mean \chid\ results for the realistic simulation 
range from 1.0146 to 1.0495.
The discrepancy is largest for horn pair LFI18/23, which we now pick up for a closer examination.

We performed yet another series of simulations, where we applied full destriping, and deconvolved the LFI18/23 horn pair.
We took as starting point the ideal simulation, and turned on, one at a time, each of the features
that distinguish the realistic simulation from the ideal one. That is: destriping resolution \nside=1024 instead of 64, 
horn-uniform weighting instead of noise weighting,  baseline length 1 s instead of 0.25 s, and destriping mask.
We generated 50 noise realizations for each case. The results of the \chid\ test for these simulations are shown 
in Fig. \ref{fig:chi2_worstcase}.  We observe that detector weighting and destriping resolution have very little effect 
on the result.
The finite baseline length and destriping mask both contribute significantly.
Both effects alone raise the \chid\ mean to 1.03.

As discussed in Sect. \ref{sec:mask}, the mask affects the residual noise in a way that we cannot model 
in the covariance matrix.  
The effect of baseline length, instead, could be cured by running mapmaking with a shorter baseline.

\subsection{Effect of \lmax}

%
\begin{figure}
\centering
\includegraphics[width=8.8cm]{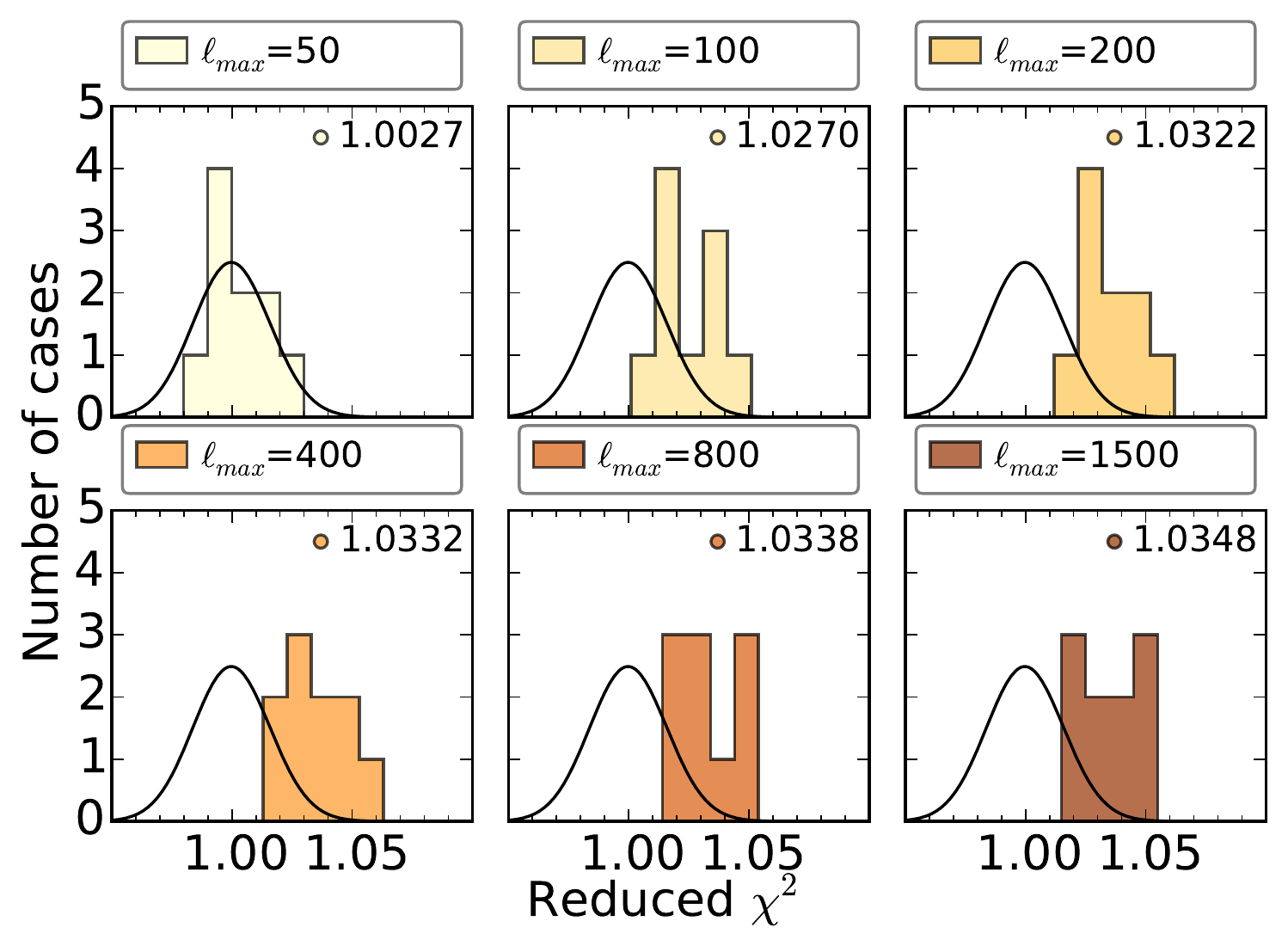}
\caption{Effect of \lmax\ parameter on \chid\ test.
We deconvolve the full 70 GHz data set (10 realizations) with \lmax=50, 100, 200, 400, 800, 1500,
and compare the lowest $\ell$ = 0--50 multipoles with the NCVM with \lmax=50.
}
\label{fig:chi2_res}
\end{figure}

The noise covariance matrix models a deconvolution process where
the harmonic coefficients are solved up to multipole \lmax.
For a good match between the matrix and the harmonic coefficients,
the matrix and the actual deconvolution must use the same value of \lmax.
One might prefer running deconvolution with a high \lmax\ to avoid aliasing effects,
and then pick the lowest multipoles for a low-resolution data set.
This, however, degrades the match between the matrix and the data.

We illustrate this in figure \ref{fig:chi2_res}.
We deconvolved the 10 realizations of our 70 GHz full noise simulation
with different \lmax, pick the lowest $\ell=0-50$ multipoles,
and ran a \chid\  test with the  \lmax=50 NCVM.
The matrix gives a good match with the coefficients obtained with \lmax=50,
but at \lmax=100 there is already a significant deviation.
Going from \lmax=200 to \lmax=1500 changes the result only little.
This is in line with results obtained for white noise, shown in Fig. \ref{fig:bias_resolution}.

It is beyond the scope of this work to answer the question
 if the low \lmax\ can cause signal distortion.
Should this be the case, we still have a few options left.
We can improve the match by constructing the NCVM to higher \lmax.
Computational cost increases rapidly with \lmax, but \lmax=50 used in this work is by no means an upper limit. 
Luckily, figure \ref{fig:chi2_res} suggests that the resolution effect has already largely saturated at \lmax=100,
so increasing \lmax\ even a little may already solve the problem.

Another option worth exploring is to construct the full matrix as combination of two components,
a white noise component,
computed to higher resolution (as we have done in Sect. \ref{sec:whitenoise})
and from a full noise component computed to lower resolution.

\subsection{Validation through noise bias}


\begin{figure*}
\includegraphics[width=18cm]{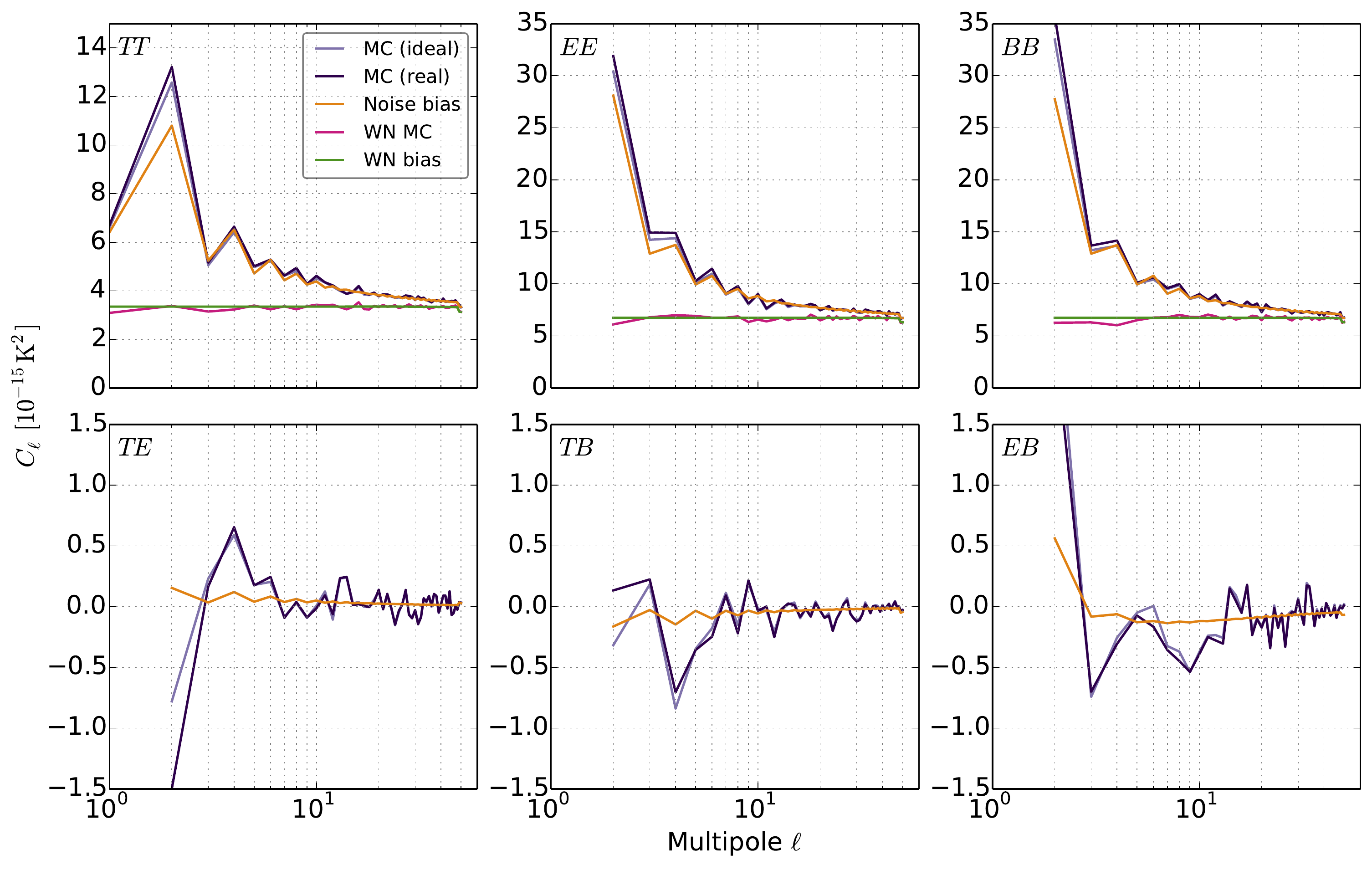}
\caption{70 GHz noise bias compared against MC simulations.
We compare the noise bias from the full NCVM (orange) against ideal (light blue) and realistic (dark blue) simulations.
For the auto spectra we show also the white noise bias (red) and corresponding MC simulations (green).
The MC spectra are averaged over 100 noise realizations.}
\label{fig:bias_validation_70GHz_full}
\end{figure*}


\begin{figure}
\includegraphics[width=8.8cm]{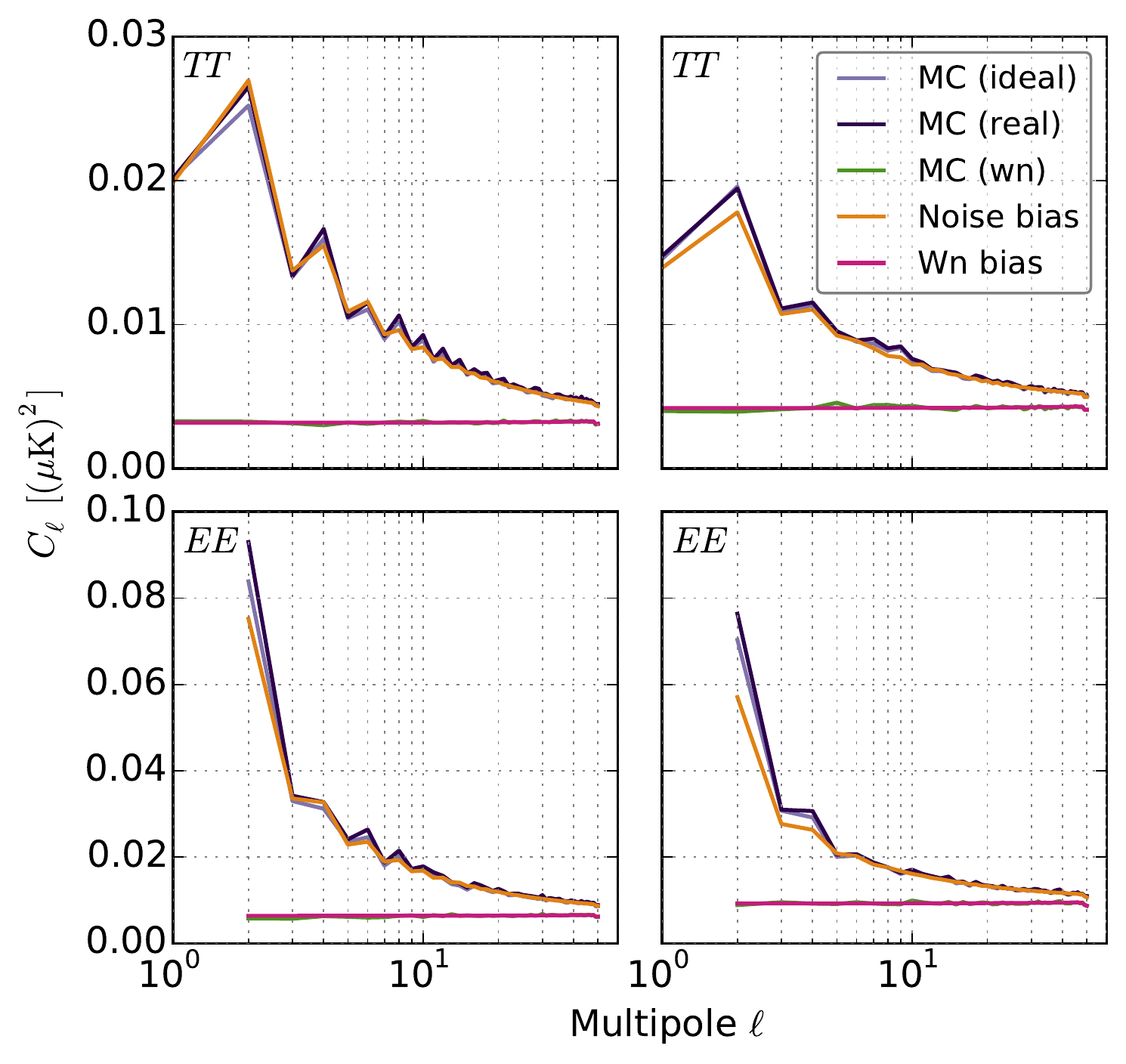}
\caption{Noise bias for 30 and 44 GHz. $TT$ and $EE$ spectra compared against MC simulations.
Line types are the same as in Fig \ref{fig:bias_validation_70GHz_full}).
}
\label{fig:bias_validation_freq}
\end{figure}


\begin{figure}
\includegraphics[width=8.8cm]{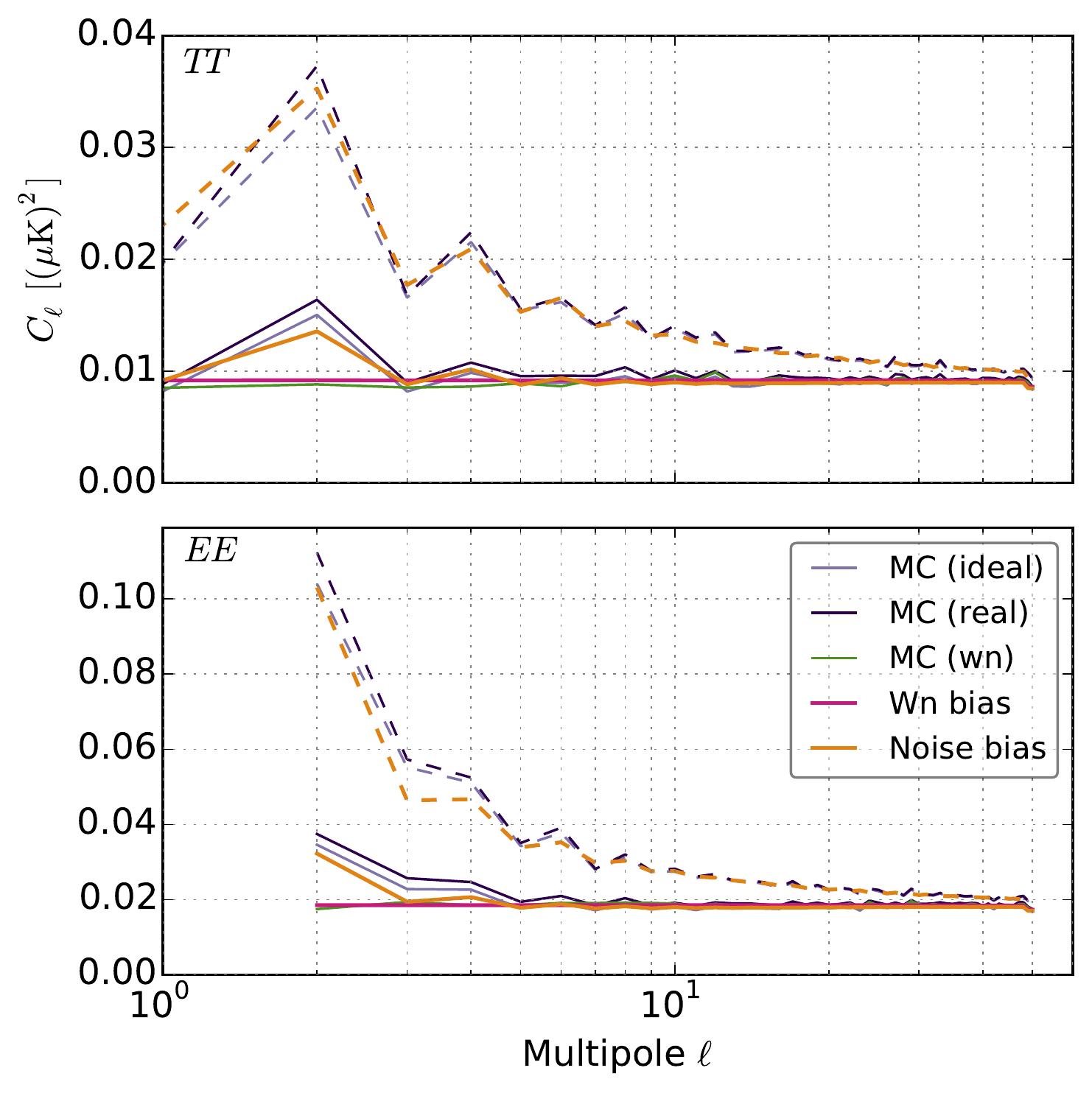}
\caption{Noise bias in $TT$ and $EE$ for horn pair LFI18/23 compared against MC simulations:
{\em Solid}:  full destriping. 
{\em Dashed}: independent destriping
}
\label{fig:bias_validation_LFI1823}
\end{figure}

Another way of validating the NCVM product is to compare the predicted noise bias of Eq. (\ref{noisebias_ncvm})
against the MC spectrum of Eq. (\ref{MC_bias}).
Figure \ref{fig:bias_validation_70GHz_full} shows 
the comparison for the full 70 GHz data.
We show all six spectral components.
The MC spectra shown are averaged over 100 noise realizations.
We show both the idealised and the realistic simulation.
For auto-spectra ($TT$, $EE$, $BB$) we show also the white noise simulation and the corresponding noise bias.

The predicted auto-spectra agree reasonably well with the simulation,
except for multipole $\ell=2$, where the noise bias underestimates the true noise level.  
In the case of the much smaller cross spectra ($TE$, $TB$, $EB$), the 100 realizations
do not provide enough statistics for a meaningful comparison.
In the subsequent analysis we therefore concentrate on $TT$ and $EE$ spectra. 

In Fig.
\ref{fig:bias_validation_freq}
we show  the TT and EE spectra for 30 and 44 GHz channels.
The agreement is again good, and the deviation at $\ell=2$
is smaller than in case of 70 GHz.
The difference between ideal and realistic simulations is small.

Figure \ref{fig:bias_validation_LFI1823} shows
a comparison for horn pair LFI18/23.
We consider again two destriping options.
Simulations show that full destriping yields a much lower residual noise level,
and this is captured by the NCVM as well.

\begin{figure}
\includegraphics[width=0.95\columnwidth]{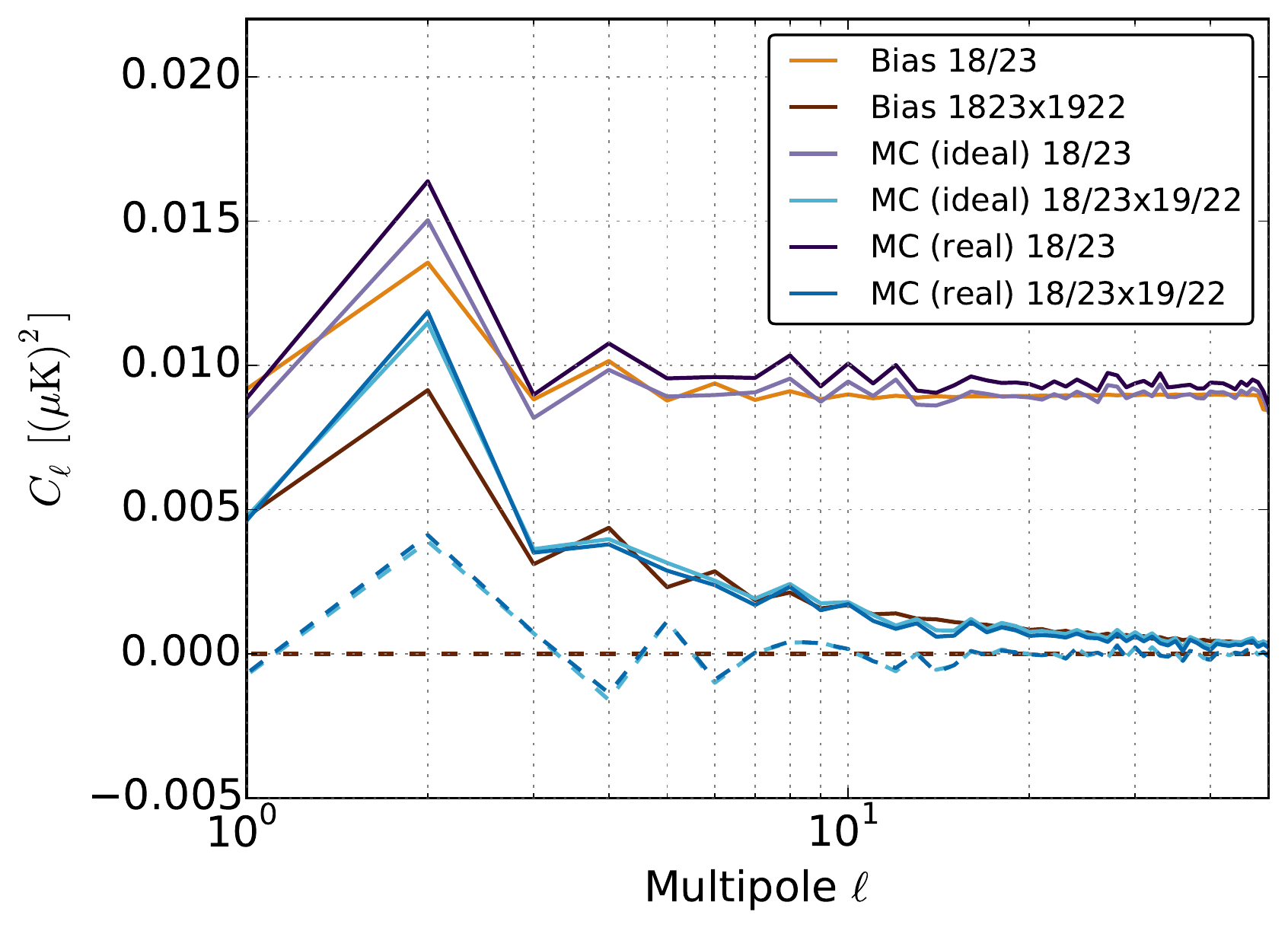}
\caption{$TT$ spectrum for 18/23 and 19/23  (70 GHz) data sets.
Each group of three lines shows a comparison between MC simulations
and corresponding NCVM prediction,
The NCVM prediction is shown in orange or brown, the MC results in blueish colours.
The topmost group shows the auto spectrum of 18/23 (blow-up of Fig. \ref{fig:bias_validation_LFI1823} ).
The group in the middle is the cross-spectrum between 18/23 and 19/22, for full destriping.
The group shown in dashed line type shows the cross-spectra for independent destriping.
}
\label{fig:crossbias}
\end{figure}

Figure \ref{fig:crossbias} shows the cross spectrum between 70 GHz horn pairs 18/23 and 19/22.
When the horn pairs are destriped independently, there is no correlation in the noise.
When the whole 70 GHz data set is destriped together, there is a significant cross-correlation at low multipoles.
The NCVM models this correctly.
We show in the same plot also the auto spectrum for horn pair 18/23 (same as in Fig. \ref{fig:bias_validation_LFI1823}).
The theoretical noise bias agrees well with the idealised simulation, but underestimates the realistic noise by a few per cent,
in agreement with the \chid\ result for the same case.


\section{Covariance in pixel domain}
\label{sec:pixelspace}

The primary output of \artdeco\ is an array of harmonic coefficients.
Consequently, the primary noise covariance matrix is constructed in harmonic space.

From the harmonic coefficients one can further construct a sky map via spherical harmonic transform.
The harmonic coefficients must be multiplied by a smoothing kernel before performing the expansion. 
Smoothing serves two purposes.
Firstly, it removes ringing artefacts that arise around point sources and other regions of high signal gradients.
Secondly, smoothing compensates for the rise in noise level towards high multipoles.


\begin{figure}
\centering
\includegraphics[width=8.8cm]{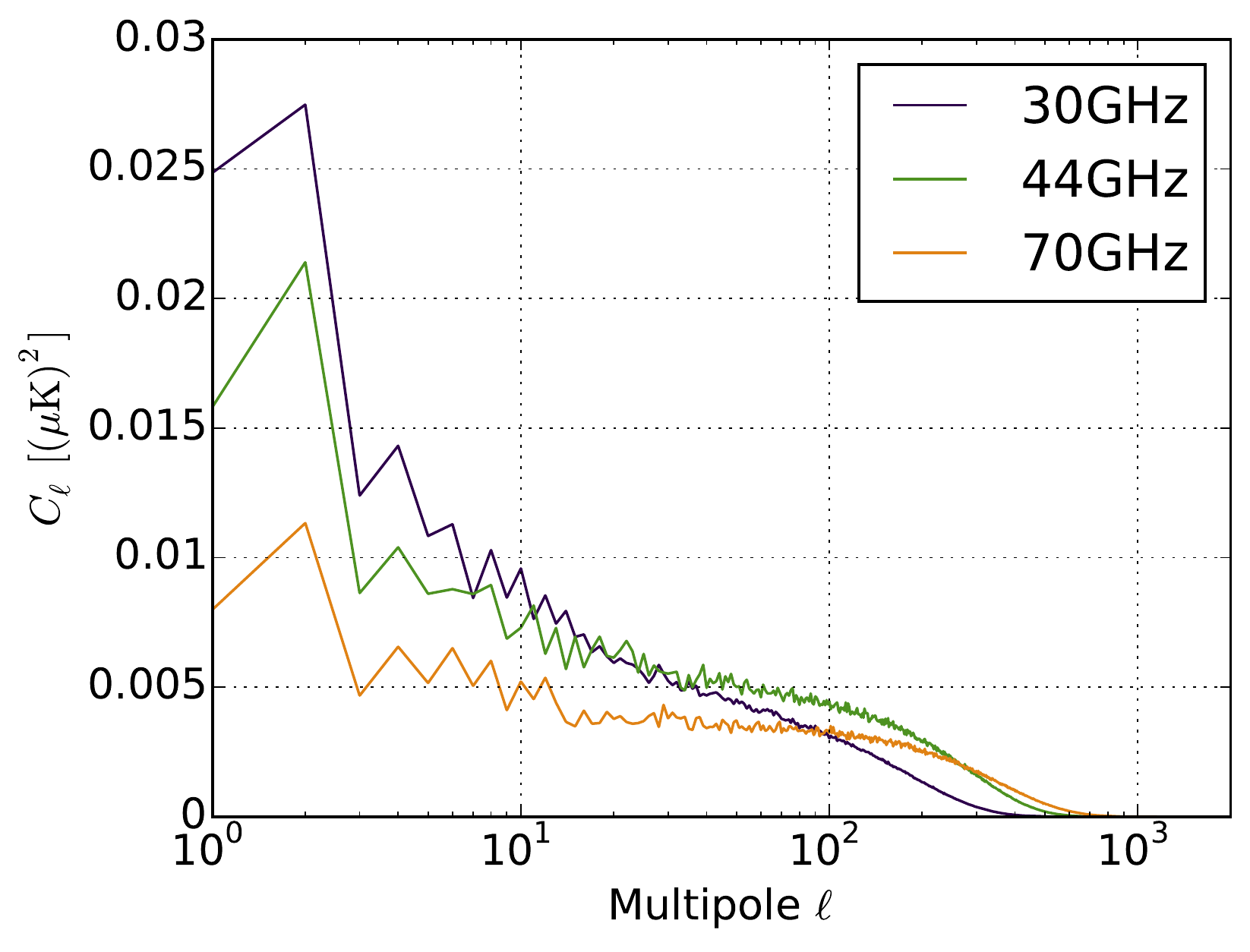}
\caption{Deconvolved $TT$ spectrum after smoothing with a Gaussian beam of FWHM=40' (30 GHz),
30' (44 GHz) or 20' (70 GHz). Smoothing suppresses the noise at high multipoles.
 }
\label{fig:noise_smoothed}
\end{figure}


\begin{figure}
 \begin{center}
 \begin{tabular}{cc}
	\includegraphics[width=40mm]{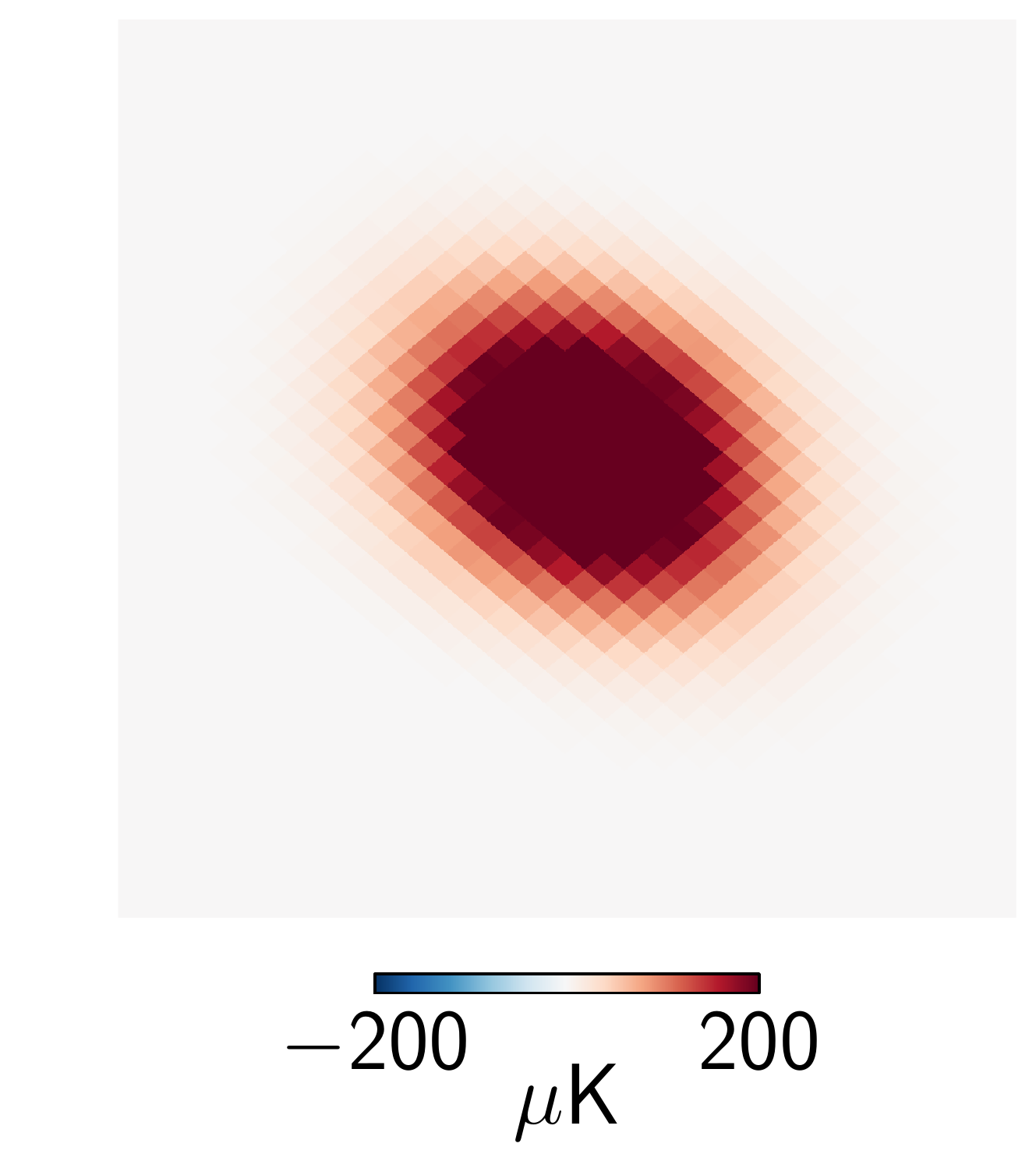} & 
	\includegraphics[width=40mm]{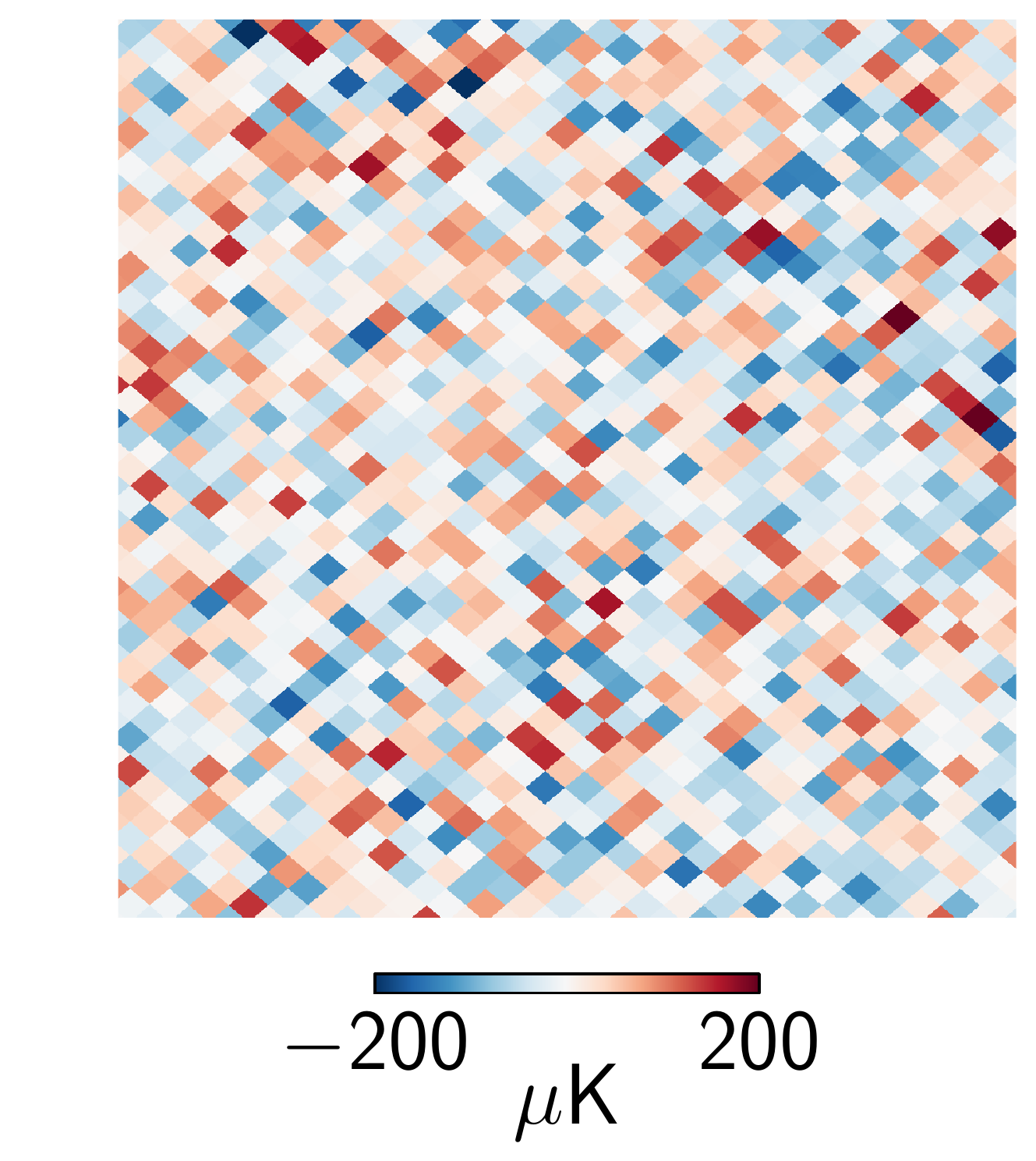} \\
	\includegraphics[width=40mm]{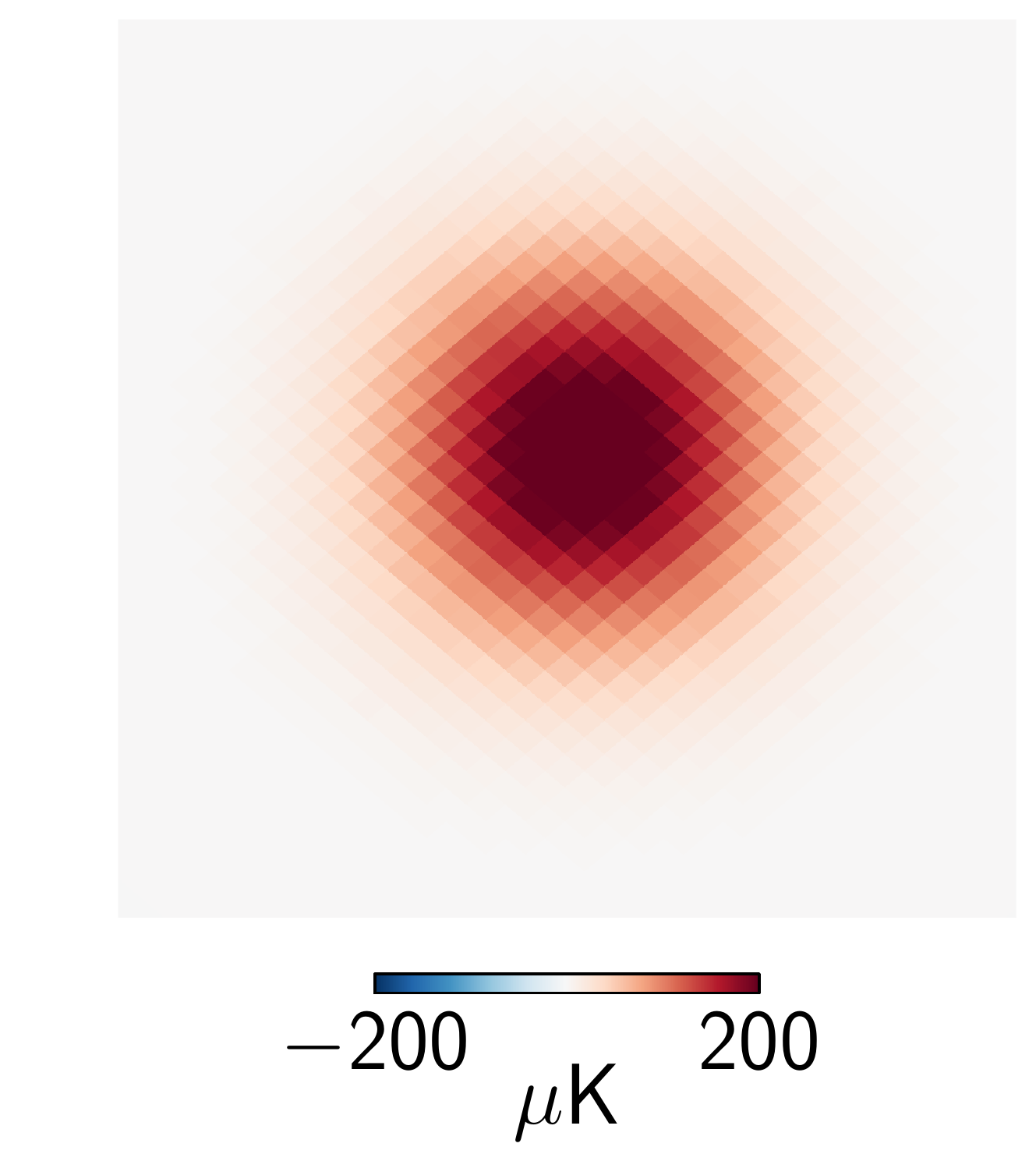} & 
	\includegraphics[width=40mm]{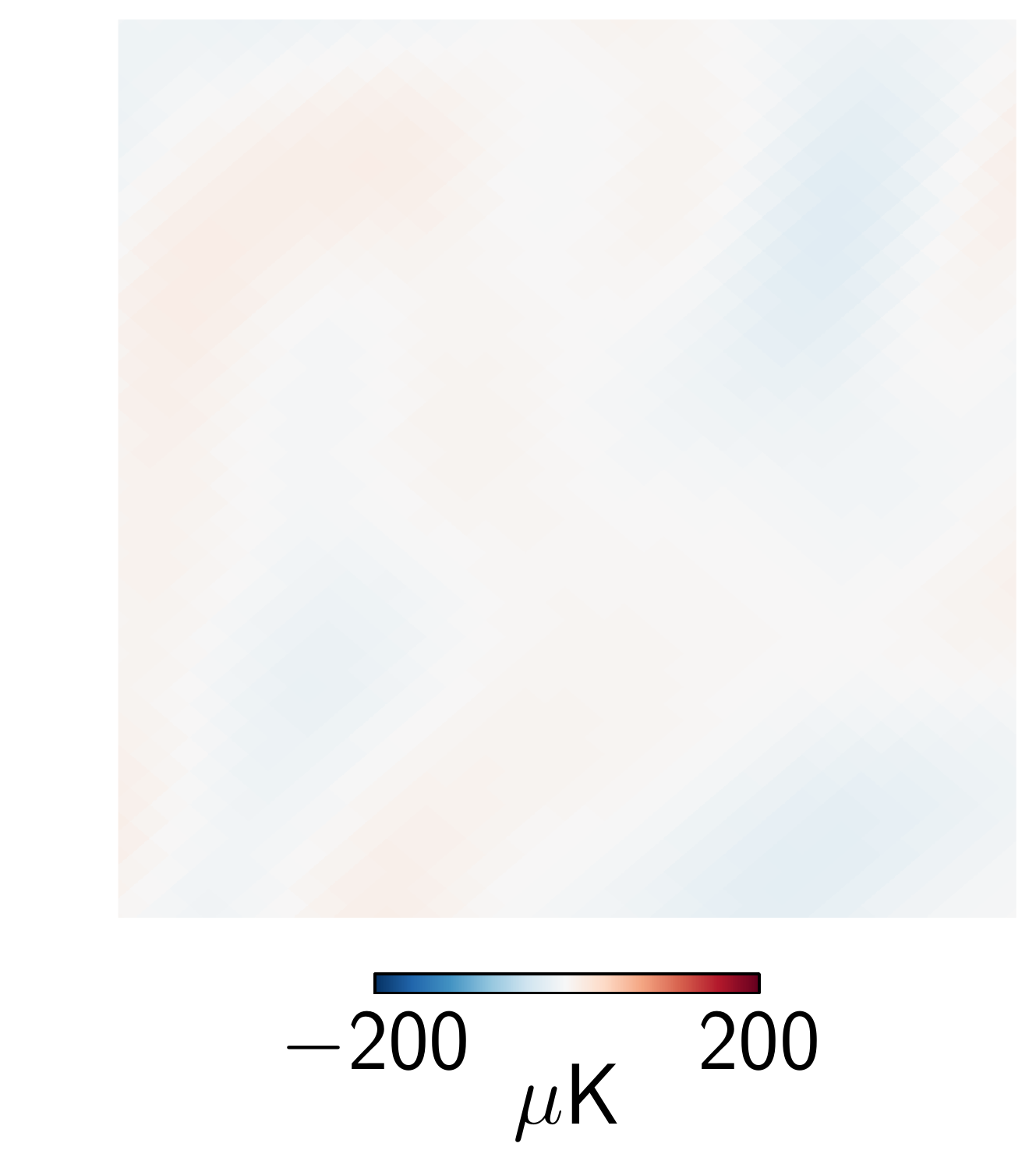} \\
 \end{tabular}
 \end{center}
 \caption{A 1 Jy point source and noise at same location (0.0373$^\circ$,0) in a simulated 30 GHz map.
 We show a patch of 2$^\circ$ around the source.
 Top: binned map.
Bottom:  deconvolved and smoothed to 40'.
Deconvolution stretches noise in direction perpendicular to beam elongation.
Smoothing suppresses the noise amplitude.}
   \label{fig:point_source}
\end{figure}

\subsection{High-resolution maps}

We consider first high-resolution maps.
We do not attempt to build a covariance matrix for high resolution,
but study the noise properties through Monte Carlo simulations.

We transformed our high-resolution simulations to pixel space.
Eliminating ringing artefacts in general requires 
a smoothing width larger than the average width of the actual beam.
The same conclusion is supported by noise analysis.
As shown in Sect. \ref{sec:whitenoise},
the deconvolved noise spectrum rises more steeply than the inverse symmetrized beam window.
Smoothing by this window will thus not be sufficient.

We applied a Gaussian smoothing kernel of width
FWHM=40' (30 GHz), 30' (44 GHz), or 20' (70 GHz).
The widths were chosen so as to remove ringing effects at each respective frequency.
Figure \ref{fig:noise_smoothed} shows the $TT$ noise spectra
with smoothing applied.  All spectra vanish at high multipoles.

Figure \ref{fig:point_source} illustrates the effect of deconvolution on noise in the pixel domain.
We show a simulated 1 Jy point source at 30 GHz, located just above the origin of the Galactic coordinate system. 
In order to have the source in the centre of an \nside=1024 pixel we place it not at the exact Galactic centre,
but at (0.0373$^\circ$,0) (lat,lon), which is the centre of the pixel just above. 
We show also as example one realization of noise at the same location.
The noise in the binned map is dominated by white noise.
The smoothing operation associated with deconvolution suppresses the noise level in the deconvolved map.


\begin{figure}
\centering
\includegraphics[width=40mm]{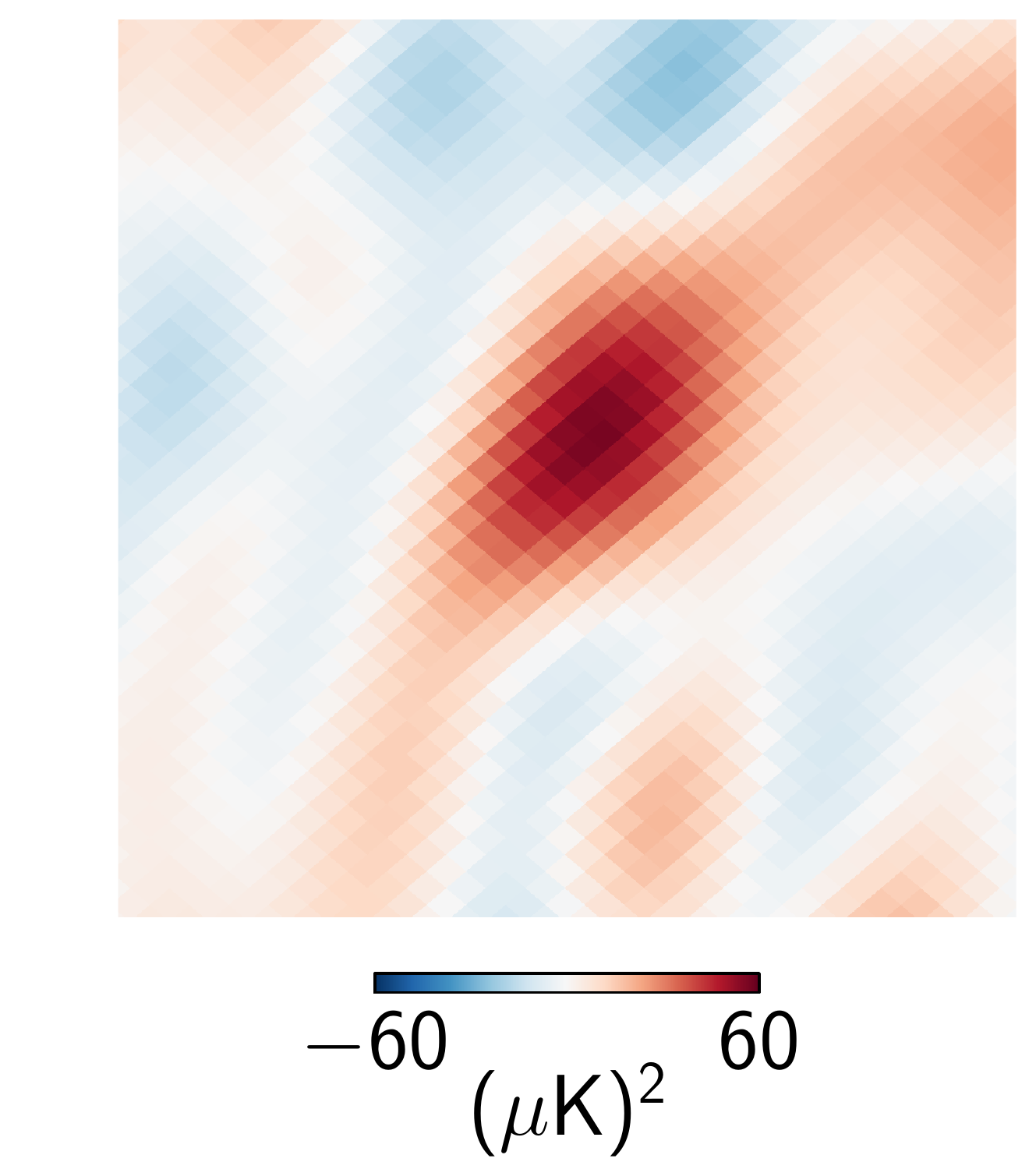}
\caption{Correlation structure between the central pixel and its surroundings, 
computed from 40 realizations.
We show the same sky region as in Fig. \ref{fig:point_source}}
\label{fig:pixel_noise}
\end{figure}

Owing to the smoothing, the image of a point source on a deconvolved map is not point-like.
It is, however, perfectly symmetric.
At the same time the deconvolution procedure generates correlations in the noise component,
in the direction perpendicular to the direction of beam main axis.
To quantify this, we compute an estimate of the correlation of the central pixel with its surroundings,
as
\begin{equation}
C_{{\rm map}XYj} = \langle n_{Xi}n_{Yj} \rangle
\end{equation}
where $X,Y=I,Q,U$ are the Stokes components, $n_j$ is the noise in pixel $j$, 
index $i$ represents the reference pixel, and $j$ labels all other pixels of the sky.
We evaluate $C_{\rm map}$ as an average over the 40 available  noise realizations.
The $II$ component map is plotted in Fig. \ref{fig:covmap}.
As expected,  noise is correlated in a direction at right angles to the beam main axis.

\subsection{Low-resolution maps and pixel NCVM}


\begin{figure}
\centering
\includegraphics[width=8.0cm]{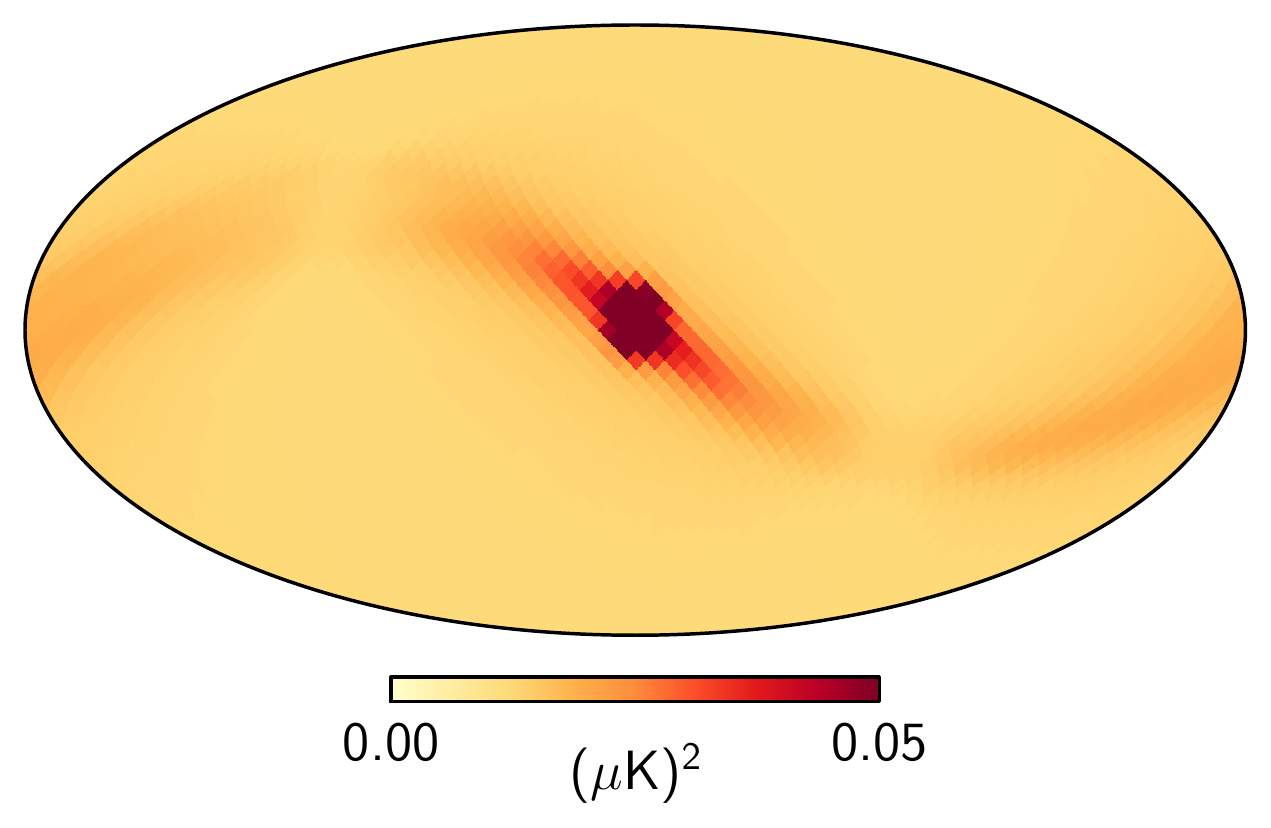}
\caption{70 GHz $II$ correlation between a reference pixel located at (2.388$^\circ$,0)
and the rest of the sky.
 }
\label{fig:covmap}
\end{figure}

We proceed to produce low-resolution maps, and related covariance matrices.
We transformed the low-resolution simulations of Sect. \ref{sec:validation} to pixel space.
We adopted the smoothing procedure that was applied to LFI 2015 release data \citep{planck2014-a07}.
We applied Gaussian smoothing with FWHM=440'  and constructed maps at resolution \nside=16 (220').

The same operations must be applied to the covariance matrix.
The smoothing operation reduces the effective number of degrees of freedom,
making the covariance matrix singular or badly scaled.
This is cured by adding a small amount of white regularization noise 
within each pixel of the map.
A corresponding constant variance was added on the diagonal of the covariance matrix.
Following the LFI procedure, we added white regularization noise with rms=2 \muK\ ($I$) or 
rms=0.02 \muK\ ($Q$ and $U$) on each pixel.

The operation on map can formally be written as
\begin{equation}
  \vec m=H \vec a  +\vec n_{\rm reg}\label{map_transform}
\end{equation}
Here $H$ represents the harmonic expansion combined with smoothing,
$\vec a$ is the vector of harmonic coefficients, as produced by \artdeco, 
$\vec m$ is the beam-deconvolved map,
and $\vec n_{\rm reg}$ represents regularization noise.

The noise in map of  Eq. (\ref{map_transform}) is
characterized by the covariance matrix
\begin{equation}
  \tens{C}_{\rm pix} = H \tens C H^{\rm T} +\tens{N}_{\rm reg} \label{pixmatrix}
\end{equation}
where $\tens{C}$ is the harmonic covariance matrix,
and $\tens{N}_{\rm reg}$ is diagonal matrix representing the regularization noise.

The operation $H \tens C H^{\rm T}$ is computationally intensive and broken down into the following individual steps:
\begin{enumerate}
\item Each row of $\tens C$ is transformed via SHT. This is done using the efficient \texttt{libsharp} implementation \citep{reinecke2013}.
Since this library only supports real-valued maps, two SHTs (for the real and imaginary part, respectively) must be carried out for each row;
depending on the parameter choice, these are either scalar or fully polarized. 
Since a matrix row is located entirely on a single MPI task, no communication is necessary for this step.

\item To perform the analogous operation on every column of the intermediate result $\tens C H^{\rm T}$,
this matrix must be effectively transposed in memory, so that columns reside on a single MPI task., as the rows did before. 
This is a communication-intensive step.

\item SHTs are carried out over the rows of $(\tens C H^{\rm T})^{\rm T}$ in a fashion analogous to step 1, resulting in
$\tens{C}_{\rm pix}^T$.

\item In another communication step, this matrix is again transposed.
\end{enumerate}

Figure \ref{fig:covmap} illustrates the structure of the pixel covariance matrix.
We show as a map one column of the 70 GHz covariance matrix,
which represents $II$ correlation between the reference pixel and the rest of the sky.
We pick the reference pixel just above the galactic centre.
The reference pixel is centred at (2.388$^\circ$,0)
and has pixel number 1440 in the ring pixeling scheme of Healpix at resolution \nside=16.

The reference pixel is strongly correlated with the neighbouring pixels.
A weaker correlation structure follows the scanning direction.
The map has a significant positive minimum of roughly 0.012 $\mu{\rm K}^2$,
reflecting the difficulty of distinguishing the map monopole from a global noise offset.
The correlation structure may vary a lot according to the reference pixel chosen.
The plot shown here represents one example.

\subsection{{\bf $\chi^2$ tests in pixel space}}


\begin{figure}
\centering
\includegraphics[width=8.8cm]{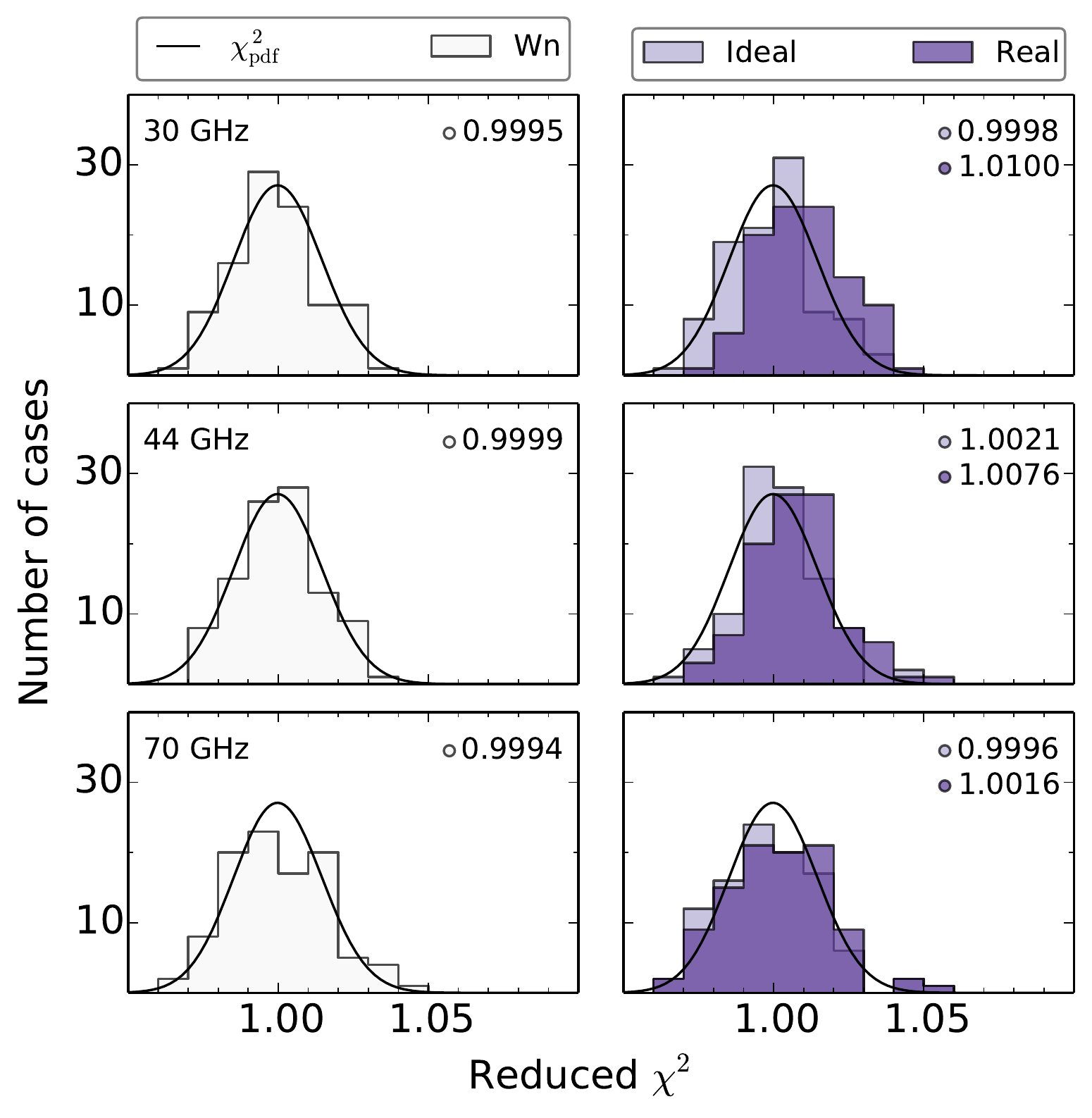}
\caption{Chi2 test in pixel space, for 30,44,70 GHz.}
\label{fig:chi2_frequency_pixel}
\end{figure}

\begin{figure}
\centering
\includegraphics[width=8.8cm]{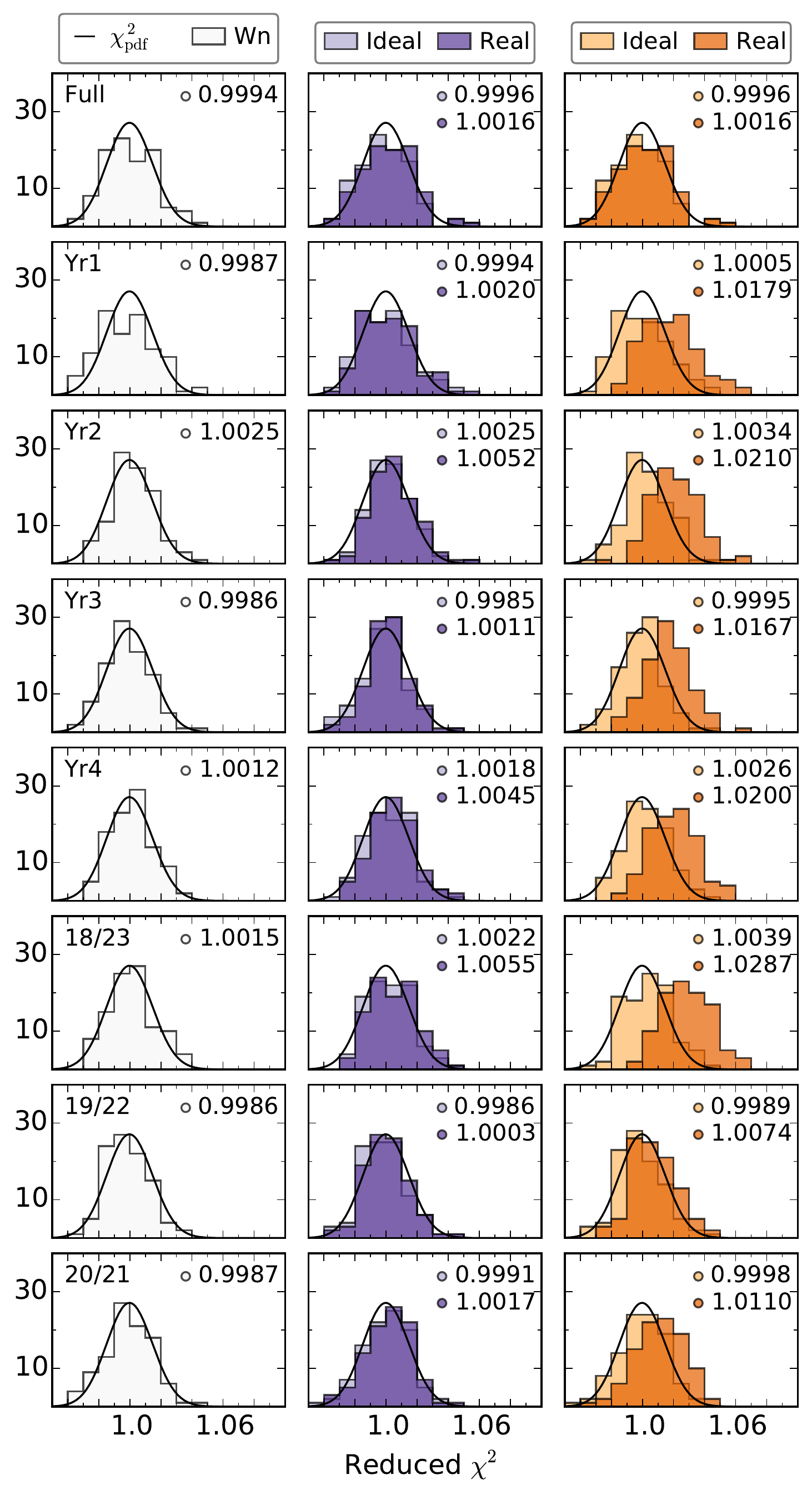}
\caption{Chi2 test in pixel space, for 70 GHz subsets.
{\em Left}: white noise. {\em Middle}: ideal and realistic noise, independent destriping. {\em Right}: ideal and realistic noise, full destriping.}
\label{fig:chi2_grid_pixel}
\end{figure}

We validate the pixel covariance matrices
through \chid\ tests, similarly to the validation tests presented in Sect. \ref{sec:validation}.
The number of degrees of freedom is now given by the number of pixels in the Healpix map,
times 3 Stokes components, which for \nside=16 gives $N_{\rm dof}$=9216.
Consequently, the one-sigma statistical variation becomes 0.0147,
and rms of mean for 100 realizations becomes 0.00147.

As in Sect. \ref{sec:validation}, we compared the white noise covariance matrix against white noise simulations,
and full covariance matrix against ideal and realistic simulations.
The \chid\ results for full four-year frequency maps and for 70 GHz subsets
are shown in Figs. \ref{fig:chi2_frequency_pixel} and \ref{fig:chi2_grid_pixel}.

The mean reduced  \chid\ values for pixel covariance are in general closer to one than 
in the case of harmonic covariance.  
This is the effect of regularization noise.
We are adding to the maps a noise component that is accurately
modelled by the extra term in covariance matrix,
thus improving the agreement with the matrix and the maps.

The \chid\ mean is 1.01 for 30 GHz 4-year data, 1.0076  for 44 GHz,
and 1.0016 for 70 GHz. The latter is only slightly larger than the 
one-sigma statistical variation.
For 70 GHz subsets the values range from 1.0017 to 1.0287.

\subsection{Pre-whitened noise}

The power of the \chid\ test is limited in the sense
that it collapses a complicated noise structure into a single number,
which is not constrained from below.
It is easy to conceive a situation where two errors cancel to produce a \chid\ 
result close to one.
For instance, a simple overestimation of the noise rms in TOI domain
leads to an artificially low \chid\ result.

We complemented the validation tests by analysis of pre-whitened noise.
The types of tests we performed here require a real-valued data set.
For this reason we can perform the test in pixel space only.

Consider a pre-whitened noise map, computed as
\begin{equation}
\vec x = \tens{M}^{-1}\vec m
\end{equation}
where $\vec m$ is the noise map, 
and $\tens{M}$ is the ``square root'' of the covariance matrix $\tens{C}$, for which
$\tens{C}=\tens{M}\tens{M}^T$.  Matrix $\tens{M}$ is can be constructed in multiple ways.
We took it to be the  Cholesky decomposition of $\tens{C}$.
If matrix $\tens{C}$ accurately describes the properties of the noise map $\vec m$,
$\vec x$ will be a vector of uncorrelated Gaussian noise with unit variance, $\langle \vec x\vec x^T\rangle =\tens{I}$.

We build histograms of values of $\vec x$ and compare them to theoretical Gaussian distribution,
in a fashion similar to the analysis presented in \cite{Stompor2002}.
We combine the 100 noise realizations into the same plot to improve statistics.
The variance of the distribution yields the usual \chid\ test.
Comparison of the full distribution against the expected Gaussian distribution
provides a more stringent test.

\begin{figure}
\centering
\includegraphics[width=8.8cm]{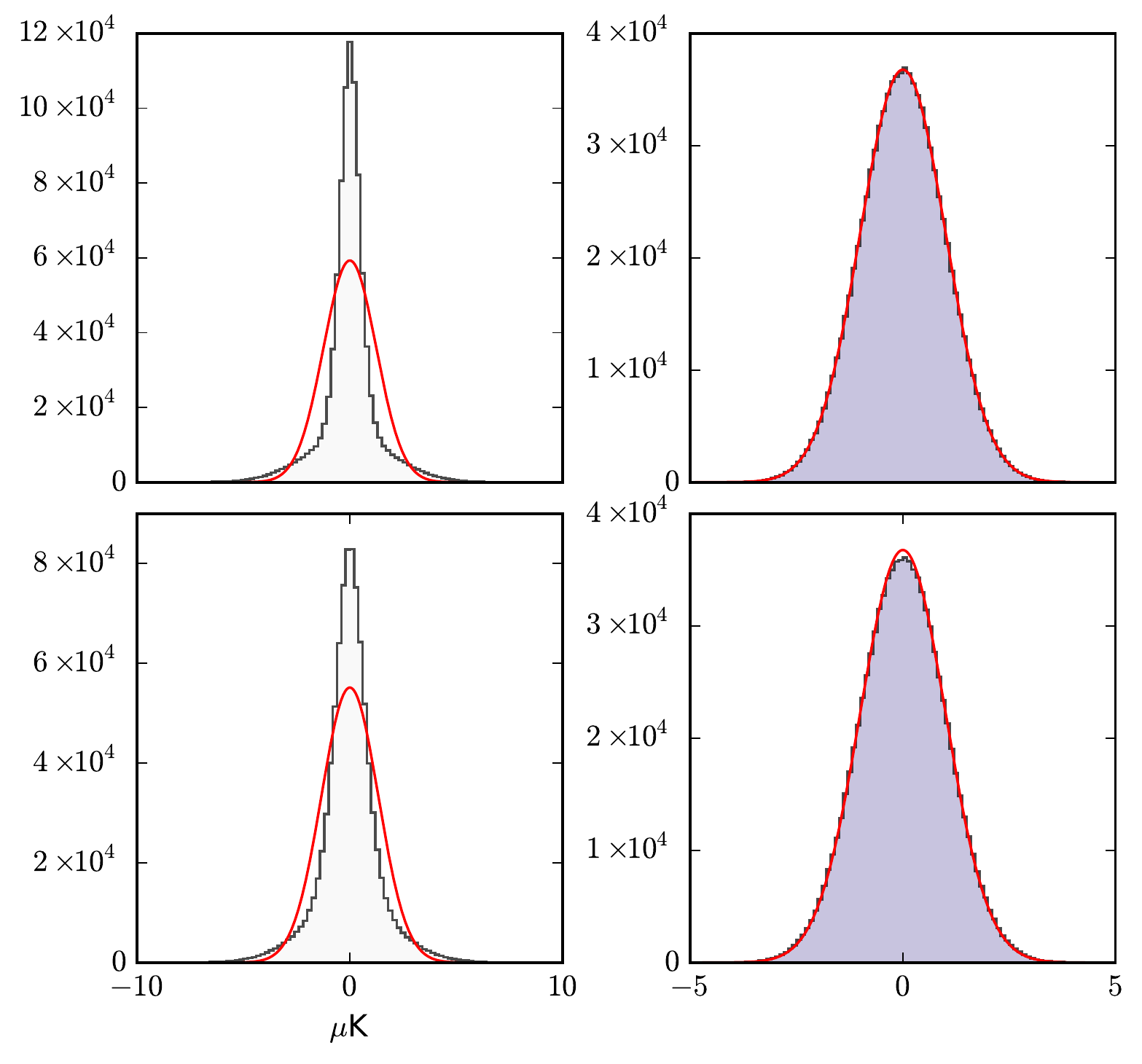}
\caption{Distribution of noise map values before ({\em left}) and after ({\em right}) whitening
by application of NCVM.  The red curve on the left gives the gaussian distribution with stdev equal to that of the data.
On the right, the red curve gives the Gaussian distribution with unity variance,
the expected ideal result.
{\em Top}: 70 GHz full mission, realistic simulation.
{\em Bottom}: Horn pair LFI18/23, full destriping, realistic simulation.
}
\label{fig:prewhitened_noise}
\end{figure}

We show the distributions for two selected simulation cases  in 
Fig. \ref{fig:prewhitened_noise}, both with realistic noise.
The first example is one where the NCVM models the noise well, the 70 GHz frequency map. 
The second example is the worst case from earlier tests, the LFI18/23 horn pair map with full destriping.
In both cases, noise before whitening is very non-uniform.  We show for comparison the Gaussian distribution with
variance equal to that of the data. The variance is dominated by the high tails of the distribution,
leading to a broad distribution that does not fit the data anywhere.
On the right we show the pre-whitened noise, together with the ideal Gaussian distribution with unit variance.
In the case of 70 GHz frequency map, the distribution of whitened noise follows the Gaussian closely,
indicating that the NCVM models the noise well.
In the horn pair case there is a visible deviation. There are no high tails in the distribution,
the deviation manifests rather as a overall broadening of the distribution.
In fact, the distribution matches very well a Gaussian shape with standard deviation of 1.015.

To be more quantitative, we performed Kolmogorov-Smirnov test
on the pre-whitened noise. 
We constructed the cumulative distribution of the values of $\vec x$,
and found the maximum difference with respect to the Gaussian distribution.
The probability to exceed (PTE) values computed according to Kolmogorov-Smirnov
statistics are given in Table \ref{tab:kolmogorov}. The number of degrees of freedom is $N_e=921600$ 
(100 maps of 9216 pixels each).
In general, the K-S test yields the same conclusions as the \chid\ test. In white noise and ideal simulation cases the NCVM
models the noise well.  The worst match is found for 70 GHz subsets where we combine realistic simulation
 with full destriping.   The 18/23 horn pair again stands out as the worst case.
 The full frequency maps and submaps with independent destriping are modelled well.
 We can identify a couple of case where the \chid\ test indicates a better match than the K-S test.
 In particular, for horn pair 19/22 with independent destriping and realistic simulation we get  the excellent result \chid=1.0003,
 but only 0.01 in the K-S test. This is likely an example of a case where two opposite errors compensate in the \chid\ test.

The distribution of pre-whitened noise for horn pair 18/23, shown in Fig. \ref{fig:prewhitened_noise},
indicates that the error made in NCVM computation when ignoring non-idealities of realistic data,
results in an overall underestimation of the noise,
and can to some degree be accounted for scaling the NCVM by an appropriate constant.
To test this hypothesis, we performed another round of K-S tests where we rescaled the pre-whitened noise $\vec x$ by its stdev value,
bringing the variance to unity.  This is equivalent to scaling the NCV matrix by the variance of $\vec x$.
The required scaling factor is directly given by the \chid\ result,
and the \chid\ result for this modified test is by construction exactly one.
The K-S test can still reveal discrepancies in the shape of the noise distribution.
The PTE values from the modified test are given in the last column in Table  \ref{tab:kolmogorov}.
We see that a simple rescaling of the covariance matrix leads to a much better agreement between the matrix and the maps.


\begin{table*}
\caption{Mean \chid\ values and PTE (probability to exceed) values from Kolmogorov-Smirnov test,
comparing the pixel-space NCVM against 100 MC realizations.
We give results 
for 30 GHz, 44 GHz, and 70 GHz full mission frequency maps,
and for various 70 GHz subsets (for  full and independent destriping options,
for pure white noise, ideal and realistic simulations).
For the rightmost column we rescaled the NCVM
by the inverse of \chid\ and re-run the K-S test for realistic simulation.  
The \chid\ result for this case is by construction exactly one.
}
\label{tab:kolmogorov}
\begin{center}
\begin{tabular}{l rr   rr rr rr}
  &  \multispan{2} white &  \multispan{2} ideal &  \multispan{2} realistic &  \multispan{2} realistic rescaled\hfil \\
\hline\hline
  \noalign{\vskip 4pt}
  Simulation &\null\quad\qquad\chid\ & PTE & \null\quad\qquad\chid\ & PTE & \null\quad\qquad\chid\ & PTE & \null\quad \chid\ & PTE \cr
  \hline
   \noalign{\vskip 4pt}
 \multispan{4}  Frequency maps\hfill & \hfil\cr
  \noalign{\vskip 4pt}
30 GHz & 0.9995  & 0.879 & 0.9998 & 0.020 & $\qquad$ 1.0100 & 0.003 & 1 & 0.033 \cr
44 GHz &  0.9999 & 0.602 & 1.0021 & 0.896 & 1.0076 & 0.110 & 1 & 0.947  \cr
70 GHz & 0.9994 & 0.311 & 0.9996 & 0.463 & 1.0016 & 0.594  & 1 & 0.583 \cr
  \noalign{\vskip 4pt}
\hline
  \noalign{\vskip 4pt}
\multispan{4} 70\,GHz subsets, full destriping\hfil\cr
  \noalign{\vskip 4pt}
Yr1 & 0.9987 & 0.110 &  1.0005  & 0.043 &  1.0179 & 252$\cdot10^{-6}$  & 1 & 0.067 \cr
Yr2 & 1.0025  & 0.077 & 1.0034   & 0.883 &  1.0210 & 5$\cdot10^{-6}$    & 1 & 0.995 \cr
Yr3  & 0.9986 & 0.814 &  0.9995  & 0.924 & 1.0167  & 39$\cdot10^{-6}$   & 1 & 0.984  \cr
Yr4  & 1.0012 & 0.108 &  1.0026  & 0.271 &  1.0200  & 33 $\cdot10^{-6}$  & 1 & 0.257 \cr

18/23  &  1.0015  & 0.185 & 1.0039 & 0.117 & 1.0287 & 4.4$\cdot10^{-12}$  & 1 &   0.436 \cr
19/22  & 0.9986  & 0.065  &  0.9989 & 0.014  & 1.0074  & 496$\cdot10^{-6}$   & 1 &   0.015 \cr
20/21   &  0.9987 & 0.399 & 0.9998  & 0.740  & 1.0110  & 0.002  & 1 &  0.745  \cr

  \noalign{\vskip 4pt}
\hline
   \noalign{\vskip 4pt}
\multispan{4} 70\,GHz subsets,  independent destriping\hfil\cr
  \noalign{\vskip 4pt}

Yr1   & 0.9987 &  0.110 & 0.9994  & 0.074 &  1.0020  & 0.012  & 1 & 0.018 \cr
Yr2   & 1.0025 &  0.077 & 1.0025  & 0.275 & 1.0052  &  0.211  & 1 & 0.410 \cr
Yr3   & 0.9986 & 0.814  & 0.9985   &  0.914 & 1.0011  & 0.955   & 1 & 0.977 \cr
Yr4   & 1.0012 & 0.108  & 1.0018  &  0.358 & 1.0045  & 0.465  & 1 &   0.419 \cr

18/23  &  1.0015 &  0.185 &  1.0022 &   0.236 &  1.0055  & 0.076  & 1 & 0.368  \cr
19/22  & 0.9986  &  0.065 &  0.9986 &  0.016  & 1.0003  & 0.010   & 1 & 0.012\cr
20/21  &  0.9987 & 0.399  &  0.9991 &  0.428  &  1.0017  & 0.180   & 1 &  0.332\cr

\hline\hline
\end{tabular}
\end{center}
\end{table*}


\section{Conclusions}
\label{sec:conclusions}

We performed an analysis of noise properties of beam deconvolved maps,
both in the harmonic and pixel domain.
The analysis is built on the formalism of the \artdeco\ deconvolution code. 
As the starting point we have a situation where a correlated $1/f$ type noise stream
is first destriped to remove the correlated component as accurately as possible
and then fed as input to the deconvolver. 

The main result of this work is the low-resolution noise covariance matrix;
this matrix describes the residual noise
in the harmonic coefficients which is the primary output of \artdeco.
The covariance matrix models the contribution of
correlated residual noise that is left in the data after destriping
and which contributes strongly to the residual at lowest multipoles.
The covariance matrix also captures the effect of the deconvolution process
itself on the noise.
From the harmonic covariance matrix we further constructed a pixel-pixel covariance matrix,
which describes the noise properties of deconvolved low-resolution maps.

As a practical application 
we computed covariance matrices for the \Planck\ LFI four-year mission,
using real LFI beams and noise spectra, and reconstructed detector pointings.
We constructed the matrices up to multipole \lmax=50.
We constructed further pixel-pixel covariance matrices at resolution \nside=16.

We compared the covariance products against Monte Carlo simulations.
For validation purposes we performed a series of ideal simulations,
which closely follow the intrinsic assumptions of the covariance matrix.
The mean reduced \chid\ values in harmonic space range from 0.9991 to 1.0071,
while the one-sigma statistic variation is 0.0016.
In pixel space the agreement is even better, ranging from 0.9986 to 1.0039.

The covariance matrices are not expected to perfectly model realistic LFI noise
due to the simplifications we have to do in matrix construction.
We performed another series of realistic simulations 
to find out how accurately the matrices model real LFI data.
The mean reduced \chid\ values in harmonic space 
are 1.0037 for 70 GHz, 1.0138 for 44 GHz, and 1.019 for 30 GHz.
In pixel space the corresponding numbers 
are 1.0016 (70 GHz), 1.0076 (44 GHz), and 1.01 (30 GHz).

Various subsets of 70 GHz data gave \chid\ results ranging from 1.0003 to 1.0495.
We studied more closely the case with the largest discrepancy,
and found that the main contributors to the discrepancy are the use of a Galactic mask,
and finite baseline length, both related to destriping.
The mask is required in LFI analysis to control the signal error,
but our NCVM does not model its effect correctly.
The inaccuracy arising from baseline length
could be removed by running the destriping step with a shorter baseline.

We considered two destriping options.  In full destriping the full data 
of a frequency channel is destriped together, and a subset of it used for deconvolution.
In independent destriping we use the same data subset in both processing steps.
The NCVM is found to model the residual noise well in the latter case.
The full destriping option yields a lower general noise level, with the side effect that
the systematic error sources become more important.
Consequently, the agreement between the NCVM and the data is poorer
and the largest \chid\ values are obtained with this option.
Fortunately, although the full destriping option yields the lowest residual noise for a particular submap,
the independent destriping option is more important from the point
of view of cosmological analysis. Noise covariance matrices are required input
in many methods of cosmological parameter estimation.
These methods typically input either the full frequency maps,
or a set of submaps with independent noise,
which we produced using the independent destriping option. 

We complemented the validation tests with Kolmogorov-Smirnov tests
in pixel space. We used the NCV matrix to pre-whiten the simulated noise maps,
and test the hypothesis that the whitened noise is indeed white.
We observe that in cases where the NCVM underestimates the noise,
the discrepancy is in most cases, at least to a first approximation,
well modelled by an overall rescaling of the matrix.
The value of the optimal scaling factor remains to be determined from simulations.

It is instructive to compare the \chid\ results for low-resolution maps
to similar tests performed on the official LFI data products,
as presented in \cite{planck2014-a07}.
The official products include pixel-pixel covariance matrices at resolution \nside=16.
\chid\ results from for full mission frequency maps are 1.042 (30 GHz), 1.027 (44 GHz),
and 1.010 (70 GHz).  For the same data sets we obtain 1.01, 1.0076, and 1.0016, respectively.
Although our tests indicate a better agreement, the results must be interpreted with care.
Our simulation procedure mimics that of official LFI pipeline.
We applied same smoothing width (440'), the same target resolution, and the same amount of regularization noise.
There are, however, also important differences.  The results of \cite{planck2014-a07}
are based on 10000 realizations with a time-dependent noise model, while in our simulations 
the noise properties remains the same throughout the mission.
A reliable comparison would require tests performed on the same simulation data,
and NCVM constructed with the same noise model.
The FFP8 simulations used by \cite{planck2014-a07} are available in the form of high-resolution maps.
For deconvolution, however, we need alternatively the TOI, or the 4D map objects constructed from it,
which are not available

At higher multipoles the residual noise is dominated by the white noise component of TOI, deformed by the deconvolution process.
Our simulations show that deconvolution leaves relatively less residual correlated noise on top of 
the white component at high multipoles than does the ordinary binning operation.
 From this it follows that (with the exception of 30 GHz where the correlated residual is still large)
 the full noise bias at high multipoles can be estimated to high accuracy from the white noise covariance matrix,
 which is cheaper to compute than the full covariance.
 Also, white noise simulations are much cheaper computationally than full noise simulations,
which require the destriping step.
A possible future development is to equip the deconvolution code with an internal noise generator,
which fills the input data structure with white noise.  This would 
avoid the bottleneck of writing the simulated TOI on disk,
thus allowing extensive MC simulations.

We computed the white noise covariance to \lmax=200.
At higher multipoles we constructed partial matrices that cover a narrow multipole range around the region of interest.
The noise bias obtained this way underestimates the noise spectrum obtained from Monte Carlo simulations by 1 -- 3\%.
Should higher accuracy be required, the recommended approach is to estimate 
the bias through Monte Carlo simulations.
That will require optimization of the MC pipeline, as described above.

In cases where a subset of available data is used for deconvolution,
there are two options for destriping.
One can either destripe the same data set,  or the full data set.
The second option significantly reduces the level of residual correlated noise,
but also leaves the different subsets correlated.
We derive a covariance matrix that describes the level of intercorrelation,
and validate it on simulations involving 70 GHz horn-pair data sets.

Finally we examined the noise properties in deconvolved high-resolution maps at a qualitative level.
We show that the residual noise is correlated in the direction perpendicular to the beam axis.
The overall noise level is low owing to the smoothing operation involved in map construction.


\begin{acknowledgements}
Some of the results in this paper have been derived using the HEALPix
\citep{gorski2005} package. 
MR is supported by the German Aeronautics Center and Space Agency (DLR), under
program 50-OP-0901, funded by the Federal Ministry of Economics and
Technology. 
EK is supported by Academy of Finland grant 253204.
VL and KK are supported by Academy of Finland grant 283497.
A.~S. has been supported by V\"ais\"al\"a (2014) and Wihuri (2015) foundations.
We acknowledge that the results of this research have been achieved using the PRACE-3IP project 
(FP7 RI-312763) resource Sisu based in Finland at CSC.
We thank CSC -- the IT Center for Science Ltd.\ (Finland) -- for computational resources.
We thank H.~Kurki-Suonio and J.~V\"aliviita for careful reading of the manuscript and helpful comments.
We thank E.~Palmgren for help with plots.
\end{acknowledgements}

\bibliographystyle{aa}
\bibliography{Planck_bib}

\end{document}